\documentclass[]{paper}
\input{amssym}
\usepackage{graphicx}
\begin{document}

\pagestyle{empty}

\title{{\Huge New Insights into Initiation of Colon and Intestinal Cancer:The Significance of Central Stem Cells in the Crypt}}
\author{ Ali Mahdipour--Shirayeh $^{1, \bigstar}$ and Leili Shahriyari $^{2, \bigstar}$\\
\vspace{0.1in}
$^{1}$ {\small Biomedical Research Group, Applied Mathematics Department, University of Waterloo, ON N2L 3G1, Canada}\\
\vspace{0.1in}
$^{2}$ {\small Mathematical Biosciences Institute, The Ohio State University, OH, USA}\\
\vspace{0.1in}
$^{\bigstar}$ {\small This authors contributed equally to this work.}}

\date{}

\maketitle
\textbf{\textit{}}

\section*{Abstract}
Rapidly dividing tissues, like intestinal crypts, are frequently chosen to investigate the process of tumor initiation, because of their high rate of mutations. To study the interplay between normal and mutant as well as immortal cells in the human colon or intestinal crypt, we developed a 4-compartmental stochastic model for cell dynamics based on current discoveries. Recent studies of the intestinal crypt have revealed the existence of two stem cell groups. Therefore, our model incorporates two stem cell groups (central stem cells (CeSCs) and border stem cells  (BSCs)), plus one compartment for transit amplifying (TA) cells and one compartment of fully differentiated (FD) cells. However, it can be easily modified to have only one stem cell group.
We find that the worst-case scenario occurs when CeSCs are mutated, or an immortal cell arises in the TA or FD compartments. The probability that the progeny of a single advantageous CeSC mutant will take over the entire crypt is more than $0.2$, and one immortal cell always causes all FD cells to become immortals.Moreover, when CeSCs are either mutants or wild-type (w.t.) individuals, their progeny will take over the entire crypt in less than 100 days if there is no immortal cell. Unexpectedly, if the CeSCs are wild-type, then non-immortal mutants with higher fitness are washed out faster than those with lower fitness. Therefore, we suggest one potential treatment for colon cancer might be replacing or altering the CeSCs with the normal stem cells.

\section*{Author Summary}
We use human cell division rate, which is similar to mouse, to develop a model for cell dynamics of colon and intestinal crypts. Because of the lack of data for human crypts, we use the mouse experimental observations to develop the model. The model indicates that the probability of the progeny of one disadvantageous mutant, like P53$^{R175H}$, stem cell taking over the entire normal crypt is small, but if it happens it occurs fast ($<1$ month). The probability of one advantageous mutant, like APC$^{-/-}$, stem cell's progeny taking over the entire crypt is high, but it is a slow process (it needs $>50$ days). If central stem cells (CeSCs) are wild-type, then non-immortal mutants will be washed out in $<100$ days. Although a single wild-type CeSC is able to wash out all P53$^{R175H}$ mutants with a probability of $0.27$ in two months, it is unable to sweep out all APC$^{-/-}$ mutants. At least two normal CeSCs are required to wash out all APC$^{-/-}$ mutants. In agreement with experimental observations, the model completely captures cell dynamics in the mouse intestinal crypt. If human crypts contain two stem cell groups like mouse crypts, then the results are also valid for human crypts.


\section*{Introduction}

Colon cancer is the second and third most common cancer respectively in women and men all over the world [4]. In most colorectal cancers, mutation occurs over the patient's lifetime, and it is not inherited. The most frequent mutation in colon cancer is the inactivation of the tumor suppressor gene Adenomatous polyposis coli (APC) [5]. Moreover, the absence of APC immediately perturbs Wnt, which causes aberrant migration [6]. Another tumor suppressor gene that becomes inactivated in many cancers as well as in some colon cancers is p53. There is evidence that p53 inactivation leads to multiple mutations during a single cell cycle, like chromothripsis [7]. Chromothripsis is a massive genomic rearrangement occurring only on a single chromosome or a few chromosomes during a single cell cycle [8]. Moreover, in the majority of colorectal carcinomas, the loss of a large portion of chromosomes 17p and 18q has been observed [9]. These massive changes on the cell's genomes can evolve the cell to become immortal like the Hela cell line [10]. In addition, the over-expression of polycomb ring finger oncogene BMI1 transforms epithelial cells and influences telomerase functionality that can lead to cell immortalization [11-13].

Telomeres, repetitive nucleotide sequences at each end of all chromosomes, get slightly shorter with each normal cell division until they shorten to a critical length. This leads to cell aging and ultimately to apoptosis, or cell death [58]. Normal cells have a maximum number of divisions, i.e. Hayflick limit, before these telomeres are depleted.
Rare mutations or transformation events, highly associated with telomerase activity, allow cells to scape from the first mortality phase checkpoint (M1 or Hayflick limit) and the second M2 crisis checkpoint and become immortal [53]. The majority of cancer cell lines and cancer biopsies contain telomerase activity or evidence of alternative mechanism for lengthening of telomeres (ALT), thus targeting telomerase becomes an attractive anti-cancer therapeutic option [54].
The death rate of the immortal cells in the normal homeostasis is almost zero in the time--scale of tumorigenesis, and they are resistant to the chemotherapy and radiotherapy [14,15].  Moreover, the metastatic potential of immortal cells is higher compered to the normal epithelial cells [11].

Immortal cells might be the result of dedifferentiation from non-stem cells to cancer stem-like cells through bidirectional conversions in stroma. Cancer stem-like cells can only exhibit symmetric divisions and presumably acquire the ability of metastasis as an aggressive malignancy [12].
The cancer stem cell hypothesis is one of the two established mechanisms in carcinogenesis where the main focus is on internal heterogeneity of cells within the population [12]. As stem cells in the colonic or intestinal crypt generate all epithelial cells of the crypt through a cellular hierarchy, the cancer stem cell hypothesis can be a convenient approach to investigate the development of cancer.
Although there are debates about the concept of cancer stem cells, there is no doubt on the existence of immortal cells [16]. Therefore, we develop a general model that is able to accommodate possible immortal cells, which never die in the model, to grow within the progenitor or fully differentiated cell compartments. Upgrading the existing models with mutational events and spatial structures [17-25] the general and relevant mechanism in colorectal/intestinal cancer will be studied in this paper.

More precisely, we model the cell dynamics of the intestinal crypt based on the available experimental data sets. We investigate the spread of one initial mutant at different locations within the crypt. Furthermore, the proposed model accommodates the possibility that a mutant cell becomes immortal in a single cell division. In this model, we vary the fitness of mutants, which defines the probability that a mutant cell divides and replaces its neighbor cells. We assume that the fitness of all w.t. cells is 1, and the relative fitness of mutants is $r$. In other words, when there are $j$ mutants and $i$ w.t. cells competing to divide and fill out the available empty space, the probability that a mutant cell divides is $\frac{rj}{rj+i}$.

Recently, the probability $P_R$ that a mutant stem cell replaces its neighbor for various mutants was empirically obtained [26]. Based on this mouse experiment, the fitness of mutant APC$^{+/-}$ is 1.6, while the fitness of mutant APC$^{-/-}$ is 3.8.  Moreover, the fitness of the dominant-negative hotspot P53$^{R172H}$ mutant, which corresponds to the human hotspot P53$^{R175H}$, has been obtained.
The P53$^{R172H}$ protein has been shown to inhibit the w.t. p53 function, and tumors expressing p53$^{R172H}$ are more metastatic than tumors deleted for p53 [27]. Surprisingly, the fitness of P53$^{R172H}$ mutants is $0.9$ in the normal colon, while the fitness of the P53$^{R172H}$ mutant is $1.4$ in the inflammatory environment. Thus, we consider a range of fitness values for $r$ such that the mutants can be disadvantageous ($r<1$), neutral ($r=1$), or advantageous ($r>1$) compared to w.t. cells.
We consider the expansion of mutant or marked cells presented at the initial time of simulations. For simplicity, we assume all mutants in the tissue have the same fitness $r$. We also assume that when a mutant cell divides, one of its children is immortal with a very small probability.

Several computational models have been developed to study the dynamics of multistage carcinogenesis [28-35]. Moreover, there are many computational and mathematical models investigating crypt cell dynamics [36-42]. However, there are several recent experimental studies on animal colon/intestinal crypts that reveal new information about the cell dynamics. Therefore, new mathematical models are required to accommodate the latest experimental discoveries. Note, because of the inability to use the cell fate mapping experimental techniques in humans, the available data for the human colon/intestinal crypt cell dynamics is very limited. Therefore, in order to obtain more realistic computational models for humans, we need to incorporate the available observed data from animal studies.

 Many cell dynamics models designed for the intestinal and colon crypts, because of their fairly simple cell dynamics comparing to the other tissues, i.e one directional movement from bottom to the top of the crypt [50]. The cell dynamics of the intestinal crypt are very similar to the colon crypt.
One of the first models is a homogenous model developed in 1992 to obtain cell division rates at each location of human crypts as well as cell cycles [46].
The time and probability of the progeny of neutral stem cells taking over the stem cell niche and the entire normal intestinal crypts have been modeled in [50]. This stochastic model, which treats all stem cells as one stem cell type, is in perfect agreement with their experimental data, and it predicts stem cells mostly divide symmetrically.
Bravo et al. [41] also provided an agent-based computer simulations for cell dynamics in normal human colon crypts.

Although there are many mathematical models for cell dynamics of normal crypts, several models have been designed to investigate the process of mutants' production and their dynamics.
Zhao and Michor [51] developed a one-dimensional homogeneous model, which includes only one column of cells, to track mutants in the crypt.  They found that most divisions should occur at the bottom of the crypt in order to maximize the time to cancer. In contrast, a recent two-dimensional model, which only contains two columns of cells,  shows that most divisions should occur at the top of the crypt to delay cancer [52].
In [51], at each time step, a cell at position $i$ is selected to divide, two daughter cells are then placed into positions $i$ and $i+1$, causing cells that previously resided in positions $i+1$ to shift by one position toward the top of the crypt.
The difference, which accounts for the discrepancy between the results of [51] and [52], is that the model developed by Shahriyari et al. [52] considers the probability of two-hit mutant
production from wild type cells. Instead, [51] starts with an APC$^{+/-}$ mutation (a 1-hit mutant) at a
given position and calculate the time to produce a APC$^{-/-}$ mutation (a 2-hit mutant). The
predictions of [51] agree with [52] when they consider the probability of second mutation production conditioning on the existence of 1-hit mutants.
Additionally, Mirams et al. [39] developed a computer model for cell dynamics of colonic crypts to obtain the probability of mutants taking over or washing out from the crypt. In this model, cells are defined by the location of their centers, and cell movement was determined by assuming that each cell exerts a linear spring force on its neighbors. They used an equation of motion to determine the centre of each cell at each time.

Tissue cells as well as colon and intestinal cells are commonly divided in three categories; stem cells (SCs), transit amplifying (TA), and fully differentiated cells (FD). Colonic and intestinal stem cells, which are located at the bottom of the crypt, generate themselves and TA cells. TA cells, which are the most dividing cells in the crypt, give rise to the specialized differentiated cells that fulfill the physiological functions of the intestine [3]. Sometimes in the response to an injury, TA cells that have some potential characteristics of stem cells  can re-obtain some stem cells' activities to regenerate the crypt [3]. Here, we do not model this scenario, because we investigate normal cell dynamics of crypts with no injuries.

Recently, it has been observed that the intestinal crypts contain two stem cell compartments: border stem cells (BSCs) and central stem cells (CeSCs) [24]. BSCs, which are located between CeSCs and transit amplifying (TA) cells, have high potential for differentiation; in contrast, the CeSCs, which are located at the very bottom of the crypt, are biased towards proliferation.
The first computational cell dynamics model, which considered the existence of two stem cells groups, was developed by Shahriyari et al. [47] to obtain the probability of second-hit mutant production in the stem cell niche.
Here, by generalizing that model, we develop a unique 4-compartmental stochastic model for the colon and intestinal crypt, which incorporates new discoveries; the existence of two stem cell groups. At the initial time of simulations, we place mutant/marked cells at one of these compartments and we calculate the probability and the time that their progeny take over the crypt or are washed out from the crypt.
The results of our model are in complete agreement with the experimental data obtained by Snippert et al. [50] and Ritsma et al. [24].  We obtain the values of the model's parameters from available published experimental data sets. Remarkably, we find that if only CeSCs are w.t. and all other cells in the crypt are mutants (but not immortal), then in less than hundred days, all mutants are washed out from the crypt. These results suggest that if we want to change the fate of the crypt, we only need to alter the central stem cells.


\section*{Materials and Methods}
\subsection*{Set-up}

Moran model and branching process have been widely used to study fixation probabilities and time to fixations in compartmental models [32].
Here, we use Moran process to model cell dynamics of the colon and intestinal crypts, because the number of cells in the crypt stays approximately constant. In the Moran models, the total number of cells remains constant at each updating time step. In most homogeneous Moran models, which include only one type of cells, it has been assumed that at each updating time step one cell dies and one cell divides. However, in order to model stem cells' symmetric divisions, some non-homogeneous Moran models assume two cell deaths coupled with two divisions [52]. Here, we consider a 4-compartmental stochastic model for the tissue architecture. This model consists of one compartment for TA cells $D_t$ and one compartment for FD cells $D_f$, and two stem cell (SC) groups $S_b$ and $S_c$, corresponding to BSCs and CeSCs, respectively. We assume at each updating time step, two FD cells die and two cells divide based on their fitness.
If stem cells divide with probability $\sigma$ the division is symmetric and with probability $1-\sigma$ is asymmetric. Asymmetric stem cells' divisions can only occur at the $S_b$ compartment.
There are two types of symmetric stem cell division; proliferation (newborn cells are two stem cells) and differentiation (newborn cells are two TA cells).
 In this model, differentiation only happens in the $S_b$ compartment, because it is close to the TA cells. However, with probability $\gamma$, proliferation happens in the $S_c$ compartment and with probability $1-\gamma$ it happens at the $S_b$ compartment. If it occurs at the $S_c$ group then one random stem cell from the $S_c$ compartment migrates to the $S_b$.
We denote the total number of non-stem cells by $D=|D_t|+|D_f|$, the total number of stem cells by $S=|S_b|+|S_c|$, and the total number of cells by $N=D+S$.
In summary, at each updating time step two FD cells die and two divisions occur based on the following algorithm:
\begin{itemize}
\item With probability $\lambda_f$, two FD cells divide. Or,
\item with probability 1-$\lambda_f$, one TA cell differentiates to two FDs, and one of the following scenarios happens.
\begin{itemize}
\item With probability $1-\lambda_s$, one $D_t$ cell proliferates to replace the differentiated $D_t$ cell. Or
\item with probability $\lambda_s$, one stem cell divides according to one of the following steps:
\begin{itemize}
 \item With probability $1-\sigma$, the division occurs in the $S_b$ asymmetrically, i.e. a $TA$ cell is generated. Or
 \item  with probability $\sigma$ the division is symmetric. With probability $\delta=\frac{S^{10}}{{S_0}^{10} + S^{10}}$ this symmetric division is differentiation and happens in the $S_b$ group, and with probability $1-\delta$ is proliferation ($S_0$ is the initial total number of SCs). If it is proliferation then with probability $\gamma$, it happens at the $S_c$ and with probability $1-\gamma$ it occurs at the $S_b$. If it occurs at the $S_c$ group, then one random cell from the $S_c$ migrates to the $S_b$.
Also, if the proliferation occurs at the $S_b$ group, with probability $\alpha$ one random cell from the $S_b$ migrates to the $S_c$, and then one random cell from the $S_c$ migrates to the $S_b$ to keep the number of cells at each compartment approximately constant.
\end{itemize}
\end{itemize}
\end{itemize}
In this model, the total number of cells stays constant. However, because of the definition of the function $\delta$, the number of stem cells varies, but its variation is very small. The function $\delta$ is a feedback function, which controls the rate of stem cells differentiation and proliferation. It does not let all stem cells differentiate and their number goes to zero, or massively proliferate and their number becomes very high.
More precisely, when the number of stem cells becomes less than the initial number of stem cells, then delta, the probability of stem cell differentiation, becomes very small to increase the probability of proliferation. When the number of stem cells is higher than the initial number of stem cells, then $\delta$ becomes very high to increase the probability of stem cell differentiation.  Since $\delta$ keeps the number of stem cells at each time point approximately the same as the initial number of stem cells ($S_0 \approx S$), in calculations we can assume $\delta=\frac{S^{10}}{{S_0}^{10} + S^{10}} \approx \frac{S^{10}}{{S}^{10} + S^{10}}=0.5$.

The model also includes the possibility of migration from the border stem cells to the central stem cell compartment. In the model, when a border stem cell proliferates with probability $\alpha$ one random BSC migrates to the $S_c$, and then one random central stem cell migrates to the $S_b$.  In other words, when a BSC migrates to the $S_c$, it shifts a CeSC cell to the border stem cell compartment. In order to keep the number of BSCs more than zero and having approximately constant number of CeSC, one CeSC must migrates to the $S_b$ when a migration occurs from the $S_b$ group to the $S_c$.

We assume, if a mutant divides, then its newborn children are mutants. When one mutant non-stem cell divides, one of its children becomes immortal with probability $u$.
At each updating time step, two FD cells are chosen to die. In the model, the progeny of immortal cells are never chosen to die. Therefore, immortal cells will never be washed out from the crypt.

In general, CeSCs can only proliferate, and BSCs are able to proliferate, differentiate, and divide asymmetrically. TA cells are able to both proliferate and differentiate, and FD cells only able to proliferate. Also for simplicity, death only happens in the FD compartment.
A summary of the model is given in Fig. 1. The details of methods and calculations are provided in the supplementary materials.
%
\begin{figure}[h!]
\centering{
\includegraphics[scale=0.48]{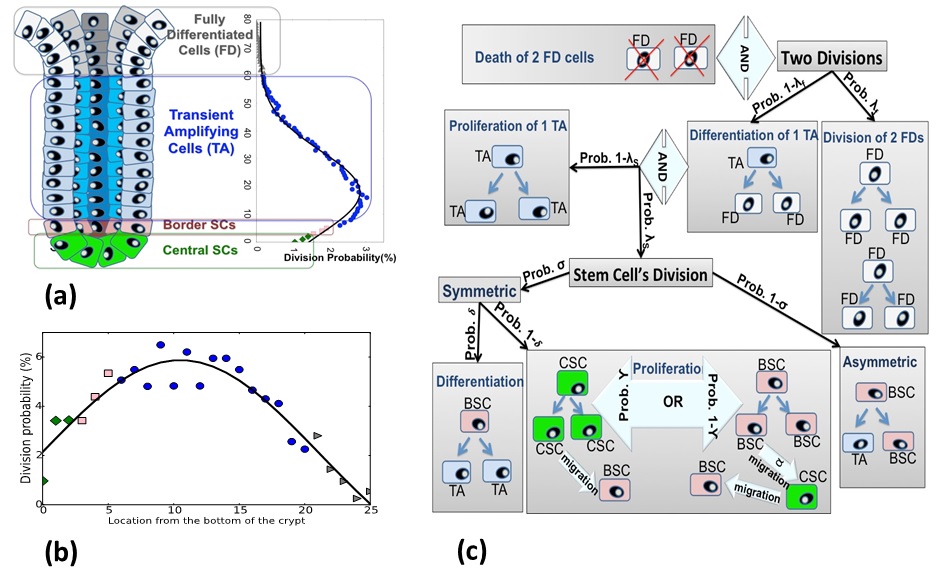}
\caption{ {\bf A schematic view of the model.} The sub-figure (a) represents the probability of cell division at each location of the human crypt obtained from [46]. The black solid line is the graph of the function $g$ defined in Parameter estimation. The sub-figure (b) shows the probability of cell division at each location of the mouse crypt obtained from [49]. The black solid line is the graph of the function $f$ defined in Parameter estimation. The sub-figure (c) represents our algorithm: at each updating step, two FD cells die and two cells divide to replace the dead cells.}
\label{fig:Model}
}
\end{figure}
\subsection*{Parameter estimation}

The number of studies on the human colon crypt's cell dynamics is very limited. However, there are some works on inferring the parameters for human colon crypts [43,44], as well as measurement of in vivo proliferation using bromodeoxyuridine (BrdU labeling) [45].  These experiments show that there are approximately 2000 cells in each human colon crypt, and the height of the crypt is around 80 cells [45].  It has been estimated that there are 5-6 stem cells at the bottom of the human colon crypt [44]. Additionally, most divisions happen in the lower part of the crypt; cells at positions 10-50 (Fig. 1), where 0 is the bottom of the crypt [45].

 Colonic crypts have the same spatial organization of cell types (stem cells, transient amplifying cells, and differentiated cells) as intestinal crypts. In addition, cell dynamics of the intestinal crypt are very similar to the colon crypt. However, at the top of the colonic crypt, the FD cells are shed into the lumen and transported away, whereas cells at the top of the intestinal crypt move up to a villus and are removed at the top of the villus.  Furthermore, mouse crypts contain all three cell compartments (stem cells, transient amplifying cells, and differentiated cells) in the same order as human crypts, with 5-7 stem cells [26]. The division rate of cells at each of these three cell compartments is very similar in mouse and human crypts [1-3]. In summary, cell dynamics of human colonic and intestinal crypts are very similar to mouse colonic and intestinal crypts.

In order to calculate the division probabilities at each compartments (FD, TA, $S_b$, and $S_c$), we normalize the division rates experimentally obtained in [46]. We find a function, which fits the normalized division rates of cells at each location $x$ in the human crypt, $g(x)=[0.974 + 24.1 \exp(-\frac{(x - 17.9)^2}{444})]/881.4$. We count cells at positions 70-80 as FD cells, therefore the probability that a division occurs in the FD compartment is given by $\lambda_f=\sum_{x=60}^{80} g(x)=0.026$. Furthermore, each crypt contains around 6 actively diving stem cells [1], they divide once every 2-3 days [46]. Hence, we assume cells at the locations 0-5 are stem cells, and TA cells are cells at the locations 6-69. Thus, the probability that a division occurs at the stem cell niche when FD cells do not replace two dead FD cells is $\lambda_s=\frac{1}{1-\lambda_f}\sum_{x=0}^{5} g(x)=0.105$. Moreover, in this model the division probability of CeSCs is given by $\lambda_s\gamma\sigma(1-\delta)$. According to Ritsma et al. [24], cells at the locations 0-2 belong to the CeSC group. Since the probability that a division occurs in the CeSCs is more than zero [24], thus $\gamma\sigma>0$. In our model, $\delta$ is approximately 0.5, hence $\sigma\gamma= \frac{2}{\lambda_s}\sum_{x=0}^{2} g(x)=0.884$.  Since $0<\sigma<1$ and $0<\gamma<1$, we conclude $\sigma$ and $\gamma$ are greater than $0.884$, which is in perfect agreement with the results obtained in [50].

Following the same technique, we also find the models' parameters for mouse intestinal crypts. We use the experimental data for the rate of cell divisions at each location of the mouse crypt, which is provided in [37]. The function $f(x)=[-0.424 + 1.09 \exp(-\frac{(x - 10.5)^2}{2 (10.6)^2})]/ 11.3292255981$ fits the normalized division rates of cells at each location $x$ in the mouse intestinal crypt (See Fig. 1(c)). The length of the mouse intestinal crypt is approximately 25 cells. We count cells at the positions 20-25 as FD cells, thus $\lambda_f= \sum_{x=20}^{25} f(x)= 0.08$. We also assume cells at the positions 0-5 are stem cells, and cells at the locations 6-19 are TA cells. Then, like before we get $\lambda_s=\frac{1}{1-\lambda_f}\sum_{x=0}^{5} f(x)=0.175$ and $\sigma\gamma= \frac{2}{\lambda_s}\sum_{x=0}^{2} f(x)=0.92$. We also vary  $\alpha$, which is the probability of migration from $S_b$ to $S_c$ when a BSC proliferates, between zero and 0.5, because the experimental observations show a small number of migrations from border stem cells to central stem cells [24]. The parameters are summarized in Table 1.

We use the parameters that we obtained for the mouse intestinal crypts to test the model. Fig. 2 shows a perfect agreement between the results of simulations and the results of experimental observations provided in [24].
This stochastic model, for the fist time incorporates the existence of two stem cell groups and the migration from the border stem cell compartment to the central stem cells to track mutants in the crypt. For this reason, this model is able to describe the biological reality better than previously published models that treated all stem cells equally.
\begin{table}[ht]
\centering
\caption{Model Parameters calculated using experimental data provided in [24,46] and [37].}
\label{parameters}
\begin{tabular}{llll}
\\ \hline
Symbol& Definition &  Human & Mouse\\ \hline
\noalign{\smallskip}
$N$ & total number of cells & 2000 &  200 \\
$\sigma$ & probability of symmetric division & 0.884-1 & 0.92-1\\
$\gamma$ & division prob. of CeSCs when SCs proliferate & 0.884-1 &  0.92-1 \\
$r_1$ & fitness of mutants & 0.9-3.8 &  0.9-3.8 \\
$\lambda_f$ & division probability of FD cells  & 0.08 &  0.08 \\
 $\lambda_s$ & division probability of stem cells & 0.175 &  0.175 \\
$\alpha$ & migration prob. from $S_b$ to $S_c$ when BSCs proliferate &  0-0.5 & 0-0.5 \\
$|S_c|$ & number of stem cells in $S_c$ (CeSC) & 4-8 &  4-8\\
$|S_b|$ & number of stem cells in $S_b$ (BSC) & 4-8 &  4-8\\
$|D_t|$ & total number of transit amplifying cells &  1500 &  150 \\
$|D_f|$ & total number of fully differentiated cells & 500 &  50 \\
\hline
\end{tabular}
\end{table}

\begin{figure}[h!]
\centering{
\includegraphics[scale=0.4]{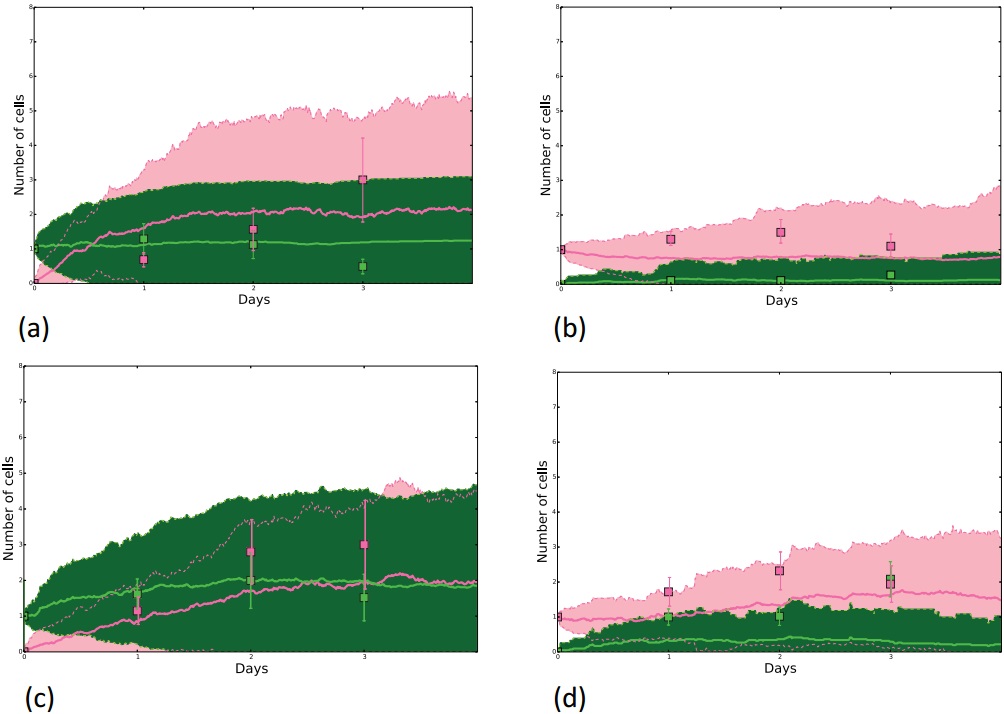}
\caption{ {\bf Comparing the model with the experimental data.} The sub-figures (a) and (c) show the average number of progeny of one marked CeSC in central (green color) and border (pink color) stem cell compartments over time. The sub-figure (b) and (d) shows the number of progeny of one marked BSC in central (green color) and border (pink color) stem cell compartments over time.  Points and bars are respectively means and standard deviations of the experimental data provided by Ritsma et al. [24]. Solid curly lines and colored areas indicate the means and standard deviations of 100 independent runs. Here, $\alpha=0.5$, $r_1=1$, and $S_c=S_b=8$, and the rest of the parameter values are given in Table 1 (mouse). The sub-figures (c-d) show only the number of the survived marked cells. In other words, they show the results for the cases that the marked cells have not been washed out from the crypt in three days.}
\label{fig:experiment}
}
\end{figure}


\section*{Results}

\subsection*{The probability of the progeny of mutant/marked CeSCs taking over the $S_c$ and the entire crypt is high}

The analytical methods reveal that the probability of the progeny of $e^*$ number of mutated/marked CeSC taking over the $S_c$ compartment is given by $\pi_{e^*}= \frac{1-\left( 1/r_1  \right)^{e^*}}{ 1-\left( 1/r_1  \right)^{S_c} }$ when no migration occurs from $S_b$ to $S_C$, where $r_1$ is the fitness of mutants and $S_c$ is the total number of CeSCs  (See SI part for the details of the calculations).
 This formula shows that the probability of the fixation in the CeSC depends only on the fitness of mutants, the number of CeSC mutants, and the total number of CeSCs.

From the definition of fitness, cells with high fitness have high proliferation rate as well as high differentiation rate. As a result, the fixation probability of advantageous CeSC mutants is higher than disadvantageous mutants, however their fixation time is also higher than disadvantageous ones (Fig. 4(a-c)). Since advantageous mutants differentiate faster than disadvantageous ones, they need more time to take over the crypt. These results  do not change when cells are able to migrate from border stem cell compartment to the central stem cells (See Fig. 3). Additionally, if the progeny of mutant CeSCs take over the FD compartment, its average occurrence time is less than 100 days (Fig. 4).

If stem cells divide only asymmetrically (i.e. $\sigma=0$), then no division happens at the CeSC compartment, thus CeSC mutants never divide. In other words, if stem cells divide fully asymmetrically then the probability that a mutant CeSC spreads to any other compartment is zero. The same scenario would occur, when the proliferation probability in the $S_c$ is zero, i.e. $\gamma=0$.


%
\begin{figure}[h!]
\centering{
\includegraphics[scale=0.4]{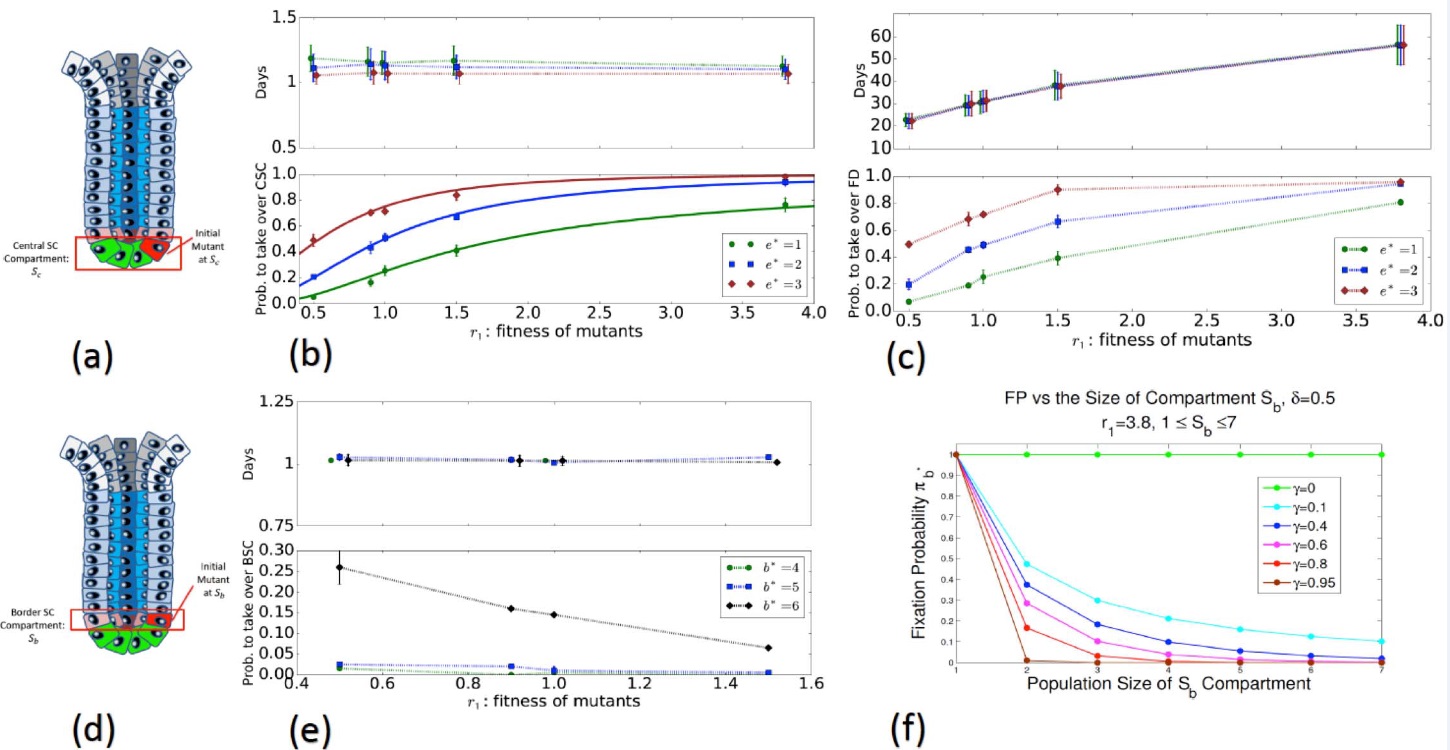}
\caption{ {\bf (a)-(c) The probability and time that mutant CeSCs will take over the $S_c$ and the FD.} The sub-figure (a) presents a schematic view of the model at the initial time. The simulations start with $e^*$ mutants in the $S_c$, while the other cells are wild-type. The sub-figure (b) indicates the average time and the probability of the progeny of mutant CeSCs taking over the CeSCs. The plot (c) shows the probability and time that the progeny of CeSC mutants will take over the FD.  In this figure $S_b=7, S_c=4$, and $u=v=0$, other parameters are given in Table 1(human). The points are the average and the bars indicate the standard deviations of 5 batches of 100 runs, and the solid lines present the results of the formula.
{\bf (d)-(f) Time and probability of the progeny of mutant BSCs taking over the $S_b$.}  The sub-figure (d) shows that there are $b^*$ number of mutants in the $S_b$ at the initial time of simulations, and (e) presents the result of simulations. The bottom sub-figure of (e) indicates the probability that the progeny of $b^*$ number of mutant BSCs will take over the entire $S_b$, and the top sub-figure shows the time of its occurrence.  The sub-figure (f) presents the analytic results, and it shows the effect of the number of BSCs, $S_b$, and the proliferation probability of CeSCs, $\gamma$, on the fixation probability, which is the probability of the progeny of mutant BSCs taking over the entire $S_b$. Here, $S_b=7$, $S_c=4$, and the rest of the parameter values are given in Table 1(human).
}
\label{fig:fix_SC}
}
\end{figure}

\begin{figure}[h!]
\centering{
\includegraphics[scale=0.6]{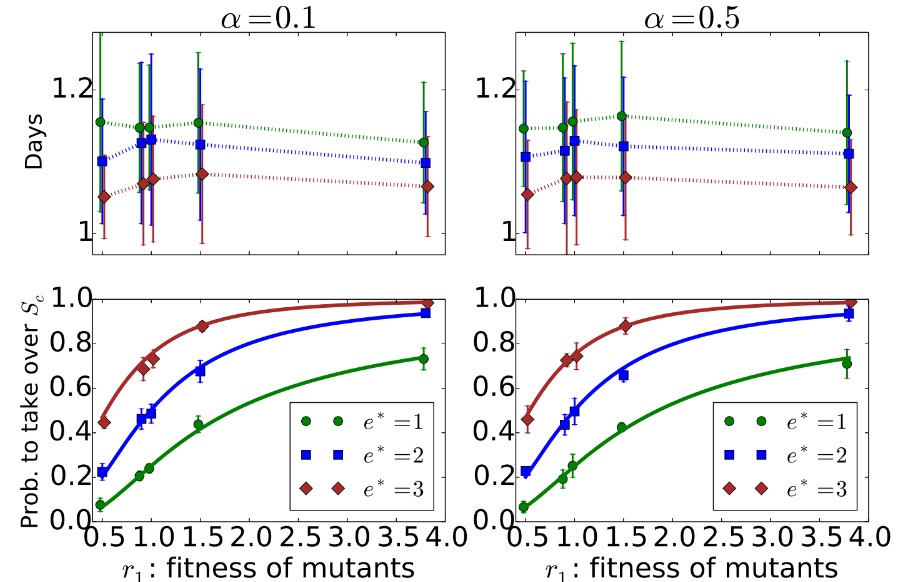}\\[2mm]
\caption{  {\bf The probability and time that mutant CeSCs will take over the $S_c$ when cells migrate from $S_b$ to $S_c$.} The simulations start with $e^*$ mutants in the $S_c$, while the other cells are wild-type. This figure indicates the average time and the probability of the progeny of mutant CeSCs taking over the CeSCs. The left sub-figures show the results for $\alpha=0.1$, and the right sub-figures indicate the results for $\alpha=0.5$. Where, $\alpha$ is the probability that one BSC cell migrates to the $S_c$ when a proliferation occurs in the $S_b$. In this figure $S_b=7, S_c=4$, and $u=v=0$, other parameters are given in Table 1(human). The points are the average and the bars indicate the standard deviations of 5 batches of 100 runs, and the solid lines present the results of the formula.}
\label{fig:fix_SC_addmig}
}
\end{figure}

\subsection*{The probability of a mutant BSC's progeny taking over the $S_b$ or $D_f$ is approximately zero}

The analytical calculations show that the probability of the progeny of $b^*$ number of BSCs taking over the entire BSC is given by $\displaystyle \pi_{b^*}= \frac{1- 1/(1-\gamma)^{b^*}}{ 1- 1/(1-\gamma)^{S_b} }$ when there is no migration from $S_b$ to $S_c$, where $\gamma$ is the proliferation probability of CeSCs when stem cells divide symmetrically.
This formula indicates that the probability of the fixation in the BSC depends only on $\delta$, $\gamma$, the number of BSC mutants, and the total number of BSCs when there is no migration from $S_b$ to $S_c$ ($\alpha=0$).
This formula and the simulations imply that the probability that the progeny of one BSC will spread over the entire BSC is almost zero, because according to the parameter estimations $\delta$ is approximately half, and $\gamma$ is between $0.884$ and one. Additionally, the progeny of BSCs are always washed out from the crypt regardless of their fitness when there is no immortal cell. If at least half of BSCs are mutants, then they might have a small chance to colonize and take over the entire BSC compartment. If this rare event happens, it occurs in one day (Fig. 4(d),(e),(f)). However, several days later the progeny of CeSCs will replace all mutant BSCs, and eventually the progeny of mutant BSCs are washed out from the crypt.

Although we expect a higher proliferation probability in the BSC compartment (i.e. smaller $\gamma$) to lead to a higher fixation probability in the BSC group, it does not increase the fixation probability $\pi_{b^*}$ very much. The reason is each proliferation is coupled with one differentiation. The differentiation of SCs increases the chance of transporting mutants from the stem cell niche to the TA group.
Additionally, advantageous mutants disappear faster from the $S_b$ than disadvantageous ones. This result is in agreement with the result of the experiments done by Ritsma et al. [24], where they observed that the probability of a BSC colonization is very small.


%
\begin{figure}[h!]
\centering{
\includegraphics[scale=0.21]{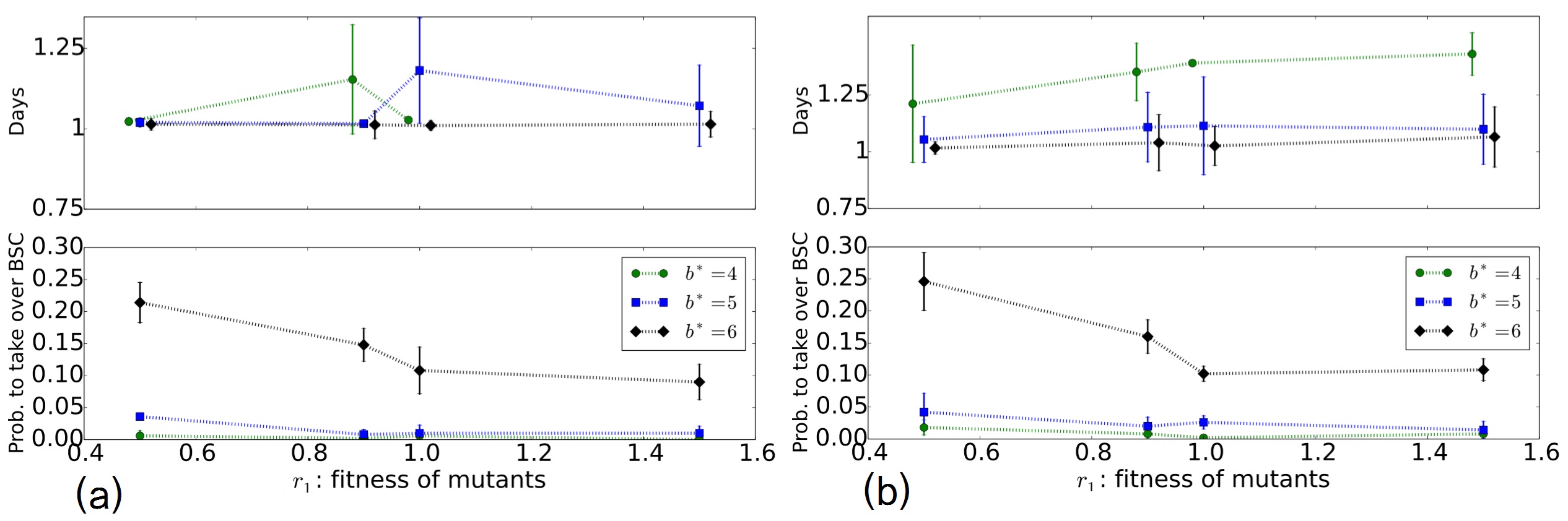}
\caption{ {\bf Time and probability of the progeny of mutant BSCs taking over the $S_b$ when migration happens from $S_b$ to $S_c$.} The bottom sub-figures indicate the probability that the progeny of $b^*$ number of mutant BSCs will take over the entire $S_b$, and the top sub-figures show the time of its occurrence.  In the sub-figure (a), the probability of migration from $S_b$ to $S_c$ when proliferation occurs in the BSC is 0.1, i.e. $\alpha=0.1$. In the sub-figure (b), $\alpha=0.5$. Here, $S_b=7$, $S_c=4$, and the rest of the parameter values are given in Table 1.
}
\label{fig:fixSb_alpha}
}
\end{figure}

\subsection*{The progeny of a small number of FD or TA mutants never take over the entire TA or FD}

Experimental data shows that FD cells do not divide much compared to TA cells. If FD cells do not divide at all ($\lambda_f=0$), then non-immortal FD mutants are always removed from the crypt. Moreover, if $\lambda_f>0$ and no immortal cell is generated in the system, the probability $\pi_{f^*=1}$ that the progeny of one FD mutant will take over the whole FD is very small (Fig. 5(b),(c)); however, it is much higher than the probability of one TA cell's progeny taking over the TA compartment, $\pi_{d^*=1}$.
When the rate of divisions in $D_t$ is approximately zero (it rarely occurs in reality), then $\pi_{d^*=1}=0$. If division rate in stem cells is small but non--zero while that for $D_t$ is high, which corresponds to the experimental observations, then the probability of one TA mutant's progeny taking over the TA compartment decreases when the mutants' fitness $r$ increases (Fig. 5(e),(f)). This means that TA cells are more capable of producing FD cells than being fixated. In the extreme scenario, when no divisions happen in SCs ($\lambda_s=0$), then  $\pi_{d^*=1}$ is $1/|D_t|$. Moreover, when no proliferation in $D_t$ is permitted and TA cells only differentiate, i.e. $\lambda_s=1$, then $\pi_{d^*}$ is zero. Note, when TA cells are only able to differentiate but not proliferate, mutants differentiate to two FD mutants (and thus extinct), therefore mutants will be removed from the TA compartment (Fig. 5(e),(f)).

We also study the probability of a TA mutant's progeny taking over the FD compartment.
In the human colon crypt, where $\lambda_f=0.011$ and $\lambda_s=0.104$ based on the parameter estimations, the probability of one TA or FD cell's progeny taking over the entire FD or TA compartment is approximately zero. This emphasizes the fact that although TA mutants generate FD mutants, the generated FD mutants are washed out from the crypt, before they get a chance to colonize.
Interestingly, mutants with high fitness will be removed from the crypt quickly, because most divisions occur in the TA compartment, and advantageous TA mutants quickly differentiate to two FD cells.

\begin{figure}[h!]
\centering{
\includegraphics[scale=0.37]{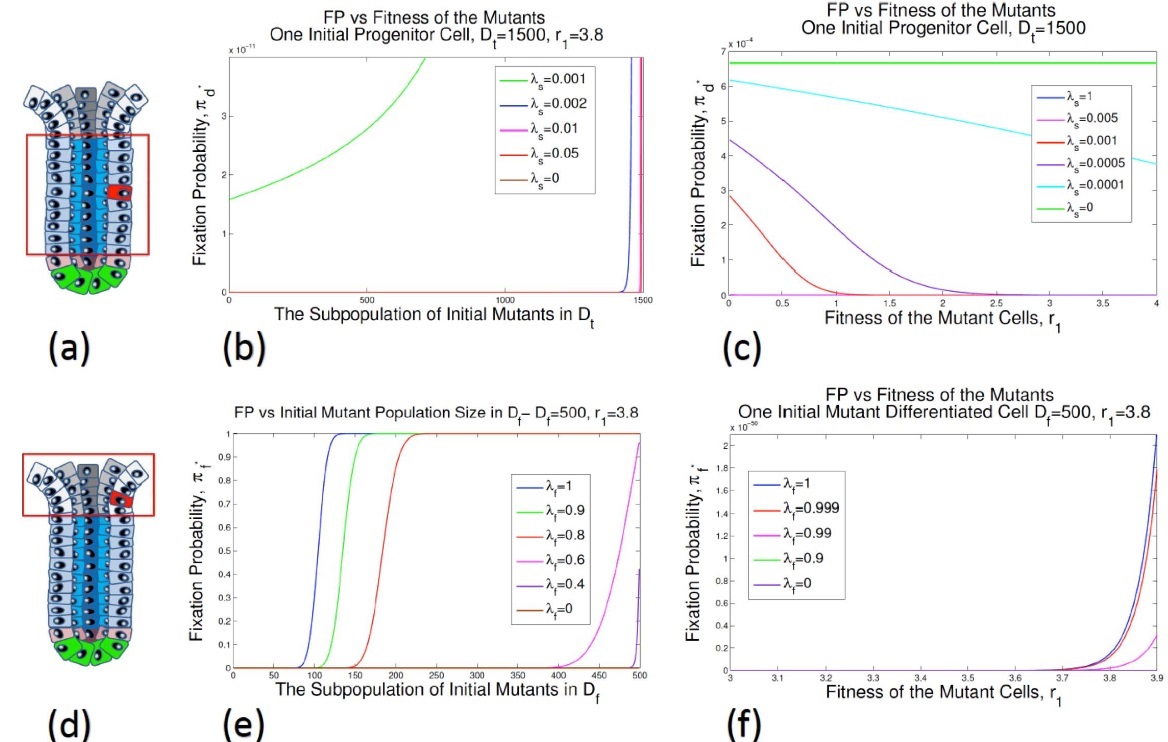}
\caption{{\bf (a)-(c) Role of TA mutants in generating FD mutants.} Figure (a) is a schematic view of the crypt at the initial time of the process. Plot (b) indicates the effect of the initial number of TA mutants and the probability of stem cells' division, $\lambda_s$, on the probability that mutants take over the entire TA. Plot (c) reveals the effect of the mutants' fitness on the fixation probability of mutants in the TA.
{\bf (d)-(f) Fixation probability in the FD compartment.} Sub-figure (d) is a schematic representation of the crypt at the initial time. (e) shows the behavior of the system for a range of the rate of divisions in the $D_f$ group, $\lambda_f$, as the initial number of mutants varies. The curves in (f) illustrate the impact of the mutants' fitness $r_1$ and the division probability of FD cells, $\lambda_f$, on the fixation probability of mutants in the FD group. This figure shows the results of the analytical formulas, when the total number of TA cells is $1500$, and the number of FD cells is $500$, and in (b) and (c), the mutant's fitness is $r_1=3.8$. }
\label{fig:fix_FD_TA}
}
\end{figure}

\subsection*{Central stem cells control the entire crypt}

The results of simulations reveal that with probability more than 0.99, the progeny of CeSCs will take over the entire human colon crypt in less than three months. In other words, if all stem cells are wild-type, while the rest of the cells in the crypt are mutants, then all crypt cells become wild-type in less than 100 days. Unexpectedly, the time that the progeny of the CeSCs need to take over the crypt decreases when the mutants' fitness increases: meaning the advantageous mutants are washed out from the crypt faster than disadvantageous ones. Although this result seems strange in the first glance and is different from the results of non-spatial models, it actually makes sense.  Advantageous mutants have higher division rates than normal cells, so they will be moved rapidly to the FD compartment, then will be removed from the crypt. We have not found any experimental data about this result, so it needs to be experimentally validated.

Moreover, the probability that the progeny of a single normal stem cell will take over the entire crypt is more than zero. This probability is more than $0.25$, if mutants are disadvantageous, and it is close to zero if mutants are advantageous. This implies that if only one of the CeSCs is wild-type, and the rest of the crypt's cells are $P53^{R172H}$ mutants, then with a probability of $0.25$ all cells will become wild-type in 1-2 months in the non-inflammatory condition. Although the probability of w.t. cells taking over the entire crypt depends on the number of normal cells in the $S_c$, its time to concurrence does not depend on the number of wild-type CeSCs (Fig. 6).

If stem cells divide fully asymmetrically ($\sigma=0$), then no division occurs in the CeSC compartment. Therefore, in this case CeSCs will not take over the crypt. Moreover, the time that the progeny of CeSCs need to take over the entire crypt is a decreasing function of $\sigma$. In other words, if stem cell divisions are mostly symmetric, then CeSCs progeny will rapidly spread over the crypt. In addition, the probability $\gamma$,  which is the probability that a CeSC divides in the case of symmetric division, is not as important as the fitness of mutants $r_1$.

\begin{figure}[h!]
\centering{
\includegraphics[scale=0.45]{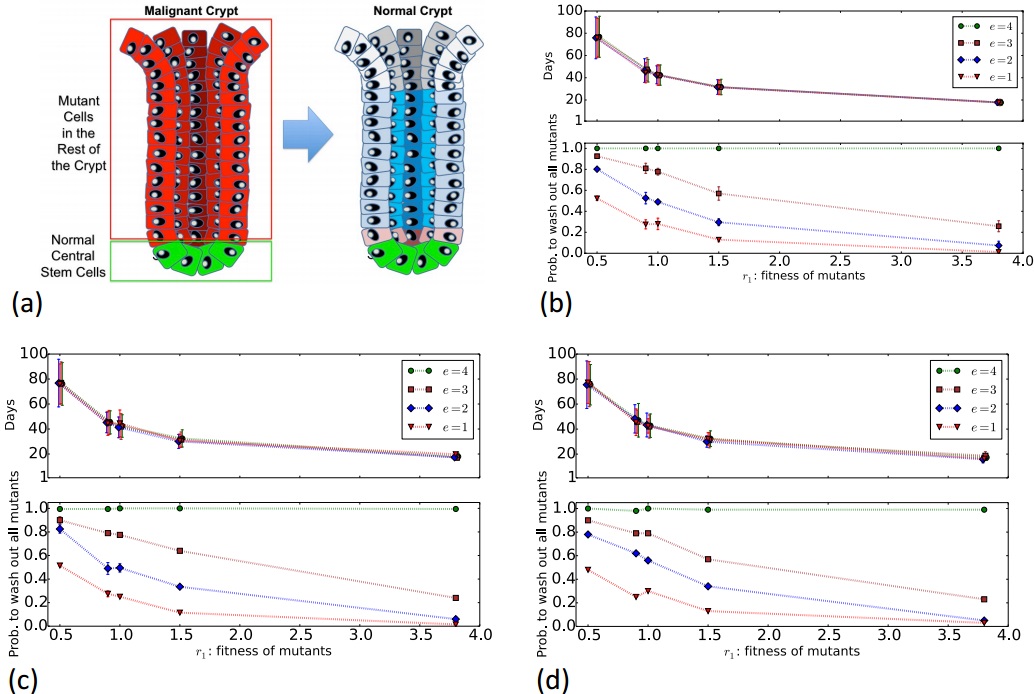}
\caption{ {\bf Probability and time that mutants are washed out from the crypt.} Cartoon picture in (a) shows a $S_c$ compartment with all normal central stem cells is able to wash out mutants in the rest of the crypt. The sub-figures (b-d) show the results of simulations indicating the average time and the probability that the all crypt's cells become wild-type, i.e. all mutants are washed out from the crypt. Sub-figure (b) shows the results for $\alpha=0.0$, (c) $\alpha=0.1$, and (d) $\alpha=0.5$, where $\alpha$ is the probability of migration from BSC to CeSC when a proliferation happens in the BSC. In this figure $S_b=7, S_c=4$, and $u=v=0$, other parameters are given in Table 1. The points are the average, and the bars indicate the standard deviation. }
\label{fig:wash}
}
\end{figure}


\subsection*{The progeny of a single immortal TA or FD cell always take over the entire FD in less than 70 days}

Some environmental conditions or genetic/epigenetic changes lead to creation of an immortal cell in the TA compartment. Therefore, here we investigate the dynamics of potential immortal cells in the crypt. We observe that the immortal TA cells have higher tendency to differentiate and generate more immortal FD cells than spreading over the TA compartment. Moreover, the progeny of even a single immortal TA or FD cell will spread over the entire FD in less than 70 days. Expectedly, the advantageous immortals spread faster than disadvantageous ones.

\begin{figure}[h!]
\centering{
\includegraphics[scale=0.27]{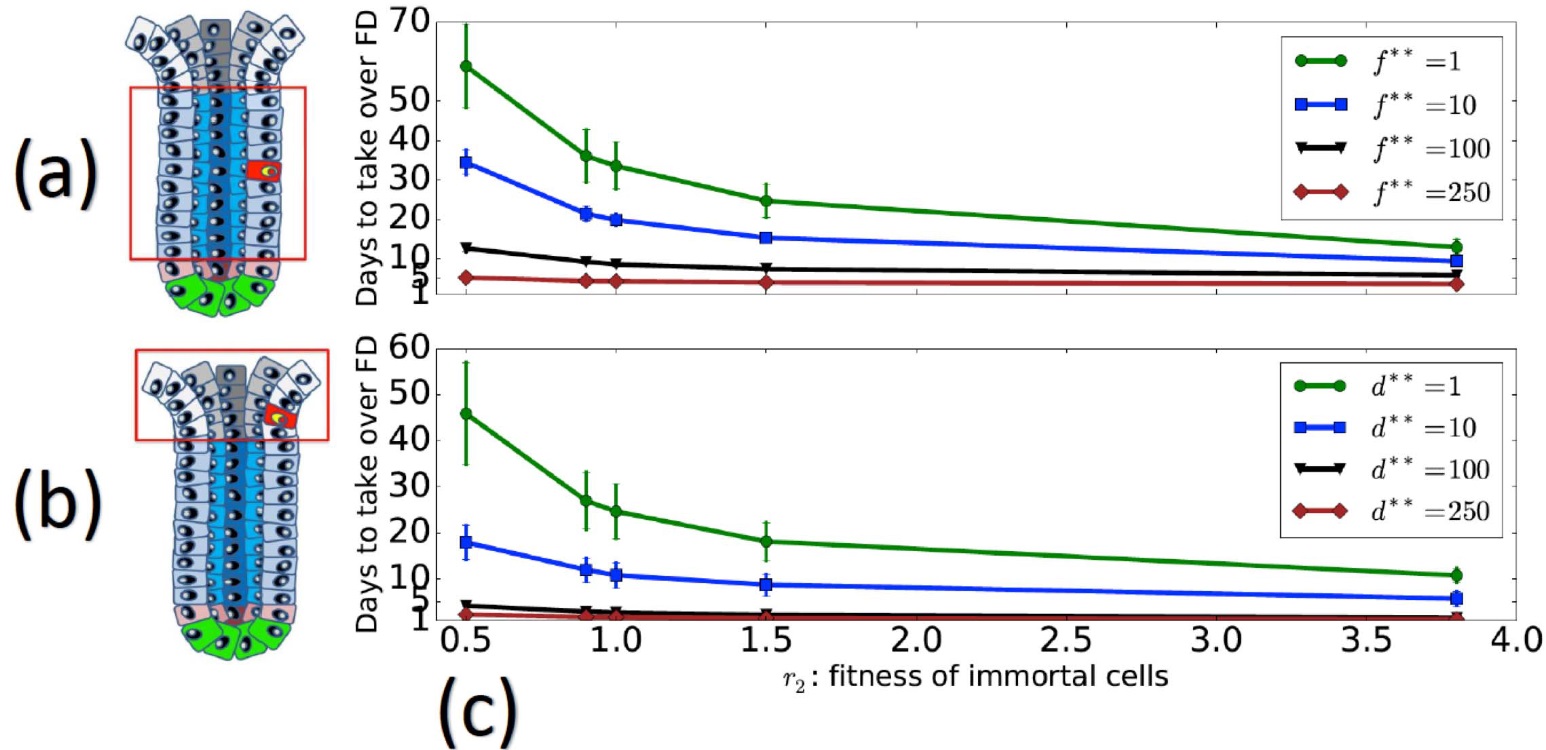}
\caption{ {\bf Fixation of immortal cells in the FD.} The sub-figures (a) and (b) represent a schematics view of the system at the initial time of simulations generating the bottom and top sub-figures of (c), respectively. In the top sub-figure (c), the process starts with $f^{**}$ immortal cells, while the rest are wild-type. In the bottom sub-figure (c), at the initial time there are $d^{**}$ immortal cells in $D_t$ and other cells are wild-type. In both sub-figures, we obtain the time that immortal cells take over the entire FD.}
\label{fig:immortal}
}
\end{figure}

\subsection*{Existence of the bi-compartmental stem cell niche has some advantages and disadvantages.}
Our 4-compartmental model can be easily modified as a 3-compartmental model with only one stem cell group by assuming there is no CeSCs ($|S_c|=0$) and the probability of division in CeSCs $\gamma$ is zero. Fig. 8 shows that the probability that the progeny of one mutant stem cell will take over the FD compartment is small (between 0.01 and 0.14) in the one stem cell compartment model. However, this probability is zero for the progeny of a mutant BSC in the 4-compartmental model.

\begin{figure}[h!]
\centering{
\includegraphics[scale=0.27]{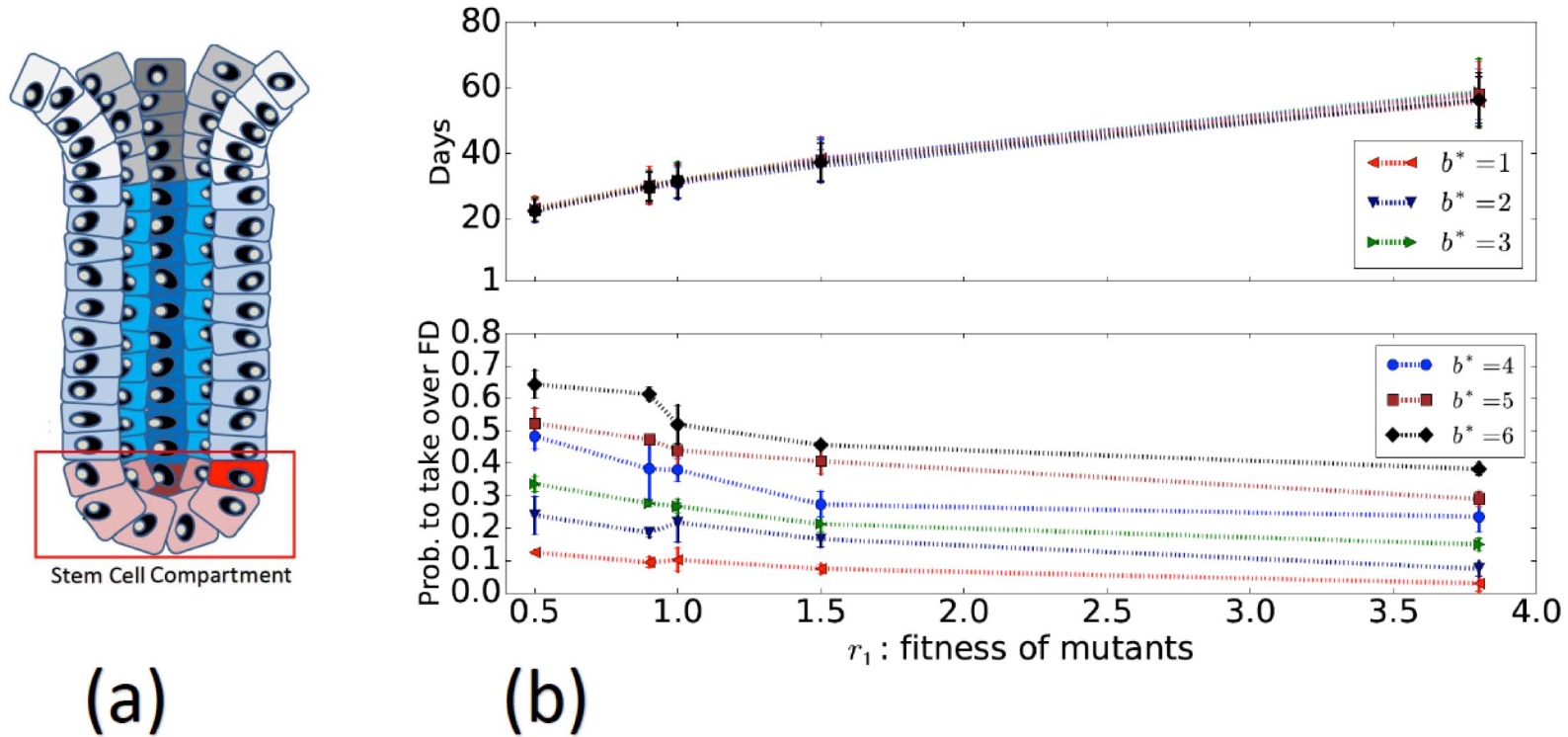}
\caption{ {\bf One stem cell group.}  The figure (a) shows that there are $b^*$ number of mutants in the one stem cell group at the initial time of the simulation, and (b) presents the result of simulations. The bottom sub-figure of (b) indicates the probability that the progeny of $b^*$ number of mutant SCs will take over the entire FD, and the top sub-figure shows the time of its occurrence. Here, $S=11$, and the rest of parameters are given in Table 1. }
\label{fig:1Stem}
}
\end{figure}

In the 4-compartmental model, if a mutant appears in the CeSCs, with a high probability it stays in the crypt, and its progeny will take over the entire crypt especially for advantageous mutants. However, in the 3-compartmental model, the possibility that an advantageous mutant stem cell differentiates to two TA cells and is removed from the crypt is high. This result may suggest the existence of only one stem cell compartment is an advantage, however if half of SCs become mutants, then mutants take over the entire crypt with a high probability in the 3-compartmental model. In the 4-compartmental model, if all BSCs are mutants (more than half of SCs) and all CeSCs are wild-type, then mutants will be washed out from the crypt in less than 3 months. Furthermore, in the work done by Shahriyari et al. [52], it has been  shown that the bi-compartmental stem cell niche delays the mutants' generation.

\subsection*{The effect of migration from border stem cells to central stem cells}
Comparing the results of simulations and the experimental observation shows that the best value for the probability of migration from BSC to CeSC, when a border stem cell proliferates, is $\alpha=0.5$ (See Fig. 2).
Simulations and analytical calculations show that the results are almost insensitive to $\alpha$ when $\alpha$ is between zero and $0.5$ (See Fig. 7 and 8).
More precisely, when $\alpha=0$ and central stem cells are normal, then the probability of washing out mutants is one. However, when $\alpha=0.5$, this probability is around $0.99$ for $0.5\leq r \leq 3.8$.
The results for higher range of $\alpha$ have been shown in the supplementary part.

Border stem cells mostly differentiate, and the proliferation of  stem cells mostly occur in the $S_c$ compartment. Therefore, with a very high probability, mutant BSCs differentiate and move to the TA compartment specially if mutants have a high fitness. However, the disadvantageous BSC mutants have a small chance to proliferate and move to the CeSC compartment, but their survival chance is very small because of their small fitness.
This phenomena causes that the probability of BSC mutants taking over BSC or the entire crypt becomes very small and almost insensitive to $\alpha$ (See Fig. 8).
%


\section*{Discussion}

The absence of APC, which causes aberrant migration [5,6], is the most frequent mutation in colon cancer [5]. If all cells in one compartment lose their APC, then a tumor initiates because cells located lower than this compartment cannot migrate to the top of the crypt. For this reason, we calculate the probability that mutants are fixated at each of the compartments. The simulations and analytical calculations, which are in perfect agreement, show that the progeny of one non-stem non-immortal cell regardless of its fitness are not able to take over any compartment. However, the progeny of one immortal neutral FD cell will take over the entire FD in less than one month.

The model shows that the time that the progeny of neutral stem cells need to take over the entire normal crypt is around 60 days, which is in perfect agreement with the experimental observations [50].
Snippert et al. [50] found that the progeny of stem cells are taking over the entire normal intestinal crypts in two months.
Our model also indicates that when all central stem cells are normal and there is no immortal cell, mutants will be washed out from the crypt with probability more than 0.99. Furthermore, it shows that a single normal central stem cell has a chance to sweep out all mutants from the crypt when there is no immortal cell.

The model suggests that only CeSCs are able to take over the entire crypt, and they are able to renew the CeSC compartment in less than 2 days, and the whole crypt in less than 3 months.
 This result is in agreement with the results of the computer simulations for colonic crypt provided in [39].  They did simulations for a list of  chosen parameters' values, which were not obtained based on biological observations. When there is no mathematical calculations, we cannot completely trust the results, because the results might be different for other parameters' values. Moreover, their model consists of only one cell type and does not consider different type of divisions, like stem cells symmetric and asymmetric divisions. Their model, which only obtains the location of cells after each division following a physical rule, predicts that mutants located more than one to two cells from the base of the crypt are unlikely to take over the entire crypt.

  Although the fitness of CeSCs does not make much difference in their fixation time in the CeSC compartment, it affects the time that they need to take over the entire crypt. The disadvantageous mutants take over the crypt quickly, but the probability of this occurrence is small. For example, the probability of one disadvantageous CeSC mutant's progeny like P53$^{R172H}$, which has a fitness of $0.9$, taking over the FD compartment is $0.2$ in the non-inflammatory environment. However, if the progeny of a disadvantageous CeSC mutant take over the FD compartment, they will spread over the entire FD in less than one month. Moreover, the probability of the progeny of one CeSC APC$^{-/-}$ mutant, which has a fitness of $3.8$, taking over the FD is $0.8$. Importantly, they need at least 40 days to spread over the entire FD compartment (Fig. 4). We conclude the probability of the progeny of the CeSC mutants with lower fitness, like P53$^{R172H}$ mutants, taking over over the crypt is small, but if it happens it is a fast process.  Additionally, the probability of the progeny of the CeSC mutants with high fitness, like APC$^{-/-}$ mutants, taking over the crypt is high, but it is a slow process.

In summary, analytic calculations and numerical simulations, which confirm the existing experimental data, show a high level of dependency on central stem cells for cell dynamics of colon and intestinal crypt. In the absence of immortal cells, a normal central stem cell compartment will sweep out all mutants from the crypt; substituting mutant stem cells in the niche with the wild-type cells will cure the crypt.
Therefore, the model suggests that stem cell therapies can be a potential treatment or preventing option for intestinal and colon cancer when there is no immortal cells.
Recently, there have been many studies implementing stem cell therapy for various diseases including cancers [57]. Most of the procedures involve local administration or direct injection of stem cells [55].
An in vivo study has revealed that injection of rat umbilical cord stem cells (rUSCs) can completely abolish rat mammary carcinomas with no evidence of metastasis or recurrence 100 days post-tumor cell inoculation [56].

The model predicts that if there are any immortal cells, they will rapidly expand and cause tumor formation.In this case, normal stem cells alone are not able to cure the crypts, and the therapeutic strategies, which are based on transforming immortal cells to mortal ones, are needed. Thus, the failure of some stem cell therapies might be associated with the existence of immortal cells.
There are some studies suggesting that inhibition of telomerase function and experimental induction of telomere shortening can reverse cell immortality and trigger apoptotic cell death [53,59]. However, there is an evidence of telomerase activity in some normal tissues such as colon and testis [54]. Therefore, the future of stem cell therapy depends on understanding each tissue's specific cell types and features as well as its cell dynamics.
%



\section*{Acknowledgments}
The authors would like to thank Dr. K. Kaveh, Dr. M. Kohandel, and Dr. S. Sivaloganathan for their helpful discussions.


%

{\Large
\textbf\newline{Supporting Information} 
}
\newline
\\

\subsection*{Analytical tools}

We denote the number of wild-type (w.t.) and mutant stem cells in the $S_b$ group respectively by $b$ and $b^*$, in the $S_c$ group by $e$ and $e^*$, in the $D_t$ group by $d$ and $d^*$, and finally in the $D_f$ compartment by $f$ and $f^*$. In addition $d^{**}$ and $f^{**}$ are respectively the number of TA and FD  immortal cells.
We also assume that the fitness of mutant cells and immortal cells are respectively $r_1$ and $r_2$.

In this model at each updating time step, two $D_f$ cells die. For each of these deaths, with probability $f^*/(f+f^*)$, one mutant $D_f$ cell dies and with probability $f/(f+f^*)$ one wild-type (w.t.) $D_f$ cell dies.
Then, two randomly chosen cells divide according to the following algorithm:

\begin{itemize}
\item With a probability of $\lambda_f$, two $D_f$ cells divide. For each of these divisions, with probability $\frac{r_1 f^*}{r_2 f^{**}+ r_1 f^*+f}$, one mutant $D_f$ cell divide and with probability $v$ one of its children becomes immortal cell. However, with probabilities $\frac{r_2 f^{**}}{r_2 f^{**}+ r_1 f^*+f}$ and $\frac{f}{r_2 f^{**}+r_1f^*+f}$, respectively one immortal and one wild-type $D_f$ cell divides. Or,
\item With a probability of $(1-\lambda_f) \frac{r_2 d^{**}}{r_2 d^{**}+r_1 d^*+d}$ or $(1-\lambda_f) \frac{d}{r_2 d^{**}+r_1 d^*+d}$, respectively one immortal TA cell differentiates to two immortal FD cells or one normal TA cell differentiates to produce two w.t. $D_f$ cells. However, with probability $(1-\lambda_f) \frac{r_1d^*}{r_2d^{**}+r_1d^*+d}$, one mutant TA cell divides and with probability $u$, one of its newborn daughter cells becomes an immortal FD cell, i.e. dedifferentiation happens, while the other offspring is a mutant TA cell. However, with probability $1-u$, both newborn individuals are mutant fully differentiated cells. Then one of below scenarios occurs
\begin{itemize}
\item With probability $(1-\lambda_s)$, one TA cell proliferates. This is the proliferation of a wild-type TA cell with a probability of $\frac{d}{r_2d^{**}+r_1d^*+d}$ to produce one w.t. TA cell, or the proliferation of immortal TA cell with a probability of $\frac{r_2 d^{**}}{r_2d^{**}+r_1 d^*+d}$ to produce one immortal TA cell. Or with probability $\frac{r_1d^*}{r_2d^{**}+r_1d^*+d}$, the proliferation of a mutant TA cell occurs, then with probability $u$ one of the newborn members is immortal TA cell and the other one is mutant TA cell and with probability $1-u$ both are mutant TA cells. Or,
\item with probability $\lambda_s$, one  stem cell divides in the following way:
\begin{itemize}
\item One mutant $S_b$ stem cell divides asymmetrically and makes one mutant TA cell with probability $(1-\sigma)\frac{r_1b^*}{r_1b^*+b}$. With probability $(1-\sigma)\frac{b}{r_1b^*+b}$, one w.t. $S_b$ stem cell divides to generate one w.t. cell in the $D_t$ compartment. Or,
\item with probability $\delta \frac{b}{r_1b^*+b}$, one wild-type $S_b$ stem cell differentiates to generate two wild-type TA cells. However, with probability $\delta\frac{r_1b^*}{r_1b^*+b}\,(1-\alpha)$ one mutant border stem cell differentiates to make two mutant TA cells. Or,
\item with probability $1-\delta$ proliferation happens in the stem cell niche.
\begin{itemize}
\item With probability $\gamma \frac{e}{r_1e^*+e}$, one normal $S_c$ cell proliferates to produce one normal stem cell in the $S_c$ group. Moreover, with probability $\gamma\frac{re^*}{r_1e^*+e}$, one mutant stem $S_c$ cell proliferates to generate one mutant $S_c$ cell. Then with probability $\frac{e}{e+e^*}$ one random w.t. stem cell from $S_c$ migrates to the $S_b$. In addition, with probability $\frac{e^*}{e+e^*}$ one mutant $S_c$ cell migrates to the $S_b$ compartment. Or,
\item with probability $(1-\gamma) \frac{b}{r_1b^*+b}\,(1-\alpha)$, one wild-type $S_b$ cell proliferates to make one wild-type stem cell in the $S_b$ compartment  and with probability $(1-\gamma)\, \frac{b}{r_1b^*+b}\,\alpha$ one of the two offsprings of a w.t. $S_b$ stem cell migrates to CsSC group. However, with probability $(1-\gamma)\frac{r_1b^*}{r_1b^*+b}\,(1-\alpha)$ one mutant $S_b$ cell proliferates to generate another mutant $S_b$ cell while $(1-\gamma)\frac{r_1b^*}{r_1b^*+b}\,\alpha$ is the probability of migration for one of the daughter cells of a malignant BSc from $S_b$ to $S_c$ group.
\end{itemize}
\end{itemize}	
\end{itemize}
\end{itemize}
\begin{figure}[h!]
\centering{
\includegraphics[width=0.5\textwidth]{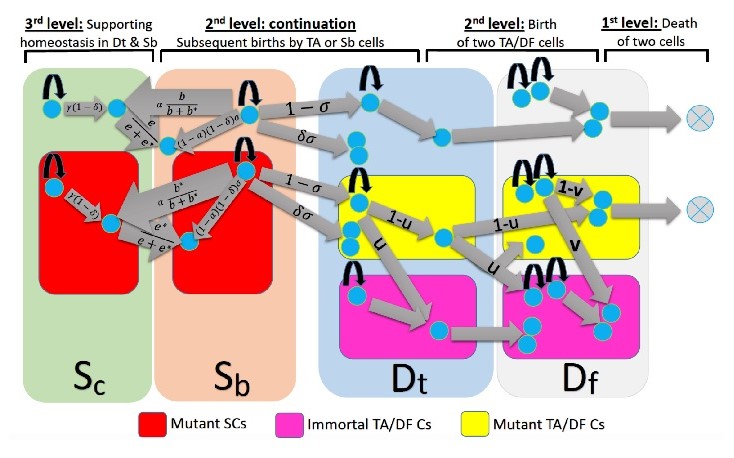}
\caption{ {\bf A Schematic representation of the model with possible pathways.} This model includes four compartments: (i) central stem cells, $S_c$, (ii) border stem cells, $S_b$, (iii) transient amplifying cells, $D_t$, and (iv) fully differentiated cells, $D_f$. Different types of proliferation and differentiation of stem and non-stem cells occur in the system in order to preserve the constant population size. The model includes the possibility of dedifferentiation; mutant $D_t$ or $D_f$ cells are able to generate immortal $D_t$ or $D_f$ cells, respectively.}
 \label{fig4}
}
\end{figure}
Briefly speaking, $\sigma$ is the probability of symmetric division in stem cell niche, while $\delta$ is the probability of differentiation in the $S_b$ compartment in the case of symmetric stem cell division. Moreover, $\gamma$ is the probability of proliferation in the $S_c$ compartment, when a stem cell proliferates. Moreover, $\lambda_f$ is the probability of choosing fully differentiated cells for birth event (see Fig. 10), while $\lambda_s$ is the probability of division in the stem cell niche. $\alpha$ is the probability of migration for BSCs to CeSC compartment. Fig. 11 and 12 reveal different steps of the procedure in details.

\section*{Evolutionary mechanism of the system}

In this model, the total number of stem cells in the $S_c$ and $S_b$ compartments (separately) remain approximately fixed, which means that homeostasis controls each stem cell compartment's size. Moreover, the other two compartments of progenitor and fully differentiated cells are subject to the same assumption and their sizes remain approximately unchanged through the evolutionary dynamics of the system. Therefore, we have a 6-dimensional multi--variable Markov model as the system of random movements over possible states $(e^*, b^*, d^*, d^{**}, f^*, f^{**})$.

\begin{figure}[h!]
\centering{
\includegraphics[scale=0.6]{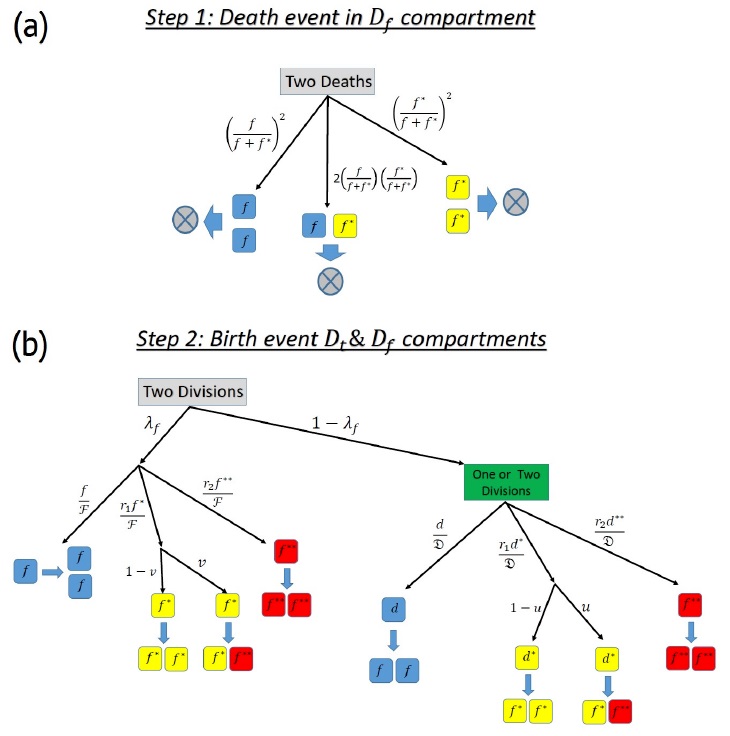}
\caption{ {\bf The cartoon figure of the possible death and birth.} The sub-figure (a) represents the three possible death in the $D_f$ compartment. The sub-figure (b) shows the probable divisions occurring in either $D_f$ or $D_t$ compartments to replace the dead cells. With a probability of $\lambda_f$ divisions occur in the $D_f$. Otherwise the replacements can be the result of divisions in $D_t$ population with a probability of $1-\lambda_f$.}
\label{fig5}
}
\end{figure}
%
\begin{figure}[h!]
\centering{
\includegraphics[scale=0.45]{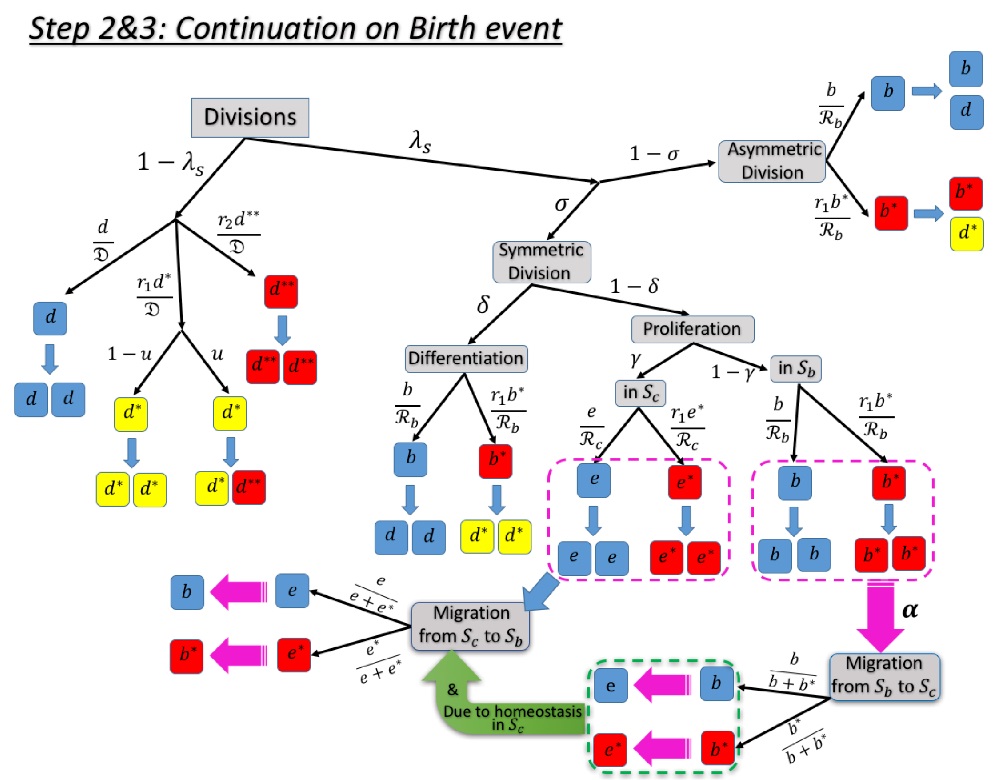}
\caption{ {\bf A representative cartoon picture representing the hierarchy of divisions occurring in the $D_t, S_b,$ and $S_c$ compartments as a continuation to the birth events.} Figure (a) reveals a continuation to the second step where all the possible cases happen in the $D_t, S_b, S_c$ compartments. A cell divides in the $D_t$ population with a probability of $1-\lambda_s$, and with a probability of $\lambda_s$ in the $S_b$ or $S_c$ compartments. The sub-figure (b) indicates the possibilities of migration from the $S_c$ compartment to the $S_b$.}
\label{fig6}
}
\end{figure}

%

We denote the probability of moving from the state $a$ to the state $b$ in one time step by $P_{a \rightarrow b}$, where $a,b\in \{(e^*, b^*, d^*, d^{**}, f^*, f^{**}\}$. For simplicity, indexes $a$ and $b$ only includes the parameter(s), which are changing. For example, the probability $P_{e^* \rightarrow e^*+1}$ is the probability of moving from the state, which has $e^*$ number of $S_c$ mutants, to the state that has $e^*+1$ number of $S_c$ mutants in one time step, while the number of the other mutants ($b^*, d^*, d^{**}, f^*, f^{**}$) has not changed. All possible non-zero transition probabilities are listed as follows.

\subsection*{Transition Probabilities}

%
\begin{itemize}
\item[(1)] $P_{f^*\rightarrow f^*+1} = $\\
$\left( \frac{f}{f+ f^*}  \right)^2 \,\left\{ 2\, \lambda_f \, \left( \frac{f}{\cal F}  \right) \,  \left(\frac{r_1\,f^*}{{\cal F}}  \right)\,(1-v) \right\} +  \frac{2 f f^*}{(f+ f^*)^2}   \,\left\{  \lambda_f \, \left[ \frac{r_1\,f^*}{{\cal F}} \,(1-v)  \right]^2 \right. $\\
$ \left. + (1-\lambda_f)\,\frac{r_1 d^*}{{\cal D}}\, (1-u)\,\left[ (1-\lambda_s)\,\frac{r_1 d^*}{{\cal D}}\,(1-u) + \lambda_s\,(1-\sigma)\,\frac{r_1 b^*}{{\cal R}_b}  \right] \ \right\}, $

\item[(2)]  $P_{f^*\rightarrow f^*-1} =$\\
$  \frac{2 f f^*}{(f+ f^*)^2} \,\left\{  \lambda_f \, \left( \frac{f}{{\cal F}} \right)^2 + (1-\lambda_f)\,\frac{d}{{\cal D}} \left[ (1-\lambda_s)\, \frac{d}{{\cal D}}  + \lambda_s\,(1-\sigma)\,\frac{b}{{\cal R}_b}  + \lambda_s\,\sigma \left( \delta\frac{b}{{\cal R}_b}  \right.\right.\right.$\\
$\left. \left. \left.   + (1-\delta)(1-\gamma)\frac{b}{{\cal R}_b} \left( (1-\alpha) + \alpha\, \frac{b}{b+b^*}\, \frac{e}{e+e^*}  + \alpha\, \frac{b^*}{b+b^*}\, \frac{e^*}{e+e^*}   \right) + (1-\delta)\gamma\,\frac{e}{{\cal R}_c}\frac{e}{e+e^*}\right) \right]  \right\}  $\\
$ + \left( \frac{f^*}{f+ f^*}  \right)^2 \left\{ 2\,\lambda_f \left(\frac{f}{{\cal F}} \right)\left(\frac{r_1 f^*}{{\cal F}}\,(1-v)\right)  \right\}, $
\item[(3)]  $P_{f^*\rightarrow f^*+2}=$\\
$\left( \frac{f}{f+ f^*}  \right)^2 \,\left\{ \lambda_f \, \left[ \frac{r_1\,f^*}{{\cal F}} \,(1-v) \right]^2 + (1-\lambda_f)\,\,\frac{r_1 d^*}{{\cal D}}\,(1-u) \left[ (1-\lambda_s)\,\frac{r_1 d^*}{{\cal D}}\,(1-u)   \right.\right.$\\
$\left.\left. + \lambda_s\,(1-\sigma)\,\frac{r_1 b^*}{{\cal R}_b}  \right] \right\},$
{
\item[(4)]  $P_{f^*\rightarrow f^*-2}=\left( \frac{f^*}{f+ f^*}  \right)^2 \,\left\{  \lambda_f \, \left( \frac{f}{{\cal F}} \right)^2 + (1-\lambda_f)\,\frac{d}{{\cal D}} \left[ (1-\lambda_s)\, \frac{d}{{\cal D}} + \lambda_s\,(1-\sigma)\,\frac{b}{{\cal R}_b} + \lambda_s\,\sigma \left( \delta\frac{b}{{\cal R}_b}    \right.\right. \right.$\\
$\left. \left. \left.   + (1-\delta)(1-\gamma)\frac{b}{{\cal R}_b} \left( (1-\alpha) + \alpha\, \frac{b}{b+b^*}\,\frac{e}{e+e^*}   + \alpha\, \frac{b^*}{b+b^*}\, \frac{e^*}{e+e^*} \right) + (1-\delta)\gamma\,\frac{e}{{\cal R}_c}\frac{e}{e+e^*}\right) \right]  \right\}, $
}
\item[(5)]  $P_{f^*,f^{**}\rightarrow f^*-1,f^{**}+1}  = \frac{2 f f^*}{(f+ f^*)^2} \, \left\{  2\, \lambda_f\, \left( \frac{f}{\cal F} \right) \left[ \frac{r_1\,f^{*}}{{\cal F}}\,v+ \frac{r_2\,f^{**}}{{\cal F}}  \right]\right\} + \left( \frac{f^*}{f+ f^*}  \right)^2\, \left\{ 2\, \lambda_f  \left( \frac{r_1\,f^*}{{\cal F}} \, (1-v)\right)\,  \left[  \frac{r_1\,f^*}{{\cal F}}\,v + \frac{r_2\,f^{**} }{{\cal F}} \right]   \right.$\\
$ \left. + (1-\lambda_f)\, \frac{r_1 d^*}{{\cal D}}\,u\,\left[ (1-\lambda_s)\, \frac{r_1 d^*}{{\cal D}}\,(1-u) + \lambda_s\,(1-\sigma)\,\frac{r_1 b^*}{{\cal R}_b}  \right] \right\}, $
\item[(6)]  $P_{f^*,f^{**}\rightarrow f^*+1,f^{**}+1}=\left( \frac{f}{f+ f^*}  \right)^2\,\left\{  2\, \lambda_f\,\left( \frac{r_1\,f^{*}}{\cal  F} \,(1-v) \right)\,\left[ \frac{r_1 f^*}{{\cal F}}\,v  + \frac{r_2 f^{**}}{{\cal F}}  \right]   \right.$\\
$\left.  + (1-\lambda_f)\,\frac{r_1 d^*}{{\cal D}}\,u\,\left[ (1-\lambda_s)\,\frac{r_1 d^*}{{\cal D}}\,(1-u) + \lambda_s\,(1-\sigma)\,\frac{r_1 b^*}{{\cal R}_b} \right]  \right\},$
\item[(7)]  $P_{f^*,f^{**}\rightarrow f^*-2,f^{**}+1}=\left( \frac{f^*}{f+ f^*}  \right)^2\,\left\{  2\, \lambda_f\,\left( \frac{f }{\cal  F}  \right)\, \left[ \frac{r_1\,f^{*}}{\cal  F}\,v + \frac{r_2 f^{**}}{{\cal F}}  \right]  \right\},$
{
\item[(8)]  $P_{d^*,f^*\rightarrow d^*-1,f^*+2}=\left( \frac{f}{f+ f^*}  \right)^2\,\left\{ (1-\lambda_f)\,\frac{r_1\,d^{*}}{\cal  D} \, (1-u) \left[ (1-\lambda_s)\, \frac{d}{{\cal D}} + \lambda_s\,(1-\sigma)\,\frac{b}{{\cal R}_b}  \right.\right. $\\
$ + \lambda_s\,\sigma\,\left( \delta\,\frac{b}{{\cal R}_b}  + (1-\delta)(1-\gamma)\frac{b}{{\cal R}_b} \left( (1-\alpha) + \alpha\,  \frac{b}{b+b^*}\,\frac{e}{e+e^*} + \alpha\, \frac{b^*}{b+b^*}\, \frac{e^*}{e+e^*}  \right) \right. $\\
$\left. \left. \left. +  (1-\delta)\gamma\,\frac{e}{{\cal R}_c}\frac{e}{e+e^*} \right) \right]  \right\},$
\item[(9)]  $P_{d^*,f^*\rightarrow d^*-1,f^*+1}= \frac{2 f f^*}{(f+ f^*)^2} \,\left\{ (1-\lambda_f)\,\frac{r_1\,d^{*}}{\cal  D} \, (1-u) \left[  (1-\lambda_s)\, \frac{d}{{\cal D}} + \lambda_s\,(1-\sigma)\,\frac{b}{{\cal R}_b}  \right.\right. $\\
$ + \lambda_s\,\sigma\,\left( \delta\,\frac{b}{{\cal R}_b}  + (1-\delta)(1-\gamma)\frac{b}{{\cal R}_b} \left( (1-\alpha) + \alpha\,  \frac{b}{b+b^*}\,\frac{e}{e+e^*} + \alpha\, \frac{b^*}{b+b^*}\, \frac{e^*}{e+e^*}  \right) \right. $\\
$\left. \left. \left. +  (1-\delta)\gamma\,\frac{e}{{\cal R}_c}\frac{e}{e+e^*} \right) \right]  \right\},$
\item[(10)]  $P_{d^*,f^*,f^{**}\rightarrow d^*-1,f^*+1,f^{**}+1}=\left( \frac{f}{f+ f^*}  \right)^2\,\left\{ (1-\lambda_f)\,\frac{r_1\,d^{*}}{\cal  D} \,u \left[   (1-\lambda_s)\, \frac{d}{{\cal D}} + \lambda_s\,(1-\sigma)\,\frac{b}{{\cal R}_b}  \right.\right. $\\
$ + \lambda_s\,\sigma\,\left( \delta\,\frac{b}{{\cal R}_b}  + (1-\delta)(1-\gamma)\frac{b}{{\cal R}_b} \left( (1-\alpha) + \alpha\,  \frac{b}{b+b^*}\,\frac{e}{e+e^*} + \alpha\, \frac{b^*}{b+b^*}\, \frac{e^*}{e+e^*}  \right) \right. $\\
$\left. \left. \left. +  (1-\delta)\gamma\,\frac{e}{{\cal R}_c}\frac{e}{e+e^*} \right) \right]  \right\},$
\item[(11)]  $P_{d^*,f^*,f^{**}\rightarrow d^*-1,f^*-1,f^{**}+1}=\left( \frac{f^*}{f+ f^*}  \right)^2 \,\left\{ (1-\lambda_f)\,\frac{r_1\,d^{*}}{\cal  D} \,u \left[  (1-\lambda_s)\, \frac{d}{{\cal D}} + \lambda_s\,(1-\sigma)\,\frac{b}{{\cal R}_b}  \right.\right. $\\
$ + \lambda_s\,\sigma\,\left( \delta\,\frac{b}{{\cal R}_b}  + (1-\delta)(1-\gamma)\frac{b}{{\cal R}_b} \left( (1-\alpha) + \alpha\,  \frac{b}{b+b^*}\,\frac{e}{e+e^*} + \alpha\, \frac{b^*}{b+b^*}\, \frac{e^*}{e+e^*}  \right) \right. $\\
$\left. \left. \left. +  (1-\delta)\gamma\,\frac{e}{{\cal R}_c}\frac{e}{e+e^*} \right) \right]  \right\},$
\item[(12)]  $P_{d^*,f^{**}\rightarrow d^*-1,f^{**}+1}=\ \frac{2 f f^*}{(f+ f^*)^2}  \,\left\{ (1-\lambda_f)\,\frac{r_1\,d^{*}}{\cal  D} \,u \left[   (1-\lambda_s)\, \frac{d}{{\cal D}} + \lambda_s\,(1-\sigma)\,\frac{b}{{\cal R}_b}  \right.\right. $\\
$ + \lambda_s\,\sigma\,\left( \delta\,\frac{b}{{\cal R}_b}  + (1-\delta)(1-\gamma)\frac{b}{{\cal R}_b} \left( (1-\alpha) + \alpha\,  \frac{b}{b+b^*}\,\frac{e}{e+e^*} + \alpha\, \frac{b^*}{b+b^*}\, \frac{e^*}{e+e^*}  \right) \right. $\\
$\left. \left. \left. +  (1-\delta)\gamma\,\frac{e}{{\cal R}_c}\frac{e}{e+e^*} \right) \right]  \right\},$
}
\item[(13)]  $P_{f^{**}\rightarrow f^{**}+1}= \left( \frac{f}{f+ f^*}  \right)^2\,\left\{ 2\,\lambda_f\,\frac{f}{{\cal F}}\,\left[ \frac{r_1 f^*}{{\cal F}}\,v + \frac{r_2 f^{**}}{{\cal F}}  \right] \right\} $\\
$ +  \frac{2 f f^*}{(f+ f^*)^2}  \,\left\{ 2\,\lambda_f\,\frac{r_1\,f^*}{\cal F}\,(1-v)\left[ \frac{r_1\,f^*}{\cal F}\,v + \frac{r_2\,f^{**} }{\cal F} \right] + (1-\lambda_f)\,\frac{r_1\,d^{*}}{\cal  D} \,u \left[ (1-\lambda_s)\,\frac{r_1 d^*}{{\cal D}}\, (1-u)  \right.\right. $\\
$\left.\left. + \lambda_s\,(1-\sigma)\,\frac{r_1 b^*}{{\cal R}_b}  \right]  \right\},$
\item[(14)]  $P_{f^{**}\rightarrow f^{**}+2}= \left( \frac{f}{f+ f^*}  \right)^2\,\left\{ \lambda_f\, \left[ \frac{r_1 f^*}{{\cal F}}\,v + \frac{r_2 f^{**}}{{\cal F}} \right]^2 + (1-\lambda_f)\,\frac{r_2 d^{**}}{{\cal D}}\,\left[ (1-\lambda_s)\,\left( \frac{r_1 d^*}{{\cal D}}\,u + \frac{r_2 d^{**}}{{\cal D}} \right)  \right]\right\},$
\item[(15)]  $P_{f^*,f^{**}\rightarrow f^*-2,f^{**}+2} = \left( \frac{f^*}{f+ f^*}  \right)^2\,\left\{ \lambda_f\, \left[ \frac{r_1 f^*}{{\cal F}}\,v + \frac{r_2 f^{**}}{{\cal F}} \right]^2 + (1-\lambda_f)\,\frac{r_2 d^{**}}{{\cal D}}\,\left[ (1-\lambda_s)\,\left( \frac{r_1 d^*}{{\cal D}}\,u + \frac{r_2 d^{**}}{{\cal D}} \right)  \right]\right\},$
\item[(16)]  $P_{f^*,f^{**}\rightarrow f^*-1,f^{**}+2}=  \frac{ 2 f f^*}{(f+ f^*)^2}  \,\left\{ \lambda_f\, \left[ \frac{r_1 f^*}{{\cal F}}\,v + \frac{r_2 f^{**}}{{\cal F}} \right]^2 + (1-\lambda_f)\,\frac{r_2 d^{**}}{{\cal D}}\,\left[ (1-\lambda_s)\,\left( \frac{r_1 d^*}{{\cal D}}\,u + \frac{r_2 d^{**}}{{\cal D}} \right)  \right]\right\},$
\item[(17)]  $P_{d^*\rightarrow d^*+1}= \left( \frac{f}{f + f^*} \right)^2 \, \left\{(1-\lambda_f)\,\frac{d}{{\cal D}}\, \left[  (1-\lambda_s)\,\frac{r_1 d^*}{{\cal D}}\,(1-u) + \lambda_s\,(1-\sigma)\,\frac{r_1 b^*}{{\cal R}_b}  \right] \right\},$
\item[(18)]  $P_{d^*,f^*\rightarrow d^*+1,f^*-1}= \frac{ 2 f f^*}{(f+ f^*)^2}  \, \left\{(1-\lambda_f)\,\frac{d}{{\cal D}}\, \left[  (1-\lambda_s)\,\frac{r_1 d^*}{{\cal D}}\,(1-u) + \lambda_s\,(1-\sigma)\,\frac{r_1 b^*}{{\cal R}_b}  \right] \right\},$
\item[(19)]  $P_{d^*,f^*\rightarrow d^*+1, f^*-2}= \left( \frac{f^*}{f + f^*} \right)^2 \, \left\{(1-\lambda_f)\,\frac{d}{{\cal D}}\, \left[  (1-\lambda_s)\,\frac{r_1 d^*}{{\cal D}}\,(1-u) + \lambda_s\,(1-\sigma)\,\frac{r_1 b^*}{{\cal R}_b}  \right] \right\},$
\item[(20)]  $P_{b^*,d^*,f^*\rightarrow b^*-1,d^*+2, f^*-2} = \left( \frac{f^*}{f + f^*} \right)^2 \, \left\{(1-\lambda_f)\,\frac{d}{{\cal D}}\, \left[\lambda_s\,\sigma\,\delta\,\frac{r_1 b^*}{{\cal R}_b}  \right] \right\},$
\item[(21)]  $P_{b^*,d^*,f^*\rightarrow b^*-1,d^*+2, f^*-1}=\frac{ 2 f f^*}{(f+ f^*)^2}  \,\left\{(1-\lambda_f)\,\frac{d}{{\cal D}}\, \left[\lambda_s\,\sigma\,\delta\,\frac{r_1 b^*}{{\cal R}_b}  \right] \right\},$
{
\item[(22)]  $P_{d^*\rightarrow d^*-1} = \left( \frac{f^*}{f+ f^*}  \right)^2 \, \left\{ (1-\lambda_f)\,\frac{r_1\,d^{*}}{\cal  D} \, (1-u) \left[  (1-\lambda_s)\, \frac{d}{{\cal D}} + \lambda_s\,(1-\sigma)\,\frac{b}{{\cal R}_b}  \right.\right. $\\
$ + \lambda_s\,\sigma\,\left( \delta\,\frac{b}{{\cal R}_b}  + (1-\delta)(1-\gamma)\frac{b}{{\cal R}_b} \left( (1-\alpha) + \alpha\,  \frac{b}{b+b^*}\,\frac{e}{e+e^*} + \alpha\, \frac{b^*}{b+b^*}\, \frac{e^*}{e+e^*}  \right) \right. $\\
$\left. \left. \left. +  (1-\delta)\gamma\,\frac{e}{{\cal R}_c}\frac{e}{e+e^*} \right) \right]  \right\},$
\item[(23)]  $P_{d^{**},f^*,f^{**}\rightarrow d^{**}-1,f^*-2,f^{**}+2}= \left( \frac{f^*}{f+ f^*}  \right)^2\,\left\{ (1-\lambda_f)\, \frac{r_2 d^{**}}{{\cal D}}\, \left[  (1-\lambda_s)\, \frac{d}{{\cal D}} + \lambda_s\,(1-\sigma)\,\frac{b}{{\cal R}_b}  \right.\right. $\\
$ + \lambda_s\,\sigma\,\left( \delta\,\frac{b}{{\cal R}_b}  + (1-\delta)(1-\gamma)\frac{b}{{\cal R}_b} \left( (1-\alpha) + \alpha\,  \frac{b}{b+b^*}\,\frac{e}{e+e^*} + \alpha\, \frac{b^*}{b+b^*}\, \frac{e^*}{e+e^*}  \right) \right. $\\
$\left. \left. \left. +  (1-\delta)\gamma\,\frac{e}{{\cal R}_c}\frac{e}{e+e^*} \right) \right]  \right\},$
}
\item[(24)]  $P_{d^{**}\rightarrow d^{**}+1} = \left( \frac{f}{f+ f^*}  \right)^2\, \left\{(1-\lambda_f)\,\frac{d}{{\cal D}}\, \left[ (1-\lambda_s)\,\left( \frac{r_1 d^*}{{\cal D}}\,u + \frac{r_2 d^{**}}{{\cal D}} \right)  \right] \right\},  $
\item[(25)]  $P_{d^{**},f^*\rightarrow d^{**}+1,f^*-1} = \frac{2 f f^*}{(f+ f^*)^2 }  \, \left\{(1-\lambda_f)\,\frac{d}{{\cal D}}\, \left[ (1-\lambda_s)\,\left( \frac{r_1 d^*}{{\cal D}}\,u + \frac{r_2 d^{**}}{{\cal D}} \right)  \right] \right\},  $
\item[(26)]  $P_{d^{**},f^*\rightarrow d^{**}+1,f^*-2} = \left( \frac{f^*}{f+ f^*}  \right)^2\, \left\{(1-\lambda_f)\,\frac{d}{{\cal D}}\, \left[ (1-\lambda_s)\,\left( \frac{r_1 d^*}{{\cal D}}\,u + \frac{r_2 d^{**}}{{\cal D}} \right)  \right] \right\},  $
\item[(27)]  $P_{b^*,d^{*} \rightarrow b^*-1,d^{*}+2} = \left( \frac{f}{f+ f^*}  \right)^2\, \left\{  (1-\lambda_f)\,\frac{d}{{\cal D}}\, \left[ \lambda_s\,\sigma\delta\,\frac{r_1 b^*}{{\cal R}_b} \right]  \right\},$
{
\item[(28)]  $P_{b^*,d^{*}, \rightarrow b^*+1,d^{*}-1} = \left( \frac{f^*}{f+ f^*}  \right)^2\, \left\{  (1-\lambda_f)\,\frac{r_1 d^*}{{\cal D}}\,(1-u) \left[ \lambda_s\,\sigma\,(1-\delta)\,\left(  (1-\gamma)\,\frac{r_1 b^*}{{\cal R}_b} \left(   (1-\alpha)   +\alpha\,  \frac{b^*}{b+b^*}\,\frac{e^*}{e+e^*} \right. \right.\right. \right. $\\
$\left.\left. \left. \left. +\alpha\,  \frac{b}{b+b^*}\,\frac{e}{e+e^*}  \right)   + \gamma\,\frac{r_1 e^*}{{\cal R}_c}\,\frac{e^*}{e+e^*}  \right)  \right]  \right\},$
}
\item[(29)]  $P_{b^*,d^{*}\rightarrow b^*-1,d^{*}+1} = \left( \frac{f^*}{f+ f^*}  \right)^2\, \left\{  (1-\lambda_f)\,\frac{r_1 d^*}{{\cal D}}\,(1-u) \left[ \lambda_s\,\sigma\,\delta\,\frac{r_1 b^*}{{\cal R}_b} \right]  \right\},$
{
\item[(30)]  $P_{b^*\rightarrow b^*+1} =  \left( \frac{f}{f+ f^*}  \right)^2\, \left\{ (1-\lambda_f)\,\frac{d}{{\cal D}}\,\left[  \lambda_s\,\sigma\,(1-\delta)\,\left(  (1-\gamma)\,\frac{r_1 b^*}{{\cal R}_b} \left(   (1-\alpha)   +\alpha\,  \frac{b^*}{b+b^*}\,\frac{e^*}{e+e^*} \right. \right.\right. \right. $\\
$\left.\left. \left. \left. +\alpha\,  \frac{b}{b+b^*}\,\frac{e}{e+e^*}  \right)   + \gamma\,\frac{r_1 e^*}{{\cal R}_c}\,\frac{e^*}{e+e^*}  \right)  \right]  \right\},$
\item[(31)]  $P_{b^*,f^*\rightarrow b^*+1,f^*-1} =   \frac{2 f f^*}{(f+ f^*)^2} \, \left\{ (1-\lambda_f)\,\frac{d}{{\cal D}}\,\left[  \lambda_s\,\sigma\,(1-\delta)\,\left(  (1-\gamma)\,\frac{r_1 b^*}{{\cal R}_b} \left(   (1-\alpha)   +\alpha\,  \frac{b^*}{b+b^*}\,\frac{e^*}{e+e^*} \right. \right.\right. \right. $\\
$\left.\left. \left. \left. +\alpha\,  \frac{b}{b+b^*}\,\frac{e}{e+e^*}  \right)   + \gamma\,\frac{r_1 e^*}{{\cal R}_c}\,\frac{e^*}{e+e^*}  \right)  \right]  \right\},$
\item[(32)]   $P_{b^*,f^*\rightarrow b^*+1,f^*-2} = \left( \frac{f^*}{f+ f^*}  \right)^2\, \left\{ (1-\lambda_f)\,\frac{d}{{\cal D}}\,\left[ \lambda_s\,\sigma\,(1-\delta)\,\left(  (1-\gamma)\,\frac{r_1 b^*}{{\cal R}_b} \left(   (1-\alpha)   +\alpha\,  \frac{b^*}{b+b^*}\,\frac{e^*}{e+e^*} \right. \right.\right. \right. $\\
$\left.\left. \left. \left. +\alpha\,  \frac{b}{b+b^*}\,\frac{e}{e+e^*}  \right)   + \gamma\,\frac{r_1 e^*}{{\cal R}_c}\,\frac{e^*}{e+e^*}  \right)  \right]  \right\},$
\item[(33)]  $P_{e^*\rightarrow e^*+1} = \left( \frac{f}{f+ f^*}  \right)^2\, \left\{ (1-\lambda_f)\,\frac{d}{{\cal D}}\, \left[  \lambda_s\,\sigma\,(1-\delta)\,\left(  (1-\gamma)\,\frac{r_1 b^*}{{\cal R}_b} \, \alpha \, \frac{b^*}{b+b^*}+ \gamma\,\frac{r_1 e^*}{{\cal R}_c} \right) \,\frac{e}{e+e^*}  \right] \right\},$
\item[(34)]  $P_{d^{**},f^{**}\rightarrow d^{**}-1,f^{**}+2}= \left( \frac{f}{f+ f^*}  \right)^2\,\left\{ (1-\lambda_f)\, \frac{r_2 d^{**}}{{\cal D}}\, \left[  (1-\lambda_s)\, \frac{d}{{\cal D}} + \lambda_s\,(1-\sigma)\,\frac{b}{{\cal R}_b}  \right.\right. $\\
$ + \lambda_s\,\sigma\,\left( \delta\,\frac{b}{{\cal R}_b}  + (1-\delta)(1-\gamma)\frac{b}{{\cal R}_b} \left( (1-\alpha) + \alpha\,  \frac{b}{b+b^*}\,\frac{e}{e+e^*} + \alpha\, \frac{b^*}{b+b^*}\, \frac{e^*}{e+e^*}  \right) \right. $\\
$\left. \left. \left. +  (1-\delta)\gamma\,\frac{e}{{\cal R}_c}\frac{e}{e+e^*} \right) \right]  \right\},$
\item[(35)]  $P_{e^*, d^*\rightarrow e^*+1,d^*-1} = \left( \frac{f^*}{f+ f^*}  \right)^2\, \left\{ (1-\lambda_f)\,\frac{r_1 d^*}{{\cal D}}\,(1-u)\, \left[  \lambda_s\,\sigma\,(1-\delta)\,\left(  (1-\gamma)\,\frac{r_1 b^*}{{\cal R}_b} \, \alpha\,  \frac{b^*}{b+b^*}+ \gamma\,\frac{r_1 e^*}{{\cal R}_c}   \right) \,\frac{e}{e+e^*}   \right] \right\},$
\item[(36)]  $P_{e^*,b^*\rightarrow e^*-1, b^*+1} = \left( \frac{f}{f+ f^*}  \right)^2\, \left\{ (1-\lambda_f)\, \frac{d}{{\cal D}}\, \left[ \lambda_s\,\sigma\,(1-\delta)\,\left(  (1-\gamma)\,\frac{b}{{\cal R}_b}\,\alpha\, \frac{b}{b+b^*} +   \gamma\,\frac{e}{{\cal R}_c}    \right) \, \frac{e^*}{e + e^*}     \right] \right\},$
\item[(37)]  $P_{e^*,b^*, d^*\rightarrow e^*-1, b^*+1,d^*-1} = \left( \frac{f^*}{f+ f^*}  \right)^2\, \left\{ (1-\lambda_f)\, \frac{r_1 d^*}{{\cal D}}\,(1-u) \left[ \lambda_s\,\sigma\,(1-\delta)\,\left(  (1-\gamma)\,\frac{b}{{\cal R}_b}\,\alpha\, \frac{b}{b+b^*}  +   \gamma\,\frac{e}{{\cal R}_c}    \right) \, \frac{e^*}{e + e^*}     \right] \right\},$
\item[(38)]  $P_{e^*,d^*, f^{**}\rightarrow e^*+1, d^*-1,f^{**}+1} = \frac{2 f f^*}{(f+ f^*)^2}  \, \left\{ (1-\lambda_f)\, \frac{r_1 d^*}{{\cal D}}\,u \left[  \lambda_s\,\sigma\,(1-\delta)\,\left(  (1-\gamma)\,\frac{r_1 b^*}{{\cal R}_b} \, \alpha\, \frac{b^*}{b+b^*} + \gamma\,\frac{r_1 e^*}{{\cal R}_c}  \right)\,\frac{e}{e+e^*}   \right] \right\},$
}
\item[(39)]  $P_{d^*,d^{**},f^*\rightarrow d^*-1,d^{**}+1,f^*+2} = \left( \frac{f}{f+ f^*}  \right)^2\, \left\{ (1-\lambda_f)\,\frac{r_1 d^*}{{\cal D}}\,(1-u) \,\left[ (1-\lambda_s)\, \left( \frac{r_1 d^{*}}{{\cal D}}\,u + \frac{r_2 d^{**}}{{\cal D}} \right)  \right] \right\},$
\item[(40)]  $P_{d^*,d^{**},f^*\rightarrow d^*-1,d^{**}+1,f^*+1} =  \frac{2 f f^*}{(f+ f^*)^2}  \, \left\{ (1-\lambda_f)\,\frac{r_1 d^*}{{\cal D}}\,(1-u) \,\left[ (1-\lambda_s)\, \left( \frac{r_1 d^{*}}{{\cal D}}\,u + \frac{r_2 d^{**}}{{\cal D}} \right)  \right] \right\},$
\item[(41)]  $P_{d^*,d^{**}\rightarrow d^*-1,d^{**}+1} = \left( \frac{f^*}{f+ f^*}  \right)^2\, \left\{ (1-\lambda_f)\,\frac{r_1 d^*}{{\cal D}}\,(1-u) \,\left[ (1-\lambda_s)\, \left( \frac{r_1 d^{*}}{{\cal D}}\,u + \frac{r_2 d^{**}}{{\cal D}} \right)  \right] \right\},$
\item[(42)]  $P_{b^{*},d^*,f^*\rightarrow b^{*}-1,d^*+1,f^*+2}= \left( \frac{f}{f+ f^*}  \right)^2\, \left\{ (1-\lambda_f)\,\frac{r_1 d^*}{{\cal D}}\,(1-u) \,\left[ \lambda_s\,\sigma\,\delta\, \frac{r_1 b^{*}}{{\cal R}_b} \right] \right\},$
\item[(43)]  $P_{b^{*},d^*,f^*\rightarrow b^{*}-1,d^*+1,f^*+1} = \frac{2 f f^*}{(f+ f^*)^2}  \, \left\{ (1-\lambda_f)\,\frac{r_1 d^*}{{\cal D}}\,(1-u) \,\left[ \lambda_s\,\sigma\,\delta\, \frac{r_1 b^{*}}{{\cal R}_b} \right] \right\},$
{
\item[(44)]  $P_{b^{*},d^*,f^*,f^{**}\rightarrow b^{*}+1,d^*-1,f^*+1,f^{**}+1} = \left( \frac{f}{f+ f^*}  \right)^2\, \left\{ (1-\lambda_f)\,\frac{r_1 d^*}{{\cal D}}\, u \,\left[  \lambda_s\,\sigma\,(1-\delta)\,\left(  (1-\gamma)\,\frac{r_1 b^*}{{\cal R}_b} \left(   (1-\alpha)   +\alpha\,  \frac{b^*}{b+b^*}\,\frac{e^*}{e+e^*} \right. \right.\right. \right. $\\
$\left.\left. \left. \left. +\alpha\,  \frac{b}{b+b^*}\,\frac{e}{e+e^*}  \right)   + \gamma\,\frac{r_1 e^*}{{\cal R}_c}\,\frac{e^*}{e+e^*}  \right)  \right]  \right\},$
\item[(45)]  $P_{b^{*},d^*,f^*,f^{**}\rightarrow b^{*}+1,d^*-1,f^*-1, f^{**}+1} = \left( \frac{f^*}{f+ f^*}  \right)^2\, \left\{ (1-\lambda_f)\,\frac{r_1 d^*}{{\cal D}}\, u \,\left[  \lambda_s\,\sigma\,(1-\delta)\,\left(  (1-\gamma)\,\frac{r_1 b^*}{{\cal R}_b} \left(   (1-\alpha)   +\alpha\,  \frac{b^*}{b+b^*}\,\frac{e^*}{e+e^*} \right. \right.\right. \right. $\\
$\left.\left. \left. \left. +\alpha\,  \frac{b}{b+b^*}\,\frac{e}{e+e^*}  \right)   + \gamma\,\frac{r_1 e^*}{{\cal R}_c}\,\frac{e^*}{e+e^*}  \right)  \right]  \right\},$
\item[(46)]  $P_{b^{*},d^*,f^{**}\rightarrow b^{*}+1,d^*-1,f^{**}+1} = \frac{2 f f^*}{(f+ f^*)^2} \, \left\{ (1-\lambda_f)\,\frac{r_1 d^*}{{\cal D}}\, u \,\left[ \lambda_s\,\sigma\,(1-\delta)\,\left(  (1-\gamma)\,\frac{r_1 b^*}{{\cal R}_b} \left(   (1-\alpha)   +\alpha\,  \frac{b^*}{b+b^*}\,\frac{e^*}{e+e^*} \right. \right.\right. \right. $\\
$\left.\left. \left. \left. +\alpha\,  \frac{b}{b+b^*}\,\frac{e}{e+e^*}  \right)   + \gamma\,\frac{r_1 e^*}{{\cal R}_c}\,\frac{e^*}{e+e^*}  \right)  \right]  \right\},$
}
\item[(47)]  $P_{b^{*},d^*,f^*,f^{**}\rightarrow b^{*}-1,d^*+1,f^*+1,f^{**}+1} = \left( \frac{f}{f+ f^*}  \right)^2\, \left\{ (1-\lambda_f)\,\frac{r_1 d^*}{{\cal D}}\, u \,\left[ \lambda_s\,\sigma\,\delta\,\frac{r_1 b^{*}}{{\cal R}_b}  \right] \right\},$
\item[(48)]  $P_{b^{*},d^*,f^*,f^{**}\rightarrow b^{*}-1,d^*+1,f^*-1,f^{**}+1} = \left( \frac{f^*}{f+ f^*}  \right)^2\, \left\{ (1-\lambda_f)\,\frac{r_1 d^*}{{\cal D}}\, u \,\left[ \lambda_s\,\sigma\,\delta\,\frac{r_1 b^{*}}{{\cal R}_b}  \right] \right\},$
\item[(49)]  $P_{b^{*},d^*,f^{**}\rightarrow b^{*}-1,d^*+1,f^{**}+1} =  \frac{2 f f^*}{(f+ f^*)^2}  \, \left\{ (1-\lambda_f)\,\frac{r_1 d^*}{{\cal D}}\, u \,\left[ \lambda_s\,\sigma\,\delta\,\frac{r_1 b^{*}}{{\cal R}_b}  \right] \right\},$
{
\item[(50)]  $P_{d^{**},f^*,f^{**}\rightarrow d^{**}-1,f^*-1,f^{**}+2}=  \frac{2 f f^*}{(f+ f^*)^2} \,\left\{ (1-\lambda_f)\, \frac{r_2 d^{**}}{{\cal D}}\, \left[  (1-\lambda_s)\, \frac{d}{{\cal D}} + \lambda_s\,(1-\sigma)\,\frac{b}{{\cal R}_b}  \right.\right. $\\
$ + \lambda_s\,\sigma\,\left( \delta\,\frac{b}{{\cal R}_b}  + (1-\delta)(1-\gamma)\frac{b}{{\cal R}_b} \left( (1-\alpha) + \alpha\,  \frac{b}{b+b^*}\,\frac{e}{e+e^*} + \alpha\, \frac{b^*}{b+b^*}\, \frac{e^*}{e+e^*}  \right) \right. $\\
$\left. \left. \left. +  (1-\delta)\gamma\,\frac{e}{{\cal R}_c}\frac{e}{e+e^*} \right) \right]  \right\},$
\item[(51)] $P_{e^*,b^*, d^*, f^{**}\rightarrow e^*-1, b^*+1, d^*-1,f^{**}+1} = \frac{2 f f^*}{(f+ f^*)^2}  \, \left\{ (1-\lambda_f)\, \frac{r_1 d^*}{{\cal D}}\,u \left[ \lambda_s\,\sigma\,(1-\delta)\,\left(  (1-\gamma)\,\frac{b}{{\cal R}_b}\,\alpha \, \frac{b}{b+b^*} +   \gamma\,\frac{e}{{\cal R}_c}    \right) \, \frac{e^*}{e + e^*}     \right] \right\},$
\item[(52)]  $P_{b^*,d^{*}, f^*\rightarrow b^*+1,d^{*}-1, f^*+1} = \frac{2 f f^*}{(f+ f^*)^2}  \,\left\{  (1-\lambda_f)\,\frac{r_1 d^*}{{\cal D}}\,(1-u) \left[ \lambda_s\,\sigma\,(1-\delta)\,\left(  (1-\gamma)\,\frac{r_1 b^*}{{\cal R}_b} \left(   (1-\alpha)   +\alpha\,  \frac{b^*}{b+b^*}\,\frac{e^*}{e+e^*} \right. \right.\right. \right. $\\
$\left.\left. \left. \left. +\alpha\,  \frac{b}{b+b^*}\,\frac{e}{e+e^*}  \right)   + \gamma\,\frac{r_1 e^*}{{\cal R}_c}\,\frac{e^*}{e+e^*}  \right)  \right]  \right\},$
\item[(53)]  $P_{b^*,d^{*}, f^*\rightarrow b^*+1,d^{*}-1, f^*+2} = \left( \frac{f}{f+ f^*}  \right)^2\,  \left\{  (1-\lambda_f)\,\frac{r_1 d^*}{{\cal D}}\,(1-u) \left[  \lambda_s\,\sigma\,(1-\delta)\,\left(  (1-\gamma)\,\frac{r_1 b^*}{{\cal R}_b} \left(   (1-\alpha)   +\alpha\,  \frac{b^*}{b+b^*}\,\frac{e^*}{e+e^*} \right. \right.\right. \right. $\\
$\left.\left. \left. \left. +\alpha\,  \frac{b}{b+b^*}\,\frac{e}{e+e^*}  \right)   + \gamma\,\frac{r_1 e^*}{{\cal R}_c}\,\frac{e^*}{e+e^*}  \right)  \right]  \right\},$
\item[(54)]  $P_{e^*,b^*,d^*, f^*\rightarrow e^*-1, b^*+1,d^*-1,f^*+1} = \frac{2 f f^*}{(f+ f^*)^2}  \, \left\{ (1-\lambda_f)\, \frac{r_1 d^*}{{\cal D}}\,(1-u) \left[ \lambda_s\,\sigma\,(1-\delta)\,\left(  (1-\gamma)\,\frac{b}{{\cal R}_b}\,\alpha \, \frac{b}{b+b^*} +   \gamma\,\frac{e}{{\cal R}_c}    \right) \, \frac{e^*}{e + e^*}     \right] \right\},$
\item[(55)]  $P_{e^*,b^*, d^*, f^*\rightarrow e^*-1, b^*+1,d^*-1,f^*+2} = \left( \frac{f}{f+ f^*}  \right)^2\, \left\{ (1-\lambda_f)\, \frac{r_1 d^*}{{\cal D}}\,(1-u) \left[ \lambda_s\,\sigma\,(1-\delta)\,\left(  (1-\gamma)\,\frac{b}{{\cal R}_b}\,\alpha \, \frac{b}{b+b^*}  +   \gamma\,\frac{e}{{\cal R}_c}    \right) \, \frac{e^*}{e + e^*}     \right] \right\},$
\item[(56)]  $P_{e^*,f^*\rightarrow e^*+1, f^*-1} = \frac{2 f f^*}{(f+ f^*)^2} \, \left\{ (1-\lambda_f)\, \frac{d}{{\cal D}}\, \left[ \lambda_s\,\sigma\,(1-\delta)\, \left(  (1-\gamma)\,\frac{r_1\,b^*}{{\cal R}_b}\,\alpha \, \frac{b^*}{b+b^*} + \gamma\,\frac{r_1\,e^*}{{\cal R}_c}  \right) \, \frac{e}{e + e^*}  \right] \right\},$
\item[(57)]  $P_{e^*,f^*\rightarrow e^*+1, f^*-2} = \left( \frac{f^* }{f+ f^*}  \right)^2\, \left\{ (1-\lambda_f)\, \frac{d}{{\cal D}}\, \left[ \lambda_s\,\sigma\,(1-\delta)\, \left(  (1-\gamma)\,\frac{r_1\,b^*}{{\cal R}_b}\,\alpha  \, \frac{b^*}{b+b^*} + \gamma\,\frac{r_1\,e^*}{{\cal R}_c}  \right) \, \frac{e}{e + e^*}  \right] \right\},$
\item[(58)]  $P_{e^*,d^*,f^*\rightarrow e^*+1,d^*-1, f^*+1} = \frac{2 f f^*}{(f+ f^*)^2} \,\left\{ (1-\lambda_f)\, \frac{r_1\,d^*}{{\cal D}}\,(1-u)\, \left[ \lambda_s\,\sigma\,(1-\delta)\, \left(  (1-\gamma)\,\frac{r_1\,b^*}{{\cal R}_b}\,\alpha  \, \frac{b^*}{b+b^*} + \gamma\,\frac{r_1\,e^*}{{\cal R}_c} \right) \, \frac{e}{e + e^*}   \right] \right\},$
\item[(59)]  $P_{e^*,d^*,f^*\rightarrow e^*+1,d^*-1, f^*+2} = \left( \frac{f }{f+ f^*}  \right)^2\, \left\{ (1-\lambda_f)\, \frac{r_1\,d^*}{{\cal D}}\,(1-u)\, \left[ \lambda_s\,\sigma\,(1-\delta)\, \left(  (1-\gamma)\,\frac{r_1\,b^*}{{\cal R}_b}\,\alpha  \, \frac{b^*}{b+b^*} + \gamma\,\frac{r_1\,e^*}{{\cal R}_c}  \right)\, \frac{e}{e + e^*}   \right] \right\},$
\item[(60)]  $P_{e^*,d^*,f^*,f^{**}\rightarrow e^*+1,d^*-1, f^*+1,f^{**}+1} = \left( \frac{f }{f+ f^*}  \right)^2\, \left\{ (1-\lambda_f)\, \frac{r_1\,d^*}{{\cal D}}\,u \, \left[ \lambda_s\,\sigma\,(1-\delta)\, \left(  (1-\gamma)\,\frac{r_1\,b^*}{{\cal R}_b}\,\alpha  \, \frac{b^*}{b+b^*} + \gamma\,\frac{r_1\,e^*}{{\cal R}_c}   \right) \, \frac{e}{e + e^*} \right] \right\},$
\item[(61)]  $P_{e^*,d^*,f^*,f^{**}\rightarrow e^*+1,d^*-1, f^*-1,f^{**}+1} = \left( \frac{f^* }{f+ f^*}  \right)^2\, \left\{ (1-\lambda_f)\, \frac{r_1\,d^*}{{\cal D}}\,u \, \left[ \lambda_s\,\sigma\,(1-\delta)\, \left(  (1-\gamma)\,\frac{r_1\,b^*}{{\cal R}_b}\,\alpha  \, \frac{b^*}{b+b^*} + \gamma\,\frac{r_1\,e^*}{{\cal R}_c}  \right)\, \frac{e}{e + e^*}   \right] \right\},$
\item[(62)]  $P_{e^*,d^{**},f^{**}\rightarrow e^*+1,d^{**}-1, f^{**}+2} = \left( \frac{f }{f+ f^*}  \right)^2\, \left\{ (1-\lambda_f)\, \frac{r_2\,d^{**}}{{\cal D}} \, \left[ \lambda_s\,\sigma\,(1-\delta)\, \left(  (1-\gamma)\,\frac{r_1\,b^*}{{\cal R}_b}\,\alpha  \, \frac{b^*}{b+b^*} + \gamma\,\frac{r_1\,e^*}{{\cal R}_c}  \right) \, \frac{e}{e + e^*}  \right] \right\},$
\item[(63)]  $P_{e^*,d^{**},f^*, f^{**}\rightarrow e^*+1,d^{**}-1, f^*-1, f^{**}+2} = \frac{2 f f^*}{(f+ f^*)^2} \, \left\{ (1-\lambda_f)\, \frac{r_2\,d^{**}}{{\cal D}} \, \left[ \lambda_s\,\sigma\,(1-\delta)\, \left(  (1-\gamma)\,\frac{r_1\,b^*}{{\cal R}_b}\,\alpha  \, \frac{b^*}{b+b^*} + \gamma\,\frac{r_1\,e^*}{{\cal R}_c}  \right)\, \frac{e}{e + e^*}   \right] \right\},$
\item[(64)]  $P_{e^*,d^{**},f^*, f^{**}\rightarrow e^*+1,d^{**}-1, f^*-1, f^{**}+2} = \left( \frac{f^* }{f+ f^*}  \right)^2\,\left\{ (1-\lambda_f)\, \frac{r_2\,d^{**}}{{\cal D}} \, \left[ \lambda_s\,\sigma\,(1-\delta)\, \left(  (1-\gamma)\,\frac{r_1\,b^*}{{\cal R}_b}\,\alpha  \, \frac{b^*}{b+b^*} + \gamma\,\frac{r_1\,e^*}{{\cal R}_c}   \right)\, \frac{e}{e + e^*}  \right] \right\},$
\item[(65)]  $P_{e^*,b^*,d^{*},f^*, f^{**}\rightarrow e^*-1,b^*+1,d^{*}-1,f^*+1, f^{**}+1} = \left( \frac{f }{f+ f^*}  \right)^2 \left\{ (1-\lambda_f)\, \frac{r_1 d^*}{{\cal D}}\,u \left[\lambda_s\,\sigma\,(1-\delta)\,\left(  (1-\gamma)\,\frac{b}{{\cal R}_b}\,\alpha  \, \frac{b}{b+b^*} +   \gamma\,\frac{e}{{\cal R}_c}    \right) \, \frac{e^*}{e + e^*}     \right] \right\},$
\item[(66)]  $P_{e^*,b^*,d^{*},f^*, f^{**}\rightarrow e^*-1,b^*+1,d^{*}-1,f^*-1, f^{**}+1} = \left( \frac{f^*}{f+ f^*}  \right)^2 \left\{ (1-\lambda_f)\, \frac{r_1 d^*}{{\cal D}}\,u \left[\lambda_s\,\sigma\,(1-\delta)\,\left(  (1-\gamma)\,\frac{b}{{\cal R}_b}\,\alpha \, \frac{b}{b+b^*}+   \gamma\,\frac{e}{{\cal R}_c}    \right) \, \frac{e^*}{e + e^*}     \right] \right\},$
\item[(67)]  $P_{e^*,b^*,f^*\rightarrow e^*-1,b^*+1,f^*-2} = \left( \frac{f^*}{f+ f^*}  \right)^2\, \left\{ (1-\lambda_f)\, \frac{d}{{\cal D}}\, \left[ \lambda_s\,\sigma\,(1-\delta)\,\left(  (1-\gamma)\,\frac{b}{{\cal R}_b}\,\alpha  \, \frac{b}{b+b^*}+   \gamma\,\frac{e}{{\cal R}_c}    \right) \, \frac{e^*}{e + e^*}     \right] \right\},$
\item[(68)]  $P_{e^*,b^*,f^*\rightarrow e^*-1,b^*+1,f^*-1} = \frac{2 f f^*}{(f+ f^*)^2} \, \left\{ (1-\lambda_f)\, \frac{d}{{\cal D}}\, \left[ \lambda_s\,\sigma\,(1-\delta)\,\left(  (1-\gamma)\,\frac{b}{{\cal R}_b}\,\alpha   \, \frac{b}{b+b^*}+   \gamma\,\frac{e}{{\cal R}_c}    \right) \, \frac{e^*}{e + e^*}     \right] \right\},$
\item[(69)]  $P_{e^*,b^*,d^{**},f^{**}\rightarrow e^*-1,b^*+1,d^{**}-1,f^{**}+2} = \left( \frac{f}{f+ f^*}  \right)^2\,\left\{ (1-\lambda_f)\, \frac{r_2\,d^{**}}{{\cal D}}\, \left[ \lambda_s\,\sigma\,(1-\delta)\,\left(  (1-\gamma)\,\frac{b}{{\cal R}_b}\,\alpha  \, \frac{b}{b+b^*}+   \gamma\,\frac{e}{{\cal R}_c}    \right) \, \frac{e^*}{e + e^*}     \right] \right\},$
\item[(70)]  $P_{e^*,b^*,d^{**},f^*,f^{**}\rightarrow e^*-1,b^*+1,d^{**}-1,f^*-1,f^{**}+2} =  \frac{2 f f^*}{(f+ f^*)^2} \, \left\{ (1-\lambda_f)\, \frac{r_2\,d^{**}}{{\cal D}}\, \left[ \lambda_s\,\sigma\,(1-\delta)\,\left(  (1-\gamma)\,\frac{b}{{\cal R}_b}\,\alpha  \, \frac{b}{b+b^*}+   \gamma\,\frac{e}{{\cal R}_c}    \right) \, \frac{e^*}{e + e^*}     \right] \right\},$
\item[(71)]  $P_{e^*,b^*,d^{**},f^*,f^{**}\rightarrow e^*-1,b^*+1,d^{**}-1,f^*-2,f^{**}+2} =  \left( \frac{f^*}{f+ f^*}  \right)^2\, \left\{ (1-\lambda_f)\, \frac{r_2\,d^{**}}{{\cal D}}\, \left[ \lambda_s\,\sigma\,(1-\delta)\,\left(  (1-\gamma)\,\frac{b}{{\cal R}_b}\,\alpha  \, \frac{b}{b+b^*}+   \gamma\,\frac{e}{{\cal R}_c}    \right) \, \frac{e^*}{e + e^*}     \right] \right\},$
\item[(72)]  $P_{b^*,d^{**},f^{**}\rightarrow b^*+1,d^{**}-1,f^{**}+2} = \left( \frac{f}{f+ f^*}  \right)^2\,\left\{ (1-\lambda_f)\, \frac{r_2\,d^{**}}{{\cal D}}\, \left[   \lambda_s\,\sigma\,(1-\delta)\,\left(  (1-\gamma)\,\frac{r_1 b^*}{{\cal R}_b} \left(   (1-\alpha)   +\alpha\,  \frac{b^*}{b+b^*}\,\frac{e^*}{e+e^*} \right. \right.\right. \right. $\\
$\left.\left. \left. \left. +\alpha\,  \frac{b}{b+b^*}\,\frac{e}{e+e^*}  \right)   + \gamma\,\frac{r_1 e^*}{{\cal R}_c}\,\frac{e^*}{e+e^*}  \right)  \right]  \right\},$
\item[(73)]  $P_{b^*,d^{**},f^*,f^{**}\rightarrow b^*+1,d^{**}-1,f^*-1,f^{**}+2} = \frac{2 f f^*}{(f+ f^*)^2} \, \left\{ (1-\lambda_f)\, \frac{r_2\,d^{**}}{{\cal D}}\, \left[   \lambda_s\,\sigma\,(1-\delta)\,\left(  (1-\gamma)\,\frac{r_1 b^*}{{\cal R}_b} \left(   (1-\alpha)   +\alpha\,  \frac{b^*}{b+b^*}\,\frac{e^*}{e+e^*} \right. \right.\right. \right. $\\
$\left.\left. \left. \left. +\alpha\,  \frac{b}{b+b^*}\,\frac{e}{e+e^*}  \right)   + \gamma\,\frac{r_1 e^*}{{\cal R}_c}\,\frac{e^*}{e+e^*}  \right)  \right]  \right\},$
\item[(74)]  $P_{b^*,d^{**},f^*,f^{**}\rightarrow b^*+1,d^{**}-1,f^*-2,f^{**}+2} = \left( \frac{f^*}{f+ f^*}  \right)^2\, \left\{ (1-\lambda_f)\, \frac{r_2\,d^{**}}{{\cal D}}\, \left[   \lambda_s\,\sigma\,(1-\delta)\,\left(  (1-\gamma)\,\frac{r_1 b^*}{{\cal R}_b} \left(   (1-\alpha)   +\alpha\,  \frac{b^*}{b+b^*}\,\frac{e^*}{e+e^*} \right. \right.\right. \right. $\\
$\left.\left. \left. \left. +\alpha\,  \frac{b}{b+b^*}\,\frac{e}{e+e^*}  \right)   + \gamma\,\frac{r_1 e^*}{{\cal R}_c}\,\frac{e^*}{e+e^*}  \right)  \right]  \right\},$
\item[(75)]  $P_{b^*,d^*,d^{**},f^{**}\rightarrow b^*-1,d^*+2,d^{**}-1,f^{**}+2} =  \left( \frac{f}{f+ f^*}  \right)^2\,\left\{ (1-\lambda_f)\, \frac{r_2\,d^{**}}{{\cal D}}\, \left[ \lambda_s\,\sigma\,\delta \,\frac{r_1 b^*}{{\cal R}_b}\ \right] \right\},$
\item[(76)]  $P_{b^*,d^*,d^{**},f^*,f^{**}\rightarrow b^*-1,d^*+2,d^{**}-1,f^*-1,f^{**}+2} =  \frac{2 f f^*}{(f+ f^*)^2} \,  \left\{ (1-\lambda_f)\, \frac{r_2\,d^{**}}{{\cal D}}\, \left[ \lambda_s\,\sigma\,\delta \,\frac{r_1 b^*}{{\cal R}_b}\ \right] \right\},$
\item[(77)]  $P_{b^*,d^*,d^{**},f^*,f^{**}\rightarrow b^*-1,d^*+2,d^{**}-1,f^*-2,f^{**}+2} =  \left( \frac{f^*}{f+ f^*}  \right)^2\,  \left\{ (1-\lambda_f)\, \frac{r_2\,d^{**}}{{\cal D}}\, \left[ \lambda_s\,\sigma\,\delta \,\frac{r_1 b^*}{{\cal R}_b}\ \right] \right\},$
\item[(78)]  $P_{e^*,b^*\rightarrow e^*+1, b^*-1} =  \left( \frac{f}{f+ f^*}  \right)^2\,\left\{ (1-\lambda_f)\, \frac{d}{{\cal D}}\, \left[ \lambda_s\,\sigma\,(1-\delta)\,(1-\gamma) \,\frac{b}{{\cal R}_b}\,\alpha\,\frac{b^*}{b+b^*}\, \frac{e}{e+e^*} \ \right] \right\},$
\item[(79)]  $P_{e^*,b^*,f^*\rightarrow e^*+1, b^*-1,f^*-1} =   \frac{2 f f^*}{(f+ f^*)^2} \,  \left\{ (1-\lambda_f)\, \frac{d}{{\cal D}}\, \left[ \lambda_s\,\sigma\,(1-\delta)\,(1-\gamma) \,\frac{b}{{\cal R}_b}\,\alpha\,\frac{b^*}{b+b^*}\, \frac{e}{e+e^*} \ \right] \right\},$
\item[(80)]  $P_{e^*,b^*,f^*\rightarrow e^*+1, b^*-1,f^*-2} =  \left( \frac{f^*}{f+ f^*}  \right)^2\,\left\{ (1-\lambda_f)\, \frac{d}{{\cal D}}\, \left[ \lambda_s\,\sigma\,(1-\delta)\,(1-\gamma) \,\frac{b}{{\cal R}_b}\,\alpha\,\frac{b^*}{b+b^*}\, \frac{e}{e+e^*} \ \right] \right\},$
\item[(81)]  $P_{e^*,b^*,d^*,f^*\rightarrow e^*+1, b^*-1,d^*-1,f^*+2} =  \left( \frac{f}{f+ f^*}  \right)^2\,\left\{ (1-\lambda_f)\, \frac{r_1 d^*}{{\cal D}}\,(1-u)\, \left[ \lambda_s\,\sigma\,(1-\delta)\,(1-\gamma) \,\frac{b}{{\cal R}_b}\,\alpha\,\frac{b^*}{b+b^*}\, \frac{e}{e+e^*} \ \right] \right\},$
\item[(82)]  $P_{e^*,b^*,d^*,f^*\rightarrow e^*+1, b^*-1,d^*-1,f^*+1} =   \frac{2 f f^*}{(f+ f^*)^2}\, \left\{ (1-\lambda_f)\, \frac{r_1 d^*}{{\cal D}}\,(1-u)\, \left[ \lambda_s\,\sigma\,(1-\delta)\,(1-\gamma) \,\frac{b}{{\cal R}_b}\,\alpha\,\frac{b^*}{b+b^*}\, \frac{e}{e+e^*} \ \right] \right\},$
\item[(83)]  $P_{e^*,b^*,d^*\rightarrow e^*+1, b^*-1,d^*-1} =  \left( \frac{f^*}{f+ f^*}  \right)^2\, \left\{ (1-\lambda_f)\, \frac{r_1 d^*}{{\cal D}}\,(1-u)\, \left[ \lambda_s\,\sigma\,(1-\delta)\,(1-\gamma) \,\frac{b}{{\cal R}_b}\,\alpha\,\frac{b^*}{b+b^*}\, \frac{e}{e+e^*} \ \right] \right\},$
\item[(84)]  $P_{e^*,b^*,d^*,f^*,f^{**}\rightarrow e^*+1, b^*-1,d^*-1,f^*+1,f^{**}+1} =  \left( \frac{f}{f+ f^*}  \right)^2\,\left\{ (1-\lambda_f)\, \frac{r_1 d^*}{{\cal D}}\,u\, \left[ \lambda_s\,\sigma\,(1-\delta)\,(1-\gamma) \,\frac{b}{{\cal R}_b}\,\alpha\,\frac{b^*}{b+b^*}\, \frac{e}{e+e^*} \ \right] \right\},$
\item[(85)]  $P_{e^*,b^*,d^*,f^*,f^{**}\rightarrow e^*+1, b^*-1,d^*-1,f^{**}+1} =   \frac{2 f f^*}{(f+ f^*)^2} \, \left\{ (1-\lambda_f)\, \frac{r_1 d^*}{{\cal D}}\, u\, \left[ \lambda_s\,\sigma\,(1-\delta)\,(1-\gamma) \,\frac{b}{{\cal R}_b}\,\alpha\,\frac{b^*}{b+b^*}\, \frac{e}{e+e^*} \ \right] \right\},$
\item[(86)]  $P_{e^*,b^*,d^*,f^*,f^{**}\rightarrow e^*+1, b^*-1,d^*-1,f^*-1,f^{**}+1} =  \left( \frac{f^*}{f+ f^*}  \right)^2\, \left\{ (1-\lambda_f)\, \frac{r_1 d^*}{{\cal D}}\, u\, \left[ \lambda_s\,\sigma\,(1-\delta)\,(1-\gamma) \,\frac{b}{{\cal R}_b}\,\alpha\,\frac{b^*}{b+b^*}\, \frac{e}{e+e^*} \ \right] \right\},$
\item[(87)]  $P_{e^*,b^*,d^{**},f^{**}\rightarrow e^*+1, b^*-1,d^{**}-1,f^{**}+2} =  \left( \frac{f}{f+ f^*}  \right)^2\,\left\{ (1-\lambda_f)\, \frac{r_2 d^{**} }{{\cal D}}\, \left[ \lambda_s\,\sigma\,(1-\delta)\,(1-\gamma) \,\frac{b}{{\cal R}_b}\,\alpha\,\frac{b^*}{b+b^*}\, \frac{e}{e+e^*} \ \right] \right\},$
\item[(88)]  $P_{e^*,b^*,d^{**},f^*,f^{**}\rightarrow e^*+1, b^*-1,d^{**}-1,f^*-1,f^{**}+2} =   \frac{2 f f^*}{(f+ f^*)^2} \,  \left\{ (1-\lambda_f)\, \frac{r_2 d^{**} }{{\cal D}}\, \left[ \lambda_s\,\sigma\,(1-\delta)\,(1-\gamma) \,\frac{b}{{\cal R}_b}\,\alpha\,\frac{b^*}{b+b^*}\, \frac{e}{e+e^*} \ \right] \right\},$
\item[(89)]  $P_{e^*,b^*,d^{**},f^*,f^{**}\rightarrow e^*+1, b^*-1,d^{**}-1,f^*-2,f^{**}+2} =  \left( \frac{f^*}{f+ f^*}  \right)^2\,\left\{ (1-\lambda_f)\, \frac{r_2 d^{**} }{{\cal D}}\, \left[ \lambda_s\,\sigma\,(1-\delta)\,(1-\gamma) \,\frac{b}{{\cal R}_b}\,\alpha\,\frac{b^*}{b+b^*}\, \frac{e}{e+e^*} \ \right] \right\},$
\item[(90)]  $P_{e^*,b^*\rightarrow e^*-1, b^*+2} =  \left( \frac{f}{f+ f^*}  \right)^2\,\left\{ (1-\lambda_f)\, \frac{d }{{\cal D}}\, \left[ \lambda_s\,\sigma\,(1-\delta)\,(1-\gamma) \,\frac{r_1 b^*}{{\cal R}_b}\,\alpha\,\frac{b}{b+b^*}\, \frac{e^*}{e+e^*} \ \right] \right\},$
\item[(91)]  $P_{e^*,b^*,f^*\rightarrow e^*-1, b^*+2,f^*-1} =    \frac{2 f f^*}{(f+ f^*)^2} \,  \left\{ (1-\lambda_f)\, \frac{d }{{\cal D}}\, \left[ \lambda_s\,\sigma\,(1-\delta)\,(1-\gamma) \,\frac{r_1 b^*}{{\cal R}_b}\,\alpha\,\frac{b}{b+b^*}\, \frac{e^*}{e+e^*} \ \right] \right\},$
\item[(92)]  $P_{e^*,b^*,f^*\rightarrow e^*-1, b^*+2,f^*-2} =    \left( \frac{f^*}{f+ f^*}  \right)^2\, \left\{ (1-\lambda_f)\, \frac{d }{{\cal D}}\, \left[ \lambda_s\,\sigma\,(1-\delta)\,(1-\gamma) \,\frac{r_1 b^*}{{\cal R}_b}\,\alpha\,\frac{b}{b+b^*}\, \frac{e^*}{e+e^*} \ \right] \right\},$
\item[(93)]  $P_{e^*,b^*,d^*,f^*\rightarrow e^*-1, b^*+2,d^*-1,f^*+2} =    \left( \frac{f}{f+ f^*}  \right)^2\, \left\{ (1-\lambda_f)\, \frac{r_1 d^* }{{\cal D}}\, (1-u)\,\left[ \lambda_s\,\sigma\,(1-\delta)\,(1-\gamma) \,\frac{r_1 b^*}{{\cal R}_b}\,\alpha\,\frac{b}{b+b^*}\, \frac{e^*}{e+e^*} \ \right] \right\},$
\item[(94)]  $P_{e^*,b^*,d^*,f^*\rightarrow e^*-1, b^*+2,d^*-1,f^*+1} =    \frac{2 f f^*}{(f+ f^*)^2} \, \left\{ (1-\lambda_f)\, \frac{r_1 d^* }{{\cal D}}\, (1-u)\,\left[ \lambda_s\,\sigma\,(1-\delta)\,(1-\gamma) \,\frac{r_1 b^*}{{\cal R}_b}\,\alpha\,\frac{b}{b+b^*}\, \frac{e^*}{e+e^*} \ \right] \right\},$
\item[(95)]  $P_{e^*,b^*,d^*,f^*\rightarrow e^*-1, b^*+2,d^*-1,f^*+2} =    \left( \frac{f^*}{f+ f^*}  \right)^2\, \left\{ (1-\lambda_f)\, \frac{r_1 d^* }{{\cal D}}\, (1-u)\,\left[ \lambda_s\,\sigma\,(1-\delta)\,(1-\gamma) \,\frac{r_1 b^*}{{\cal R}_b}\,\alpha\,\frac{b}{b+b^*}\, \frac{e^*}{e+e^*} \ \right] \right\},$
\item[(96)]  $P_{e^*,b^*,d^*,f^*,f^{**}\rightarrow e^*-1, b^*+2,d^*-1,f^*+1,f^{**}+1} =    \left( \frac{f}{f+ f^*}  \right)^2\, \left\{ (1-\lambda_f)\, \frac{r_1 d^* }{{\cal D}}\, u\,\left[ \lambda_s\,\sigma\,(1-\delta)\,(1-\gamma) \,\frac{r_1 b^*}{{\cal R}_b}\,\alpha\,\frac{b}{b+b^*}\, \frac{e^*}{e+e^*} \ \right] \right\},$
\item[(97)]  $P_{e^*,b^*,d^*,f^{**}\rightarrow e^*-1, b^*+2,d^*-1,f^{**}+1} =    \frac{2 f f^*}{(f+ f^*)^2} \, \left\{ (1-\lambda_f)\, \frac{r_1 d^* }{{\cal D}}\, u\,\left[ \lambda_s\,\sigma\,(1-\delta)\,(1-\gamma) \,\frac{r_1 b^*}{{\cal R}_b}\,\alpha\,\frac{b}{b+b^*}\, \frac{e^*}{e+e^*} \ \right] \right\},$
\item[(98)]  $P_{e^*,b^*,d^*,f^*,f^{**}\rightarrow e^*-1, b^*+2,d^*-1,f^*-1,f^{**}+1} =    \left( \frac{f^*}{f+ f^*}  \right)^2\, \left\{ (1-\lambda_f)\, \frac{r_1 d^* }{{\cal D}}\, u\,\left[ \lambda_s\,\sigma\,(1-\delta)\,(1-\gamma) \,\frac{r_1 b^*}{{\cal R}_b}\,\alpha\,\frac{b}{b+b^*}\, \frac{e^*}{e+e^*} \ \right] \right\},$
\item[(99)]  $P_{e^*,b^*,d^{**},f^{**}\rightarrow e^*-1, b^*+2,d^{**}-1,f^{**}+2} =    \left( \frac{f}{f+ f^*}  \right)^2\, \left\{ (1-\lambda_f)\, \frac{r_2 d^{**} }{{\cal D}}\,\left[ \lambda_s\,\sigma\,(1-\delta)\,(1-\gamma) \,\frac{r_1 b^*}{{\cal R}_b}\,\alpha\,\frac{b}{b+b^*}\, \frac{e^*}{e+e^*} \ \right] \right\},$
\item[(100)]  $P_{e^*,b^*,d^{**},f^*,f^{**}\rightarrow e^*-1, b^*+2,d^{**}-1,f^*-1,f^{**}+2} =     \frac{2 f f^*}{(f+ f^*)^2} \, \left\{ (1-\lambda_f)\, \frac{r_2 d^{**} }{{\cal D}}\,\left[ \lambda_s\,\sigma\,(1-\delta)\,(1-\gamma) \,\frac{r_1 b^*}{{\cal R}_b}\,\alpha\,\frac{b}{b+b^*}\, \frac{e^*}{e+e^*} \ \right] \right\},$
\item[(101)]  $P_{e^*,b^*,d^{**},f^*,f^{**}\rightarrow e^*-1, b^*+2,d^{**}-1,f^*-2,f^{**}+2} =    \left( \frac{f^*}{f+ f^*}  \right)^2\, \left\{ (1-\lambda_f)\, \frac{r_2 d^{**} }{{\cal D}}\,\left[ \lambda_s\,\sigma\,(1-\delta)\,(1-\gamma) \,\frac{r_1 b^*}{{\cal R}_b}\,\alpha\,\frac{b}{b+b^*}\, \frac{e^*}{e+e^*} \ \right] \right\},$
}
\end{itemize}
where ${\cal F}, {\cal D}, {\cal R}_b,{\cal R}_c$ are defined in the following
\begin{eqnarray}
{\cal F} &=& r_2\,f^{**} + r_1\,f^*+ f,\\
{\cal D} &=& r_2\,d^{**} + r_1\,d^*+ d,\\
{\cal R}_b &=& r_1\,b^*+ b,\\
{\cal R}_c &=& r_1\,e^*+ e.
\end{eqnarray}

\subsection*{Fixation Probability}

The structure of the model comprises a wide variety of different scenarios which may occur in the system. Therefore, to conclude a specific probability of fixation, different mechanisms may need to be taken into account. Signifying the long-term behavior of the system after initiation of a mutant within compartments, would assist us to understand the epithelial cell dynamics in colorectal and intestinal cancer.
Even though there might be a chance for new mutations, according to the long time-scaling of mutations in compare to the the period required for the dynamics of mutants as late of absorption.
Moreover, since the washed out mechanisms within the crypt eliminates new mutants, one may investigate dynamics of newborn mutants in a specific compartment without any new mutations (forward or backward mutations), i.e. $u=v=0$. Such a scenario, in turn, will help us to understand the survival chance of one mutant in each compartment as well as the fixation probability of mutants, which may subsequently appear in the other compartments as a result of mutants' division. Therefore, in this section we study the fixation probability of a mutant in a given compartment.
To investigate the survival probability of a mutant in a particular population, we calculate the probability that the progeny of mutants will take over the whole compartment. We denote the probability of absorption of $j$ number of mutants in a population of size $N>j$ by $\pi_j$, then the probability of one mutant's progeny taking over the entire population ($\pi_1$) can be obtained using the following system of equations.
\begin{eqnarray}
&& \pi_j =  \displaystyle\sum_{m} P_{j \to m} \pi_m, \hspace{0.5cm} 1<j<N-1, \nonumber \\
&& \pi_1 =  \displaystyle\sum_{m\geq 1} P_{j \to m} \pi_m,  \label{kol}\\
&& \pi_{N-1} = \displaystyle P_{N-1 \to N} + \sum_{m\leq N-1} P_{j \to m} \pi_m. \nonumber\\
&& \pi_N = 1.
\end{eqnarray}
Here, the initial condition is to have only one mutant of a specific type while there exist no other type of mutants in the system.

The analytic results, which are based on the transition probabilities derived in the previous section, are in a perfect agreement with the simulation results. Therefore, we can rely on simulation results to investigate more complicated scenarios for our generalized multi--compartmental model.

\subsection*{A. The probability of fixation for the central stem cells ($S_c$)}

The general system introduces a multi--variable Markov chain of dependent random processes. Let us consider a particular case of one-dimensional multi--variable Moran process in which there exist only one mutant stem cell in $S_c$ compartment where $(e^*, b^*, d^*, d^{**}, f^*, f^{**}) = (1, 0,0,0,0,0) $.
In other words, in this part we only investigate the cell dynamics in the $S_c$ compartment. Moreover, regarding the large time-scale of mutations in the system compered with that for proliferation, we assume that no extra mutation is allowed within the fixation process of the new mutant, that is, $u=v=0$.
{
Therefore, the non--zero transition probabilities for the cell dynamics in the $S_c$ compartment are transitions (33), (35)-(38), (51)-(61), (62)-(71), (78)-(83), and (90)-(95) from the list given in the previous section. }
%
 %
%
 %
 %
%
%

{
Let us denote the probability of fixation starting from $e^*$ mutants located at the central stem cell compartment by $\pi_{e^*}$. Based on the above transition probabilities associated to increase and decrease in the number of $S_c$ mutants at each time step, when stem cells divide only asymmetrically (i.e. when $\sigma=0$), then $\pi_{e^*}$ is zero. Note, when stem cells divide asymmetrically, no division occurs in the CeSC compartment, thus the number of CeSC mutants does not change. The same result can be obtained when divisions only occur in the $D_t$ compartment ($\lambda_s=0$), and when stem cells do not proliferate ($\delta=1$).
However, when $\sigma, \lambda_s > 0$ and $\delta<1$, the following system of equations can be derived (for any $0 \leq b^* \leq S_b,$)
}
{
\begin{eqnarray}
&& \hspace{-0.3in}
P^+(e^*,b^*)\,\pi_{e^*+1} + P^-(e^*, b^*)\,\pi_{e^*-1} - \left( P^+(e^*,b^*) + P^-(e^*, b^*) \right)\,\pi_e^*=0, \nonumber\\
&& \hspace{10cm}   1< e^* <S_c-1,  \nonumber \\
&&\hspace{-0.3in}
P^+(1,b^*)\,\pi_{2}  - \left( P^+(1,b^*) + P^-(0, b^*) \right)\,\pi_1=0,  \label{CeSCBSC}
\\
&&\hspace{-0.3in}
P^+(S_c-1,b^*)  +  P^{-}(S_c-1, b^*)\,\pi_{S_c-2}  - \left( P^+(S_c-1,b^*) + P^-(S_c-1, b^*)   \right)\,\pi_{S_c-1}=0,
\nonumber
\end{eqnarray}
}
{
where the resemble increasing and decreasing probabilities respectively are
}
{
\begin{eqnarray}
&&
P^+(e^*,b^*)  =   \left(  (1-\gamma)\,\alpha\,\frac{b^*}{b+b^*}  + \gamma\, \frac{r_1 e^*}{{\cal R}_c}  \right)\,\frac{e}{e+e^*},  \nonumber \\[-2mm]
&& \label{AAA} \\ [-2mm]
&& P^-(e^*, b^*)  =      \left(  (1-\gamma)\,\alpha\,\frac{b}{b+b^*}   + \gamma\, \frac{ e }{{\cal R}_c}  \right)\,\frac{e^*}{e+e^*} .    \nonumber
\end{eqnarray}
}

For the case in which symmetric division is only happening in the $S_c$ group but not in the $S_b$ group (when $\gamma=1$) as well as the case  for $\alpha\ll 1$, the probability of migration from $S_b$ group to $S_c$ group,  while $\gamma>0$, we can restrict the last system to the following form
\begin{eqnarray}
&& \hspace{-0.3in} r_1\,\pi_{e^*+1} + \pi_{e^*-1} - (1+r_1)\,\pi_e^*=0, \hspace{0.5cm} 1< e^* <S_c-1, \nonumber \\
&&\hspace{-0.3in} r_1\,\pi_{2} - (1+r_1)\,\pi_1=0, \\
&&\hspace{-0.3in} r_1 + \pi_{S_c-2} - (1+r_1)\,\pi_{S_c-1}=0. \nonumber
\end{eqnarray}
The solution to this system signifies that for $1\leq e^* \leq S_c-1$:
\begin{eqnarray}
\pi_{e^*} = \frac{1-\left( \frac{1}{r_1}  \right)^{e^*}}{ 1-\left( \frac{1}{r_1}  \right)^{S_c} }.
\end{eqnarray}
Therefore, the fixation probability of one mutant central stem cell in the $S_c$, i.e. the probability of the progeny of one CeSC mutant taking over the entire $S_c$ compartment, is
\begin{eqnarray}
\pi_{1} = \frac{1-\left( \frac{1}{r_1}  \right)}{ 1-\left( \frac{1}{r_1}  \right)^{S_c} }.
\end{eqnarray}
%

\subsection*{ B. The fixation probability for mutant $S_b$ stem cells}

To obtain the fixation probability in the $S_b$ group, i.e. the probability of progeny of a mutant border stem cell taking over the entire $S_b$ compartment, we only consider the cell dynamics in the BSC compartment. We assume the system has one mutant stem cell at the initial time in the $S_b$ compartment, while no more mutants exist elsewhere, i.e. $(e^*, b^*, d^*, d^{**}, f^*, f^{**}) = (0,1, 0,0,0,0,0) $). Then, we obtain the transition probabilities $p^+( b^*)$ and $p^-( b^*)$, which are the probabilities of transforming from $b^*$ number of mutant BSCs to $b^*+1$ and $b^*-1$ number of mutant BSCs, respectively when no more mutation is occurring in the system ($u=v=0$). We denote the probability of $b^*$ number of mutants taking over the $S_b$ by $\pi_{b^*}$.

Similar to what we had in the previous section, if the probability of stem cell division $\lambda_s$ is zero, when the $D_t$ subpopulstion has no contribution in substituting the two  eliminated cells in $D_f$ group ($\lambda_f=1$), or when the probability of symmetric division $\sigma$ is zero, then the number of $S_b$ mutants does not alter that leads to $\pi_{b^*}=0$.
{
When $\delta=1$, then only symmetric differentiation occurs in the BSC compartment. Hence, the mutant in the BSC group will be differentiated to two mutant TA cells, and thus $\pi_{b^*}=0$. Also, when $\delta<1, \lambda_s\neq 0, \lambda_f<1,$ and $\sigma\neq 0$  which both symmetric and asymmetric division can occur, the fixation probability in the $S_b$ is obtained using the following system of equations: }

{
\begin{eqnarray}
&&\hspace{-1cm}  Q^+( e^*,b^*)\pi_{b^*+1} +Q^{+2}( e^*,b^*)\pi_{b^*+2} + Q^-( e^*,b^*) \pi_{b^*-1}  \nonumber  \\
&&\hspace{2cm} - \left( Q^+( e^*,b^*) +Q^{+2}( e^*,b^*) + Q^-( e^*,b^*) \right)\pi_{b^*} =0, \,\,\,\, 1< b^* < S_b-2, \nonumber   \\
&&\hspace{-1cm} Q^+( e^*,1) \,\pi_{2}+Q^{+2}( e^*,1)\pi_{3}  - \left(P^+( e^*,1) +Q^{+2}( e^*,1) + Q^-(e^*,1) \right)\,\pi_1=0, \nonumber \\
&& \hspace{-1cm}Q^+(e^*, S_b-2)\,\pi_{S_b-2} + Q^{+2}( e^*,S_b-2)  + Q^-(e^*, S_b-2) \, \pi_{S_b-2}  \\
&& \hspace{2cm}  - \left( Q^+(e^*,S_b-2) +Q^{+2}( e^*,S_b-2) + Q^-(e^*,S_b-2) \right) \, \pi_{S_b-2}=0, \nonumber \\
&& \hspace{-1cm}Q^+(e^*, S_b-1) + Q^-( S_b-1) \, \pi_{S_b-1} - \left( Q^+(e^*,S_b-1)  + Q^-(e^*,S_b-1) \right) \, \pi_{S_b-2}=0, \nonumber
\end{eqnarray}
}
where the total sum of transition probabilities of increase by one and two, and decrease by one in the number of border stem cells located in $S_b$,  $Q^+( e^*,b^*), Q^{+2}( e^*,b^*)$ and $Q^-(e^*,b^*)$, are respectively defined by the following formulas:

{
\begin{eqnarray}
Q^+( e^*, b^*)  &= &
(1-\delta)\,\left[  (1-\gamma)\,\frac{r_1 b^*}{{\cal R}_b}\,(1-\alpha)  +(1-\gamma)\,\alpha\,\left( \frac{r_1 b^*}{{\cal R}_b}\,\frac{b^*}{b+b^*} \right. \right.  \nonumber  \\
&&\left. \left.  +   \frac{b}{{\cal R}_b}\,\frac{b}{b+b^*}   \right)\,\frac{e^*}{e+e^*}   +(1-\gamma)\,\alpha\,\frac{r_1 b^*}{{\cal R}_b}\,\frac{b}{b+b^*} \,\frac{e}{e+e^*}    + \gamma\, \frac{e^*}{e+e^*}  \right]   \nonumber \\[-1mm]
&&  \label{BBB}\\[-2mm]
Q^{+2}( e^*, b^*)  &=&
(1-\delta)\,(1-\gamma)\,\frac{r_1 b^*}{{\cal R}_b}\, \alpha\, \frac{b}{b+b^*}\,\frac{e^*}{e+e^*},   \nonumber \\
Q^-( e^*,b^*) &= &
\delta\,  \frac{r_1 b^*}{{\cal R}_b} + (1-\delta)\,(1-\gamma)\,\frac{b}{{\cal R}_b}\, \alpha\, \frac{b^*}{b+b^*}\,\frac{e}{e+e^*}, \nonumber
\end{eqnarray}
}

{
The $\alpha\ll 1$ results in the small frequency of mutant CeSCs, $e^*\approx 0$. Also, the above system of equations will be reduced to the following recurrence system in the absence of immortal cells ($u=0$):}
\begin{eqnarray}
&& (1-\delta) (1-\gamma)\,\pi_{b^*+1} + \delta\,\pi_{b^*-1} - ( (1-\delta) (1-\gamma) + \delta )\pi_{b^*}=0, \hspace{0.1in} 1< b^* <S_c-1, \nonumber \\
&& (1-\delta) (1-\gamma),\pi_{2}  -   ( (1-\delta) (1-\gamma) + \delta ) \pi_1=0, \\
&&(1-\delta) (1-\gamma)+ \delta\,\pi_{S_c-2} -  ( (1-\delta) (1-\gamma) + \delta )\pi_{S_c-1}=0. \nonumber
\end{eqnarray}
This system of equations reveals that
\begin{eqnarray}
\pi_{b^*} = \frac{1-\left( \frac{\delta}{(1-\delta)(1-\gamma)}  \right)^{b^*}}{ 1-\left( \frac{\delta}{(1-\delta)(1-\gamma)}   \right)^{S_b} }.
\end{eqnarray}
Therefore, the fixation probability of a single mutant border stem cell is given by
\begin{eqnarray}
\pi_1 = \frac{1-\left( \frac{\delta}{(1-\delta)(1-\gamma)}  \right)}{ 1-\left( \frac{\delta}{(1-\delta)(1-\gamma)}   \right)^{S_b} }.
\end{eqnarray}
The dependency of the fixation probability on the initial number of $S_b$ mutants has been shown in Fig. 13.


\begin{figure}[h!]
\centering{
\includegraphics[scale=0.35]{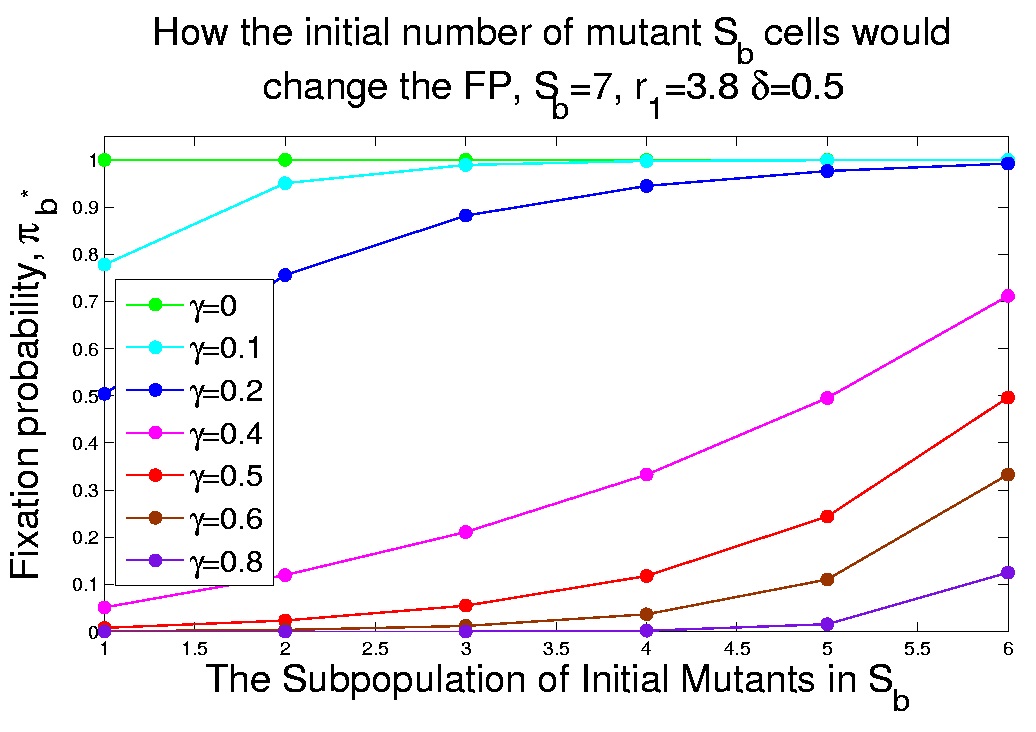} 
\caption{ {\bf Homeostasis in the number of border stem cells manages the compartmental growth via crucial factors $\delta$ and $\gamma$.} One mutant border stem cell arises in the $S_b$ compartment and no more mutations are allowed in the system. We assume that $\lambda_s\neq 0$ and $\sigma\neq 0$ which means that both symmetric and asymmetric division can occur and  $S_b=7$. This figure shows how the fixation probability $\pi_{b^*}$ varies w.r.t. the changes in the population size of mutants in the $S_b$ compartment as $\gamma$ takes various values and $\delta=0.5, r_1=3.8$.}
\label{CaseB}
}
\end{figure}
\subsection*{ C. The probability of fixation for mutant progenitor $D_t$ cells}

In this case, we assume there exist only one mutant transient amplifying cell in the $D_t$ compartment, i.e. $(e^*, b^*, d^*, d^{**}, f^*, f^{**}) = (0,0, 1,0,0,0)$. When at the initial time, there is only one mutant in the compartment of progenitor cells and no naive mutation appears in the system, $e^*$ and $b^*$ will have no chance to arise, however, the number of $d^{**}, f^*,$ and $ f^{**}$ might change. If we assume no new mutations occur ($u=v=0$), then $d^{**}$ and $ f^{**}$ will stay zero over time. Denoting  the fixation probability of $d^*$ number of mutant progenitor cells by $\pi_{d^*}$, we get
\begin{eqnarray}
&&\hspace{-1cm}  \left(  P^+( d^*)  +  P^-(d^*) \right)\,\pi_{d^*} =  p^+( d^* ) \,\pi_{d^*+1} +  P^-(d^*)\,\pi_{d^*-1} \hspace{0.2in} 1< d^* < D_t-1,  \\
&&\hspace{-1cm}  P^+( 1)\,\pi_{2}  =    \left(P^+( 1)+  P^-(1)   \right)\,\pi_1, \\
&&\hspace{-1cm}  P^+( D_t-1)+  P^-(D_t-1)\,\pi_{D_t-2} = \left(P^+( D_t-1)+ P^-(D_t-1)  \right)\,\pi_{D_t-1}.
\end{eqnarray}
Here, the coefficient of $\pi_{d^*+1}$ ($p^+( d^*)$) is, in fact, the sum of all transition probabilities which tend to an increase by one in the number of mutant TA cells, and $p^-(d^*)$ as the coefficient of $\pi_{d^*-1}$ is the sum over all possible transition probabilities leading to a decrease by one in the number of mutant progenitor cells. Under the assumptions of this case and when no more mutation is occurring in the system, there would only exist normal and mutant cells in $D_t$ compartment and no immortal progenitor cells can arise or be produced by mutant cells in this compartment (i.e. $d^{**}$ remains zero). Hence the coefficients of the system can be reduced to the following form when $\lambda_f\neq 1$, which means that the division is also allowed to occur in $D_t$ compartment.
%
\begin{eqnarray}
P^+( d^* ) &=& (1-\lambda_s)\, \frac{D_t - d^* }{ D_t +(r_1- 1)d^* } ,  \\
P^-( d^* ) &=&  \lambda_s + \,(1-\lambda_s)\,\frac{D_t - d^*  }{ D_t +(r_1- 1)d^*  },
\end{eqnarray}
and the solution to this new system is in the following form
\begin{eqnarray}
\pi_{d^*} =  \pi_1\, \sum_{k=1}^{d^*-1} \, \left( \frac{ \lambda_s \, r_1-1}{\lambda_s   }  \right)^k \frac{  \Gamma \left( \frac{ D_t + ( \lambda_s \, r_1 -1)(k+1) }{ \lambda_s\, r_1 -1} \right)\,  \Gamma( 1-D_t)     }{  \Gamma \left( \frac{ D_t +  \lambda_s \, r_1 -1 }{ \lambda_s\, r_1 -1} \right)\,  \Gamma( k+1 - D_t) },
\end{eqnarray}
where $\Gamma(t)=\int_0^{\infty} x^{t-1}\,{\rm e}^x\, dx$ is the gamma function. Then the fixation probability $\pi_1$ can be derived from the following relation
\begin{eqnarray}
\pi_1 = \left[ 1+ \sum_{j=2}^{D_t-1} \sum_{k=1}^{j-1} \, \left( \frac{ \lambda_s \, r_1-1}{\lambda_s   }  \right)^k \frac{  \Gamma \left( \frac{ D_t + ( \lambda_s \, r_1 -1)(k+1) }{ \lambda_s\, r_1 -1} \right)\,  \Gamma( 1-D_t)     }{  \Gamma \left( \frac{ D_t +  \lambda_s \, r_1 -1 }{ \lambda_s\, r_1 -1} \right)\,  \Gamma( k+1 - D_t) } \right]^{-1}.
\end{eqnarray}
%

Comparing the coefficients $p^+(d^*)$ and $p^-(d^*)$, when $\lambda_f, \lambda_s\neq 0$, the chance of decreasing by one in  the number of mutant progenitor cells is higher than that of increasing by one ($ p^+( d^* )<  p^-( d^* )$ for $1\leq d^* \leq D_t-1$), which implies that as reproduction rate takes larger values, the probability of fixation declines (See Fig. 6(c) in the paper for more details). This means progenitor cells are more capable of producing differentiated cells than being fixated. Given the assumptions of this section, if no division happen in the stem cell niche, i.e. $\lambda_s=0$, then $\pi_{d^*}=\displaystyle{\frac{1}{D_t}}$. Fig. 6(b) in the results section, represents the role of initial mutant cell population size in the absorption mechanism for this case.
\subsection*{ D. The fixation probability of immortal $D_t$ cells in the TA compartment}

In this part, we calculate the fixation probability of a single immortal TA cell in the TA compartment. Therefore, we assume at the initial time, the is only one mutant immortal cell, i.e. $(e^*, b^*, d^*, d^{**}, f^*, f^{**}) = (0,0, 0,1,0,0 )$. Base on the model, in this case the number of $e^*, b^*$, and $d^*$ will stay zero, while the rest of the variables might change.  Assuming again $u=v=0$, the system of equations for the fixation probability $\pi_{d^{**}}$, which is the probability of the progeny of $b^{**}$ number of immortal cells taking over the entire TA compartment, is
\begin{eqnarray}
&&  \hspace{-0.7cm}  P^+( d^{**}) \,\pi_{d^{**}+1} +  P^-(d^{**})\,\pi_{d^{**}-1} - \left(  P^+( d^{**})  +  P^-(d^{**}) \right)\,\pi_{d^{**}}=0,  \hspace{0.5cm}  1< d^{**}< D_t-1,  \nonumber \\
&&  \hspace{-0.7cm}  P^+( 1)\,\pi_{2}  -    \left( P^+( 1) +  P^-(1)  \right)\,\pi_1=0, \\
&&  \hspace{-0.7cm}  P^+( D_t-1)+  P^-(D_t-1)\,\pi_{D_t-2} - \left(  P^+( D_t-1)+ P^-(D_t-1)  \right)\,\pi_{D_t-1}=0.  \nonumber
\end{eqnarray}
where the coefficients $ p^+( d^{**})$ and $ p^-(d^{**})$ are respectively the probabilities that the number of immortal TA cells increases by one and decreases by one in one time step. The above system of equations can be simplified to the following system,  when $\lambda_f\neq 1$:
\begin{eqnarray}
P^+( d^{**} )= (1-\lambda_s)\, \frac{D_t - d^{**}  }{ D_t +(r_2- 1) d^{**}  } , \hspace{0.5cm}   P^-(d^{**}) =  \lambda_s + \,(1-\lambda_s)\,\frac{D_t - d^{**}  }{ D_t +(r_2- 1) d^{**}  },
\end{eqnarray}
Thus, the probability of fixation of $d^{**}$ immortal TA cells in the TA compartment is given by
\begin{eqnarray}
\pi_{d^{**}}  =  \pi_1\, \sum_{k=1}^{d^{**}-1} \, \left( \frac{ \lambda_s \, r_2-1}{\lambda_s   }  \right)^k \frac{  \Gamma \left( \frac{ D_t + ( \lambda_s \, r_2 -1)(k+1) }{ \lambda_s\, r_2 -1} \right)\,  \Gamma( 1-D_t)     }{  \Gamma \left( \frac{ D_t +  \lambda_s \, r_2 -1 }{ \lambda_s\, r_2 -1} \right)\,  \Gamma( k+1 - D_t) },
\end{eqnarray}
where $\Gamma(x)$ is the gamma function. Then the fixation probability $\pi_1$ can be derived as
\begin{eqnarray}
\pi_1 = \left[ 1+ \sum_{j=2}^{D_t-1}\sum_{k=1}^{j-1} \, \left( \frac{ \lambda_s \, r_2-1}{\lambda_s   }  \right)^k \frac{  \Gamma \left( \frac{ D_t + ( \lambda_s \, r_2 -1)(k+1) }{ \lambda_s\, r_2 -1} \right)\,  \Gamma( 1-D_t)     }{  \Gamma \left( \frac{ D_t +  \lambda_s \, r_2 -1 }{ \lambda_s\, r_2 -1} \right)\,  \Gamma( k+1 - D_t) } \right]^{-1}.
\end{eqnarray}
This result is similar to the one obtained for the mutant progenitor cells in section C. The behavior of the system is also the same as those given in Fig. 6 parts (b) and (c). Therefore, the crucial role of immortal progenitor cells can be explained mainly by producing immortal differentiated cells. It might be worthy to remark that a small increase in the number of initial immortal $D_t$ cells would not significantly affect the fixation probability (See Fig. 6(c)).

\subsection*{ E. The fixation probability of mutant $D_f$ cells in the FD compartment}

Here, we investigate the survival probability of a mutant $D_f$ cell while environment imposes no further mutations, i.e. $u=v=0$. Assuming the initial state is $(e^*, b^*, d^*, d^{**}, f^*, f^{**}) = (0,0, 0,0,1,0 )$, only the number of $f^*$ can vary. Therefore, the fixation probability of $ f^*$ number of mutant $D_f$ cells, $\pi_{f^*}$, can be obtained from the following system of equations.

{\small \begin{eqnarray}
&& \hspace{-1cm}  P^+(f^* ) \,\pi_{ f^*+2} + Q^+( f^*) \,\pi_{f^*+1} + P^-( f^*)\,\pi_{f^*-1} + Q^-(f^* )\,\pi_{f^*-2} \\
&&  \hspace{1cm}  - \left( P^+(f^*) +Q^+(f^*) + P^-(f^*)+ Q^-(f^*)\right)\,\pi_{f^*}=0, \,\,\,  2< f^* < D_f-2, \\
&&\hspace{-1cm} P^+(1) \,\pi_{3} + Q^+(1) \,\pi_{2} - \left(P^+(1) +Q^+(1) + P^-(1)\right)\,\pi_1=0, \\
&&\hspace{-1cm}  P^+(2) \,\pi_{4} + Q^+(2) \,\pi_{3} +P^-(2)\,\pi_{1}- \left(P^+(2) +Q^+(2) + P^-(2)+Q^-(2)\right)\,\pi_2=0, \\
&&\hspace{-1cm}  P^+(D_f-2)+ Q^+(D_f-2)\,\pi_{D_f-1}  +  P^-(D_f-2)\,\pi_{D_f-3} + Q^-(D_f-2 )\,\pi_{f^*-4} \nonumber \\
&& \hspace{3cm}  - \left( Q^+(D_f-1) +  P^-(D_f-1) + Q^-(D_f-1)\right)\,\pi_{D_f-2}=0,\\
&&\hspace{-1cm}  Q^+(D_f\!-1)  +  P^-(D_f\!-1)\,\pi_{D_f-2} + Q^-(D_f-1 )\,\pi_{f^*-3} \\
&&\hspace{3cm}   - \left( Q^+(D_f\!-1) +  P^-(D_f-1) + Q^-(D_f-1)\right)\,\pi_{D_f\!-1}=0.
\end{eqnarray}}
Whe{
\begin{eqnarray*}
&&P^+(f^* ) = \lambda_f\,\left(\frac{r_1\,f^*}{D_f+(r_1-1)f^*}\right)^2,\\
&& Q^+(f^* ) = 2\,\lambda_f\, \frac{r_1\,f^*}{ D_f+(r_1-1)f^*} ,\\
&& P^-(f^* ) =  2\,\lambda_f\, \left(\frac{(D_f-f^*)}{D_f+(r_1-1)f^* } \right)^2+ 2(1-\lambda_f),\\
&& Q^-(f^* ) = \lambda_f\,\left(\frac{D_f-f^*}{D_f+(r_1-1)f^*} \right)^2 + 2(1-\lambda_f),
\end{eqnarray*}
}
where we assumed that the probability of death for w.t. and malignant FD cells are equal. If divisions never occur in the FD compartment, i.e. $\lambda_f=0$, then the number of FD mutants ($d^*$) remains constant during the process, implying
\begin{eqnarray}
\pi_{f^*} =0, \hspace{1cm} \mbox{for }\,\,\,1\leq f^* \leq D_f-1.
\end{eqnarray}
On the other hand, when $0< \lambda_f \leq 1$, the coefficients $p^{\pm}, q^{\pm}$ lead to a more complicated system. In Fig. 6-(e),(f) the solutions to this system are given for some particular values of $\lambda_f$ where the other parameters have chosen from Table 1: $D_f=500, r_1=3.8$. In this figure, as $\lambda_f$ tends to zero, a dramatic change will occur in the graph. In part (f) of this figure, the graphs reveal the fact that, even for a large value of the relative fitness of mutants, the survival chance of mutants remains very small.

\subsection*{ F. The fixation probability of mutant $D_t$ cells in the FD compartment}

In this part, we investigate the probability of the progeny of one mutant TA cell taking over the entire FD compartment. Again, we assume no more mutation is expected to occur through the whole procedure, i.e. $u=v=0$. Therefore, we assume that initially there is only one mutant TA cell, i.e. $(e^*, b^*, d^*, d^{**}, f^*, f^{**}) = (0,0, 1,0,0,0 )$. This case is more complicated compared to the previous cases because of dependency of the system on both mutant progenitor and differentiated cells. Let   $\pi^{\rm FD}_{(d^*,f^*)}  $ be the fixation probability of starting from $d^*$ TA cells and $f^*$ mutant differentiated cells (initial state is $(d^*,f^*)$) inside the FD compartment. Here, we consider a bi--variable Markov chain to explore the cell dynamics in the FD and TA compartments. There are 10 corresponding transition probabilities of possible changes to the state $(d^*,f^*)$. The initial state is $(1,0)$ while the initial conditions are $\pi^{\rm FD}_{0,0}=0$ and $\pi^{\rm FD}_{d^*, D_f}=1$ for any $d^*$.

Since there are $D_f (D_t+1) -1$ different states (where $(d^*,f^*)\neq (0,0)$ and $0\leq f^* <D_f$), the transition matrix is a $D_f (D_t+1) -1$ by $D_f (D_t+1) -1$ dimensional matrix $A$, where each entry of $A$ corresponds to one of the states $(d^*,f^*)$ and includes the coefficients of representing $\pi^{\rm FD}_{(d^*,f^*)}$ in terms of all possible fixation probabilities $\pi_{\tilde{d^*},\tilde{f^*}}$:
\begin{eqnarray}
\pi^{\rm FD}_{(d^*,f^*)} = \sum_{{\tilde{d^*},\tilde{f^*}}}\, P_{(d^*,f^*)\rightarrow (\tilde{d^*},\tilde{f^*})}\,\pi^{\rm FD}_{(\tilde{d^*},\tilde{f^*})}.
\end{eqnarray}

Let $B=\Big[\,(d^*,f^*) \Big]_{0\leq d^* \leq D_t,0\leq f^* \leq D_f-1}$ be the matrix of all possible states $(d^*,f^*)$ (for $(d^*,f^*)\neq (0,0)$ and $0\leq f^* <D_f$). The matrix $B$ is isomorphic to a vector in $\Bbb{R}^{D_f (D_t+1) -1}$ by considering the subsequent rows in an ordered array as coordinates of this vector when the first array $(0,0)$ is ignored. More precisely, we have the following isomorphism
{
\begin{eqnarray*}
(d^*,f^*) \!\! \longrightarrow \!\!
\left\{\!\!
\begin{array}{ll}
f^*,   &  1\leq f^*\leq D_f-1, d^*=0,\\
(D_f-1) + (d^*-1)D_f + f^*+1, & 1\leq d^*\leq D_t, 0\leq f^*\leq D_f\!-\!1.
\end{array}
\right.
\end{eqnarray*}
}
Thus, using the vector representation of matrix $B$ (after dropping the entry $(0,0)$), we label rows and columns of the matrix $A$ with entries of the isomorphic vector to matrix $B$.
Fig. 14(c) reveals how the fixation probabilities depend on all possible states having the landscape of changes for possible states $(d^*,f^*)$.

\begin{figure}[h!]
\centering{
\includegraphics[scale=0.6]{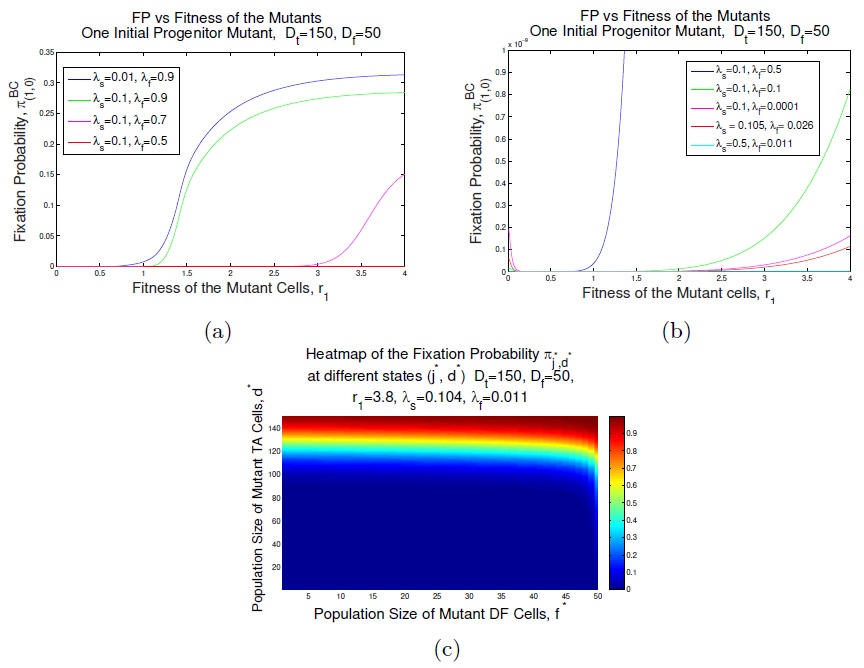}
\caption{ {\bf Multi--variable Markov chain of mutants in non-stem cell compartments.} In the absence of mutation and plasticity, when a mutant cell appears in either $D_t$ or $D_f$ compartments, we calculate the probability of fixation for mutant differentiated cells. Assuming $D_t=150$ and $D_f=50$ we  investigate three different approaches as $\lambda_s$ and $\lambda_f$ alter. Firstly, (a) represents the probability $\pi_{1,0}$ of starting from one initial mutant $D_t$ cell for lower values of $0.01 \leq \lambda_s \leq 0.1$ and higher values of $0.5\leq \lambda_f \leq 0.9$. We conclude that lower values for $\lambda_s$ and higher values for $\lambda_f$ tends to higher fixation probabilities. In contrast, changing the values of $\lambda_s$ to lower values as well, leads to a huge drop in the survival property. (c) depicts a landscape for the fixation probability for possible initial states $(d^*,f^*)$ (for $0 \leq d^*\leq 150, 0\leq f^* \leq 50$). A dramatic increase in the probability of fixation can be obtained by starting from larger initial mutant population of TA cells where $\lambda_s=0.105, \lambda_f=0.026$, and $r_1=3.8$.}}
\label{CaseF}
\end{figure}
%

If all divisions occur in the FD compartment, i.e. $\lambda_f=1$, then there is no chance that mutant TA cells divide, thus $\pi^{\rm FD}_{d^*,0}=0$ for all $1\leq d^* \leq D_t$. Moreover, when $\lambda_s$ decreases and $\lambda_f$ increases, the fixation probability dramatically climbs as these two conditions reinforce the chance for proliferation in progenitor and differentiated cells rather than divisions within other types of cells. For lower rates of $\lambda_f$, a much higher mutants' fitness is required to slightly develop the fixation probability. However, when mutants are disadvantageous, the optimal fixation probability can occur for restrained birth rates (see Fig. 14).

\subsection*{G. The fixation probability of immortal differentiated cells in the $D_f$ compartment}

Here, we calculate the probability of the progeny of $f^{**}$ number of mutant FD cells taking over the entire FD compartment, $\pi_{f^{**}}$. For this reason, we assume the initial state of the system is $(e^*, b^*, d^*, d^{**}, f^*, f^{**}) = (0,0, 0,0,0,1 )$, we also assume no new mutations or immortal cells arise ($u=v=0$). The fixation probability $\pi_{f^{**}}$ satisfies the following system of equations, when $\lambda_f\neq 0$.
\begin{eqnarray}
&&\hspace{-0.7cm}  2(D_f \!-\! f^{**}) \,\pi_{f^{**}\!+\!1} \!+\! (r_2\,f^{**})\,\pi_{f^{**}+2} \! -\!  \left( 2(D_f \!-\! f^{**}) \!+\! r_2\,f^{**}\right)\,\pi_{f^{**}} \!=\!0, \,\, 1 \leq  f^{**} < D_f-1, \nonumber \\
&&\hspace{-0.7cm}    4\,\pi_{D_f-1} + r_2\,(D_f-2)- \left( 4+ r_2\,(D_f-2)\right)\,\pi_{D_f-2}=0, \\
&&\hspace{-0.7cm}    2-  2 \,\pi_{D_f-1} = 0. \nonumber
\end{eqnarray}
The above recurrence system implies that
\begin{eqnarray}
 \pi_{f^{**}}  = 1, \hspace{0.5cm} 1\leq  f^{**} \leq D_f-1.
\end{eqnarray}
When $\lambda_f \neq 0$, a similar approach will be achieved as the case for $\lambda_f=1$ since there exist no supporting divisions from immortal $D_t$ cells to increase immortal $D_f$ population. Therefore, immortal differentiated cells, in the absence of apoptosis, can exponentially grow and an onset of a minor population of these type of cells take over the whole population of $D_f$ compartment. Another possible scheme is when $\lambda_f = 0$ which results in no chance for immortal cells to fixate even starting from $D_f-1$ number of initial cells.

\subsection*{ H. The fixation probability of immortal $D_t$ cells in the $D_f$ compartment}

Now, we obtain the probability of the progeny of $d^{**}$ number of immortal $D_t$ cells taking over the FD, $\pi^{\rm FD}_{( d^{**},f^{**})}$, while no more mutations is expected in the whole system. Assuming the initial state $(e^*, b^*, d^*, d^{**}, f^*, f^{**}) = (0,0, 0,0,1,0 )$, we investigate the probability of fixation $\pi_{ d^{**},f^{**}}$ of $d^{**}$ immortal progenitor cell(s) and $f^{**}$ immortal differentiated cell(s) in the FD compartment.
When $\lambda_f=1$, then there is no chance for any migration of immortal cells from $D_t$ compartment into $D_f$ compartment and thus
\begin{eqnarray}
\pi^{\rm FD}_{(d^{**},0)} =0, \hspace{1cm} 0\leq d^{**} \leq D_t.
\end{eqnarray}
Now if we consider the other extreme in which $\lambda_f=0$, the only resource for population growth of immortal $D_f$ cells is the immortal $D_t$ compartment. In this case, when $\lambda_s=1$, there is a chance for the only immortal cell in $D_t$ to divide symmetrically to two immortal differentiated daughter cells but these cells will not growth or decay in $D_f$ group and the the divided immortal cell in $D_t$ will not be substituted. Thus $\pi^{\rm FD}_{(d^{**},0)} =0$ for small values $d^{**}$ compared with the population size of $D_f$. In this situation, if $\lambda_s < 1$ then proliferation can occur in $D_t$ and each divided immortal cell in $D_t$ will be replaced with a certain chance when $d^{**} \geq 2$ at the beginning. But we conclude again $\pi^{\rm FD}_{(1,0)} =0$.

Now, let us assume that $0 < \lambda_f < 1$. As we explained in the above, the onset of any immortal cell in $D_f$ compartment given the assumptions of this case will tend to fixation (although the time to fixation may vary). So the fixation probability equals to the probability of having first immortal cell in $D_f$. Therefore, if $\lambda_s=1$, considering the probability of division for the initial immortal cell in $D_t$, we obtain
\begin{eqnarray}
\pi^{\rm FD}_{(d^{**}=1,f^{**}=0)} ={\displaystyle \frac{r_2}{D_t+r_2-1}}.
\end{eqnarray}
This probability will increase linearly as the initial number of immortal $D_t$ cells increases. Finally if $\lambda_s < 1$, we conclude that
\begin{eqnarray}
\pi^{\rm FD}_{(d^{**},f^{**})} =1, \hspace{1cm} 0\leq d^{**} \leq D_t,\, 0 \leq f^{**} \leq D_f-1.
\end{eqnarray}
In summary, these calculations reveal how different mechanisms would influence the system, the appearance of an immortal cell either in $D_t$ or $D_f$ compartments can trigger a cancer, and how fast this may develop to take over the whole normal population via maintaining the structured stability in the crypt.


\subsection*{ { I. The fixation probability of mutants inside CeSC group  for $\alpha>0$ }   }

{
In the present section, we generalize our investigation on the fixation process of mutants in the CeSC compartment when migration is also supposed to occur from BSC to the CeSC group. More precisely, assuming $\alpha \neq 0$ and starting from the initial state $(e^*,b^*,d^*,d^{**},f^*,f^{**})=(1,0,0,0,0,0)$, we try to understand the behavior of the system in the presence of two--way migrations between CeSC and BSC groups while other types of divisions and mutations can also occur. Again, we assume that no immortal cell is allowed in the system during the fixation procedure ($u=v=0$).}

{
Similar to the general method applied in Sec. F,  we suppose that $\pi^{\rm Ce}_{(e^*,b^*)}$ be the survival chance of a mutant eventually in CeSC compartment beginning from $e^*$ number of CeSCs and $b^*$ number of mutant BSCs. Briefly, the initial state is $(e^*,b^*)=(1,0)$ and a bi--variable Markov chain is taken into account to understand some of main features of the mechanism occurring within the crypt focusing on the SC class to explore the cell dynamics in the CeSC and BSC compartments leading to the fixation mechanism in CeSC group.  The initial conditions in this case are  $\pi^{\rm Ce}_{(0,0)}=0$ and  $\pi^{\rm Ce}_{(S_c,b^*)}=1$ for any $0\leq b^* \leq S_b$.}

{
According to the mechanism described in Sec. F,  we envisage the analytic calculation of the fixation probability $\pi^{\rm Ce}_{(1,0)}$ of starting from one mutation in the CeSC group through a two--variable Markov process of all possible states $(e^*,b^*)$ where $(e^*,b^*)\neq(0,0)$ and $0\leq e^*\leq S_c-1, 0\leq b^*\leq S_b$. Collectively, there exist $S_c (S_b+1) -1$ different states and the transition matrix $A$ is a $S_c (S_b+1) -1$ by $S_c (S_b+1) -1$ dimensional with entries as the coefficients of $\pi^{\rm Ce}_{e^*,b^*}$ in the following Kolmogorov equation: }

{
\begin{eqnarray}
\pi^{\rm Ce}_{(e^*,b^*)} = \sum_{{\tilde{e^*},\tilde{b^*}}}\, P_{(e^*,b^*)\rightarrow (\tilde{e^*},\tilde{b^*})}\,\pi^{\rm Ce}_{(\tilde{e^*},\tilde{b^*})}. \label{CaseKol}
\end{eqnarray}
}

{
Similarly, we suppose that $B=\Big[\,(e^*,b^*) \Big]_{0\leq e^* \leq S_c-1, 0\leq b^* \leq S_b}$, which is the matrix of all possible states $(e^*,b^*)$, where $(e^*,b^*)\neq (0,0)$. Again we apply the isomorphism between the matrix $B$ and a vector in $\Bbb{R}^{S_c (S_b+1) -1}$ introduced in below (where the first array $(0,0)$ is ignored:}
{
\begin{eqnarray*}
(e^*,b^*) \longrightarrow
(S_c-1) + (b^*-1)S_c+ e^*+1, \hspace{0.5cm} 1\leq e^*\leq S_c-1, 0\leq b^*\leq S_b.
\end{eqnarray*}
}

{
To calculate the fixation probability $\pi^{\rm Ce}_{(1,0)}$, the vector representation of matrix $B$  (where the entry $(0,0)$ is dropped) can be applied to solve the the system of equations (13).  Particularly, considering the initial conditions are $\pi^{\rm Ce}_{(0,0)}=0$ and $\pi^{\rm Ce}_{(S_c, b^*)}=1$ for any $0\leq b^*\leq S_b$, in this case we derive the following system of equations:  }

{
{\small \begin{eqnarray}
&& \hspace{-0.75cm}  P^{+,0}\, \pi^{\rm Ce}_{(e^*+1,b^*)} + P^{0,+}, \pi^{\rm Ce}_{(e^*,b^*+1)}   +  P^{0,-}\, \pi^{\rm Ce}_{(e^*,b^*-1)} + P^{+,-}\, \pi^{\rm Ce}_{(e^*+1,b^*-1)} + P^{-,+}\, \pi^{\rm Ce}_{(e^*-1,b^*+1)}  \nonumber \\
&&\hspace{0.5cm}  + P^{-,+2}\, \pi^{\rm Ce}_{(e^*-1,b^*+2)}  - \left[  P^{+,0}+ P^{0,+} +P^{0,-}+P^{+,-} +P^{-,+} + P^{-,+2}\right]\,\pi^{\rm Ce}_{(e^*,b^*)}   =0,   \nonumber\\
&&\hspace{6.5cm}  1 \leq e^* \leq S_c-2,  1 \leq b^* \leq S_b-2,    \label{III1}\\
&& \hspace{-0.75cm}  P^{+,0}\, \pi^{\rm Ce}_{(e^*+1,0)} + P^{0,+}, \pi^{\rm Ce}_{(e^*,1)}  + P^{-,+}\, \pi^{\rm Ce}_{(e^*-1,1)}  + P^{-,+2}\, \pi^{\rm Ce}_{(e^*-1,2)}  \nonumber \\
&&\hspace{1.75cm}   - \left[  P^{+,0}+ P^{0,+} +P^{-,+} + P^{-,+2}\right]\,\pi^{\rm Ce}_{(e^*,0)}   =0, \hspace{0.5cm}  1 \leq e^* \leq S_c-2,   \label{III2}\\
&& \hspace{-0.75cm}  P^{+,0}\, \pi^{\rm Ce}_{(e^*+1,S_b-1)} + P^{0,+}, \pi^{\rm Ce}_{(e^*,S_b)}   +  P^{0,-}\, \pi^{\rm Ce}_{(e^*,S_b-2)} + P^{+,-}\, \pi^{\rm Ce}_{(e^*+1,S_b-2)} + P^{-,+}\, \pi^{\rm Ce}_{(e^*-1,S_b)}  \nonumber \\
&&\hspace{0.5cm}   - \left[  P^{+,0}+ P^{0,+} +P^{0,-}+P^{+,-} +P^{-,+} \right]\,\pi^{\rm Ce}_{(e^*,S_b-1)}   =0,   \hspace{0.5cm}  1 \leq e^* \leq S_c-2,   \label{III3}\\
&& \hspace{-0.75cm}  P^{+,0}\, \pi^{\rm Ce}_{(e^*+1,S_b)}  +  P^{0,-}\, \pi^{\rm Ce}_{(e^*,S_b-1)} + P^{+,-}\, \pi^{\rm Ce}_{(e^*+1,S_b-1)}    - \left[  P^{+,0}+P^{0,-}+P^{+,-}  \right]\,\pi^{\rm Ce}_{(e^*,S_b)}   =0,   \nonumber\\
&& \hspace{9cm}  1 \leq e^* \leq S_c-2,   \label{III4}\\
&& \hspace{-0.75cm}  P^{+,0}\, \pi^{\rm Ce}_{(1,b^*)} + P^{0,+}, \pi^{\rm Ce}_{(0,b^*+1)}   +  P^{0,-}\, \pi^{\rm Ce}_{(0,b^*-1)} + P^{+,-}\, \pi^{\rm Ce}_{(1,b^*-1)}  \nonumber \\
&&\hspace{2.25cm}   - \left[  P^{+,0}+ P^{0,+} +P^{0,-}+P^{+,-} \right]\,\pi^{\rm Ce}_{(0,b^*)}   =0,  \hspace{0.5cm}   2 \leq b^* \leq S_b-1,    \label{III5}\\
&& \hspace{-0.75cm}  P^{+,0}\, \pi^{\rm Ce}_{(1,S_b)}   +  P^{0,-}\, \pi^{\rm Ce}_{(0,S_b-1)} + P^{+,-}\, \pi^{\rm Ce}_{(1,S_b-1)}   - \left[  P^{+,0} +P^{0,-}+P^{+,-} \right]\,\pi^{\rm Ce}_{(0,S_b)}   =0,   \label{III6}\\
&& \hspace{-0.75cm}  P^{+,0} + P^{0,+}, \pi^{\rm Ce}_{(S_c-1,1)}  + P^{-,+}\, \pi^{\rm Ce}_{(S_c-2,1)} + P^{-,+2}\, \pi^{\rm Ce}_{(S_c-2,2)}  - \left[  P^{+,0}+ P^{0,+} +P^{-,+}   + P^{-,+2}\right] \nonumber\\
&& \hspace{7cm}  \times \pi^{\rm Ce}_{(S_c-1,0)}   =0,     \label{III7}\\
&& \hspace{-0.75cm}  P^{+,0}\, \pi^{\rm Ce}_{(1,1)} + P^{0,+}, \pi^{\rm Ce}_{(0,2)}   + P^{+,-}\, \pi^{\rm Ce}_{(1,0)}  - \left[  P^{+,0}+ P^{0,+} +P^{0,-}+P^{+,-} \right]\,\pi^{\rm Ce}_{(0,1)}   =0,  \label{III8}\\
&& \hspace{-0.75cm}  P^{+,0} + P^{0,+}, \pi^{\rm Ce}_{(S_c-1,b^*+1)}   +  P^{0,-}\, \pi^{\rm Ce}_{(S_c-1,b^*-1)} + P^{+,-} + P^{-,+}\, \pi^{\rm Ce}_{(S_c-2,b^*+1)} + P^{-,+2}\, \pi^{\rm Ce}_{(S_c-2,b^*+2)}   \nonumber \\
&&  - \left[  P^{+,0}+ P^{0,+} +P^{0,-}+P^{+,-} +P^{-,+} + P^{-,+2}\right]  \, \pi^{\rm Ce}_{(S_c-1,b^*)}   =0,    \,\,\,\,   1 \leq b^* \leq S_b-2,    \label{III9}\\
&& \hspace{-0.75cm}  P^{+,0} + P^{0,+}, \pi^{\rm Ce}_{(S_c-1,S_b)}   +  P^{0,-}\, \pi^{\rm Ce}_{(S_c-1,S_b-2)} + P^{+,-} + P^{-,+}\, \pi^{\rm Ce}_{(S_c-2,S_b-1)}  \nonumber \\
&&\hspace{3.5cm}   - \left[  P^{+,0}+ P^{0,+} +P^{0,-}+P^{+,-} +P^{-,+} \right]\,\pi^{\rm Ce}_{(S_c-1,S_b-1)}   =0,  \label{III10}\\
&& \hspace{-0.75cm}  P^{+,0}   +  P^{0,-}\, \pi^{\rm Ce}_{(S_c-1,S_b-1)} + P^{+,-} - \left[  P^{+,0} +P^{0,-}+P^{+,-} \right]\,\pi^{\rm Ce}_{(S_c-1,S_b-1)}   =0.  \label{III11}\\
\end{eqnarray}}
}
{
Where $P^{+,0}{(e^*,b^*)}$ is the probability of increase by one on the number of CeSCs, $P^{0,+}{(e^*,b^*)}$ is the probability of increase by one in the number of BSCs, $P^{0,-}{(e^*,b^*)}$ is the probability of decrease by one in the number of BSCs, $P^{+,-}{(e^*,b^*)}$ is the probability of increase by one and decrease by one in the number of CeSC and BSCs respectively, $P^{-,+}{(e^*,b^*)}$ is the probability of decrease by one and increase by one in the number of CeSC and BSCs respectively and finally $P^{-,+2}{(e^*,b^*)}$ is the chance of reduction by one in the number of CeSCs and increase by two in the number of BSCs ( These probabilities can be compared with those introduced as the total probabilities of increase or decrease in $e^*$  (8) of Sec. A, and  those of change in $b^*$  as mentioned in Sec. B given in relations (13)).  }
{
\begin{eqnarray}
P^{+,0}(e^*,b^*)  &=& {\rm Prob}_{(e^*,b^*)\rightarrow (e^*+1,b^*)} \nonumber \\
&=& (1-\delta)\, \left(  (1-\gamma)\,\frac{r_1 b^*}{{\cal R}_b} \, \alpha \, \frac{b^*}{b+b^*} + \gamma\, \frac{r_1 e^*}{{\cal R}_c}  \right)\,\frac{e}{e+e^*}, \nonumber  \\
P^{0,+}( e^*, b^*)   &=& {\rm Prob}_{(e^*,b^*)\rightarrow (e^*,b^*+1)}  \nonumber \\
&=& (1-\delta)\, \left[  (1-\gamma)\,\frac{r_1 b^*}{{\cal R}_b}\,\left( (1-\alpha) +  \alpha\, \frac{b^*}{b+b^*}\, \frac{e^*}{e+e^*} +  \alpha\, \frac{b}{b+b^*}\, \frac{e}{e+e^*} \right)  \right.  \nonumber \\
&&  \left.+ \gamma\, \frac{r_1 e^*}{{\cal R}_c}\,\frac{e^*}{e+e^*}  \right] ,  \nonumber \\
P^{0,-}( b^*)  &=& {\rm Prob}_{(e^*,b^*)\rightarrow (e^*,b^*-1)}  = \delta\, \frac{r_1 b^*}{{\cal R}_b},  \nonumber \\[-2mm]
&& \label{Ppm}\\[-2mm]
P^{+,-}( e^*, b^*)   &=& {\rm Prob}_{(e^*,b^*)\rightarrow (e^*+1,b^*-1)}  \nonumber  \\
&=& (1-\delta)\,(1-\gamma)\,\frac{b}{{\cal R}_b}\, \alpha\, \frac{b^*}{b+b^*}\,\frac{e}{e+e^*}, \nonumber \\
 P^{-,+}( e^*, b^*)   &=& {\rm Prob}_{(e^*,b^*)\rightarrow (e^*-1,b^*+1)} \nonumber \\
&=& (1-\delta)\, \left[  (1-\gamma)\,\frac{b}{{\cal R}_b}\, \alpha\,\frac{b}{b+b^*} + \gamma\, \frac{e}{{\cal R}_c}  \right] \,\frac{e^*}{e+e^*} , \nonumber  \\
P^{-,+2}( e^*, b^*)   &=& {\rm Prob}_{(e^*,b^*)\rightarrow (e^*-1,b^*+2)}  \nonumber  \\
&=&(1-\delta)\,(1-\gamma)\,\frac{r_1 b^*}{{\cal R}_b}\, \alpha\, \frac{b}{b+b^*}\,\frac{e^*}{e+e^*} . \nonumber
\end{eqnarray}
}

{
where we assumed that $\lambda_f<1, \lambda_s\neq 0,$ and $\sigma\neq 0$ (for more details see Secs. A and B). Fig. 15(a) represents the dependency of the survival probability on a range of net reproduction rates for mutant cells.  Similar calculation can be performed to derive the survival chance of mutants to fixate in the CeSC group but starting with a recently born mutant in the BSC group.  See Fig. 15-(b) which reveals the survivability of a new mutant in the BSC class inside the CeSC compartment for a variety of different parameter values of $\alpha$. In general, when there is a chance of migration from the BSC group to the CeSC group ($\alpha>0$), the fixation probability of mutants (eventually fixated in CeSC group) slightly reduces of having the initial mutations in CeSC group (see the parameter values of Fig. 15(a)for such a trend). However, it is enhanced in the second case ( Fig. 15(b)) in which the progeny of one BSC mutant become fixated in the CeSC
compartment. Increasing $\alpha$ depicts more effect on the survival chance of mutants starting from the initial mutation within the CeSC group compared with the survivability of mutants in the CeSC compartment when the initial mutant arises in the BSC compartment (compare subfugures (a) and (b) in Fig. 15). Moreover, as the net reproduction rate increase for mutants, the fixation probability's behavior shows a slight decay after a sharp increase and reaching to a maximum level. Collectively, the fixation probability of a new mutant inside CeSCs is much larger than that for the appearance of an imposed mutant inside BSCs.  }

\begin{figure}[h!]
\label{CaseI}
\centering{
\includegraphics[scale=0.6]{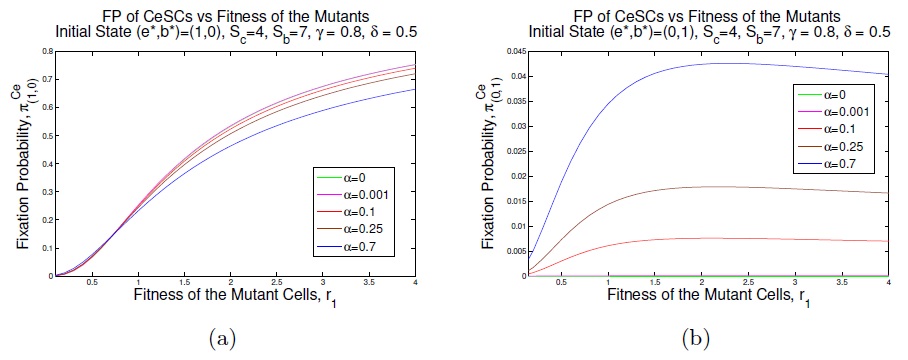}
\caption{ {
{\bf  Fixation mechanism in CeSC compartment  as a consequence of mutation in SC group.}  Assuming $\gamma=0.8, \delta=0.5, S_c=4, S_b=7$ and starting from a newborn malignant cell within (a) CeSC compartment  and (b) BSC compartment, the fixation probabilities $\pi^{\rm Ce}_{(1,0)}$ and $\pi^{\rm Ce}_{(0,1)}$ of the absorption mechanism of mutants inside the CeSc group are given. The calculations are based on equations (43)-(53). The initial conditions $(e^*,b^*)=(1,0)$  and $(e^*,b^*)=(0,1)$ respectively reveal a decrease and an increase in the trend of the fixation probability as $\alpha$, the probability of migration from the BSC group to the CeSC group (when a proliferation occurs within the BSC group) increases. Increasing $\alpha$ may have more influence on the survival chance of mutants inside the CeSC compartment starting with one mutant in the same compartment. in (b), as the reproduction rate of mutants increases, the survival probability decays after reaching to a maximum value. Overall, the fixation probability of a CeSC mutant in the CeSC group is much higher than the fixation probability of a BSC mutant in the CeSC compartment. }  \label{CaseI}}}
\end{figure}
%

\subsection*{ { J. The fixation process of a newborn mutant CeSC or BSC inside  BSC class} }

{
Another interesting scenario to investigate is the fixation procedure in the BSC compartment when the first mutant arises either in CeSC or BSC compartments. To capture the impact of transitions between these two compartments as well as the role of symmetric and asymmetric divisions when there are no immortal cells, we follow the method described in the latter section. Similarly, we investigate the probability $\pi^{\rm BC}_{(e^*, b^*)}$ starting from the states $(1,0)$ and $(0,1)$.  }

{
Our method is very similar to the previous section, where we have considered the initial state as either $(e^*,b^*)=(1,0)$ or $(e^*,b^*)=(0,1)$ to derive the fixation probabilities $\pi^{\rm BC}_{(1, 0)}$ and $\pi^{\rm BC}_{(0, 1)}$ in the BSC group. Thus, we suppose a bi--variable Markov chain to understand the evolutionary dynamics of mutants within CeSC and BSC compartments that eventually tends to the fixation of mutants in BSC group. }

{
Now we find the analytic result for the fixation probabilities $\pi^{\rm BC}_{1,0}$ and $\pi^{\rm BC}_{1,0}$ of starting from one mutation in CeSC and BSC group respectively. Applying a two--variable Markov process of all possible states $(e^*,b^*)$ for $(e^*,b^*)\neq(0,0)$ and  $0\leq e^*\leq S_c-1, 0\leq b^*\leq S_b$, the total number of different possible states is $S_c (S_b+1) -1$. The transition matrix $A$ is then a $S_b (S_c+1) -1$ by $S_b (S_c+1) -1$ dimensional where the entries are the coefficients of $\pi^{\rm BC}_{(e^*,b^*)}$ in the following system: }

{
\begin{eqnarray}
\pi^{\rm BC}_{(e^*,b^*)} = \sum_{{\tilde{e^*},\tilde{b^*}}}\, P_{(e^*,b^*)\rightarrow (\tilde{e^*},\tilde{b^*})}\,\pi^{\rm BC}_{(\tilde{e^*},\tilde{b^*})}. \label{CaseJKol}
\end{eqnarray}
}

{
Considering the matrix $B=\Big[\,(e^*,b^*) \Big]_{0\leq e^* \leq S_c-1, 0\leq b^* \leq S_b}$ as the matrix of all possible states $(e^*,b^*)$ where $(e^*,b^*)\neq (0,0)$, we use an isomorphism between the matrix $B$ and a vector in $\Bbb{R}^{S_b (S_c+1) -1}$ (where the first array $(0,0)$ is ignored) as follows:  }
{
\begin{eqnarray*}
(e^*,b^*) \longrightarrow
(S_b-1) + (e^*-1)S_c+ b^*+1, \hspace{0.5cm} 1\leq b^*\leq S_b-1, 0\leq e^*\leq S_c.
\end{eqnarray*}
}
{
The vector representation of matrix $B$ --where the entry $(0,0)$ is dropped-- can be applied to solve the the system of equations (13).  The initial conditions are $\pi^{\rm BC}_{(0,0)}=0$ and $\pi^{\rm BC}_{(e^*, S_b)}=1$ for any $0\leq e^*\leq S_c$. Calculations show that the corresponding Kolmogorov system of equations comprises similar relations to (43)- (53) by changing the role of variables $e^*$ and $b^*$, and values $S_c$ and $S_b$. Then, this system of equations implies interesting results that are depicted in Figure 4-(a) and (b), where the dependency of the survival probabilities $\pi^{\rm BC}_{(1,0)}$ on a range of net reproduction rates and various migration potentials from BSCs to CeSCs are given.  }

%
Analytic calculation reveals that in the present bi-variable Markov change, the fixation probability of a BSC mutant in BSCs or CeSCs is negligible. The results of the fixation probability in the BSC compartment varying the initial population size of mutants ($b^*=4,5,6$ where $S_b=7$) have been shown in Figure 4 in the paper.
In this figure, when $\alpha$ is large and mutants are disadvantageous ($r_1<1$), mutants migrate to CeSC. Also when $\gamma$ is large enough (as we estimated in Table 1), proliferation mostly occurs in the CeSC compartment. Then, since the fitness of mutant is lower than the fitness of normal cells (the fitness of normal cells is normalized to 1), the probability of division for a mutant  in CeSC becomes very small. Therefore, the mutant will remain in the CeSC group without dividing, or it might migrate to BSC (since each proliferation in CeSC is followed by a migration from CeSC to BSC). This causes delay and decrease in the probability of fixation of mutants in the BSC subpopulation. Note, when $\alpha$ is small, mutants will reside in the BSC and when proliferation occurs in the BSC, then they will have more chance to divide. Furthermore, when one advantageous mutant BSC divides, the probability that the number of mutant BSC increases (when $\delta\approx 0.5$) is less than the probability that the number of mutant BSC
decreases. Consequently, for large values of $r_1$, the difference between probabilities of decrease and increase in the number of mutant BSC will
enforce a reduction in the fixation probability of BSC mutants (see Figure 4, in the paper, for more details).

Figure 4 shows that the increase in the migration rate of BCSs decreases the fixation probability of mutants in the BSC group starting with an imposed mutant in the CeSC population. The major reduction in the fixation probability of malignant individuals can be achieved for lower fitness of mutants. As the fitness grows the difference between curves for various $\alpha$~s deteriorates as all the fixation probabilities decay towards zero. A similar behavior can be observed in Figure 4 for the probability of fixation for mutants starting with a newborn mutant in BSC compartment and when the fitness $r_1 < 1$ is small enough. The trends become reversed as the fitness $r_1 > 1$ increases, where the migration from BSC to the CeSC compartment enhances the survivability of mutants in the BSC compartment in this case.

\subsection*{Time to Fixation and potential therapeutic treatments}

An important concept corresponding to the fixation probability of a given finite Markov chain is the time to fixation. This quantity measures the time that the progeny of a single mutant cell require to take over an entire compartment. The time to fixation can be very important when it approximates the tumor growth period, the time needed for tumor initiation, or clonal conversion. Moreover, it seems crucial to have an estimation for the time of metastasis for an invasive mutant population when the epithelial markers divert to mesenchymal markers in a somatic cancer.

In the current study, assuming the Moran process for a four--compartmental model as described in the analytic tools section, we performed a wide variety of time estimations to maintain some critical features of tumor development within the crypt. We focus our attention on the fixation time of some initial mutant(s) in the central stem cell compartment. In Fig. 6(a) and (b) the average fixation time is given  for different percentages of mutants in the central stem cell compartment opposed to the probability of symmetric division ($\sigma$) for parameters obtained based on the experimental data summarized in Table 1 {where we assume that  the probability of migration from $S_b$ to $S_c$ is negligible ($\alpha\approx 0$)}.

Another interesting result is represented in Fig. 6(c) and (d) in which the fixation time (washed-out time) of central stem cells depicts the number of days it takes for a central stem cell compartment full of wild-type individuals to completely sweep out the rest of the crypt covered by mutants. In these figures different regimes have been considered for neutral and advantageous mutants (various values for $r_1$) and different probabilities of proliferation in the central stem cell group ($\gamma$). Other important observations can be found in the results section of the paper.
\begin{figure}[h!]
\centering{
\includegraphics[scale=0.5]{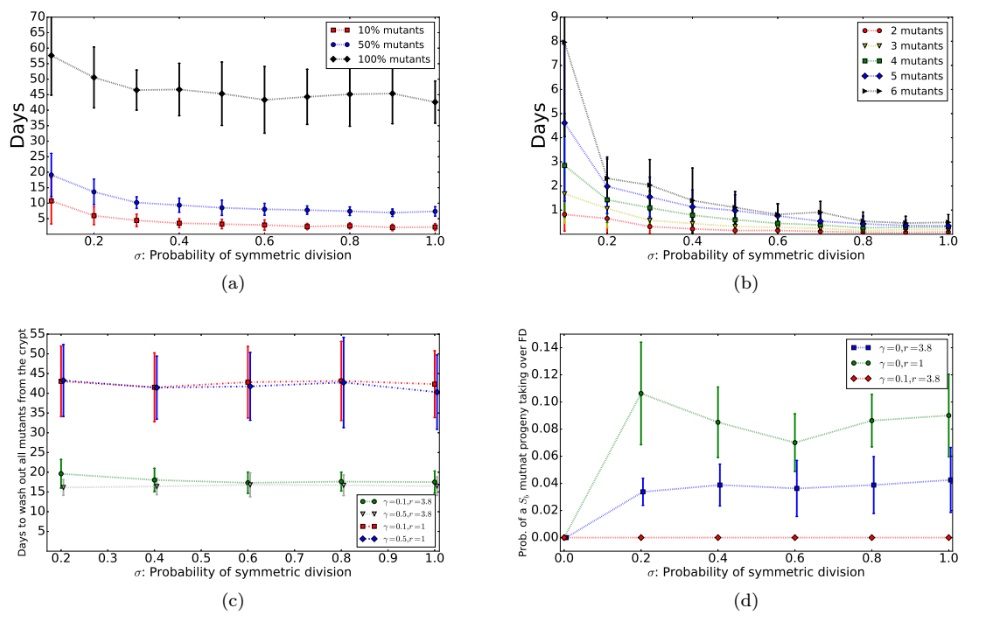}
\caption{ {\bf  (a)-(b) The average spreading time of one mutant central stem cell.} The sub-figure (a) shows the average time that the progeny of one mutant central stem cell will need to take over $10\%$, half, and the entire crypt. The sub-figure (b) shows the average time that one mutant central stem cell needs to generate 2-6 mutant central stem cells. In this figure $S_b=S_c=6$, and $u=v=0$, other parameters are given in Table 1. The points are the average time, and the bars indicate the standard deviations.
{\bf (c) The average time that the progeny of central stem cells need to take the entire crypt.}  At the initial time of this simulation all cells are mutants except central stem cells. We calculate the average time that the crypt evolves, and all cells become wild-type. In this figure $S_b=S_c=6$, and $u=v=0$, other parameters are given in Table 1. The points are the average time, and the bars indicate the standard deviations.
{\bf (d) The probability that the progeny of one mutant stem cell takes over the FD in one and two stem cell compartment models.} In this plots circles and squares indicate the results of simulation for the one-stem cell compartment model, and diamonds are the results of two stem cell compartment model. In these simulations we start the system with one border stem cell mutant, and we obtain the probability that the progeny of the mutant cell takes over the FD group.}
\label{fig:fix_1Sb_in_FD}
}
\end{figure}
%

\subsubsection*{Numerical simulation}

In order to obtain the fixation probability and time to fixation through simulation, we set the maximum updating time $T$ equal to 10,000,000. Then, we run the algorithm for 100 times, and we calculate the ratio of the fixation occurrence number out of 100. We repeat this process for 5 times to obtain the mean and the standard deviation. Moreover, to achieve the time of occurrence, the occurrence time collected for each single run whenever the fixation appeared. Then, we obtained the average and standard deviation of these times. To convert the simulation time to be in terms of day, we assumed that the average cell cycle time of the crypt be equal to one day [48]. This means that having the total number of cells equal to $N$, then the time step $t$ is equivalent to $t/N+1$ days.

\end{document}